\documentclass[a4paper,10pt,notitlepage]{article}
\usepackage[a4paper, left=3.5cm, right=3.5cm, top=2cm]{geometry}
\usepackage{amsfonts,amssymb,amsmath,amsthm,hyperref,colonequals}
\usepackage{caption}
\usepackage{cite}
\usepackage{color}
\usepackage{tabularx}
\usepackage{graphicx}
\usepackage{bbm}
\usepackage{cleveref}
\usepackage[export]{adjustbox}
\usepackage{shuffle}
\usepackage[nice]{nicefrac} 
\usepackage{url}
\usepackage{mathtools}
\theoremstyle{definition}
\newcommand{\bea}{\begin{eqnarray}}
\newcommand{\eea}{\end{eqnarray}}
\newcommand{\be}{\begin{equation}}
\newcommand{\ee}{\end{equation}}

\newcommand\tr{\text{tr}}

\newcommand\id{{\bf 1}}

\begin{document}

\thispagestyle{empty}
\setcounter{page}{0}
\begin{flushright}\footnotesize
\vspace{0.5cm}
\end{flushright}
\setcounter{footnote}{0}

\begin{center}
{\LARGE{
\textbf{Kinetic theory and classical limit for real scalar quantum field in curved space-time} 
}}

{\sc 
Pavel Friedrich$^*$, Tomislav Prokopec$^\dagger$ }\\[5mm]

{\it Institute for Theoretical Physics, Spinoza Institute and the
Center for Extreme Matter and Emergent Phenomena (EMME$\Phi$),\\ 
Utrecht University, Buys Ballot Building,
Princetonplein 5, 3584 CC Utrecht, The Netherlands
}\\[5mm]
\let\thefootnote\relax\footnotetext{\\\texttt{$^*$Electronic address:p.friedrich@uu.nl}\\$^\dagger$
Electronic address:\texttt{t.prokopec@uu.nl}
}

\textbf{Abstract}\\[2mm]
\end{center}
Starting from a real scalar quantum field theory with quartic self-interactions and non-minimal coupling to classical gravity, we define four equal-time, spatially covariant phase-space operators through a Wigner transformation of spatially translated canonical operators within a 3+1 decomposition. A subset of these operators can be interpreted as fluctuating particle densities in phase-space whenever the quantum state of the system allows for a classical limit.
We come to this conclusion by expressing hydrodynamic variables through the expectation values of these operators and moreover, by deriving the dynamics of the expectation values within a spatial gradient expansion and a 1-loop approximation which subsequently yields the Vlasov equation with a self-mass correction as a limit. 
 We keep an arbitrary classical metric in the 3+1 decomposition which is assumed to be determined semi-classically. Our formalism allows to systematically study the transition from quantum field theory in curved space-time to classical particle physics for this minimal model of self-interacting, gravitating matter. 
As an application we show how to include relativistic and self-interaction corrections to existing dark matter models in a kinetic description by taking into account the gravitational slip, vector perturbations and tensor perturbations. 
 

\newpage
\setcounter{page}{1}


\tableofcontents
\addtolength{\baselineskip}{5pt}

\section{Introduction}
For its own sake, it is interesting to understand how the so far most fundamental theory of quantum fields is related to kinetic theory - a description of physics in terms of momentum distributions that is closer to the physics of classical particles. The relation of quantum field theory and kinetic theory has mainly been studied in flat space-times \cite{ trove.nla.gov.au/work/9783845} via the generalization of the non-relativistic Wigner transformation \cite{Hillery:1983ms} to special relativity. 
However, there are few publications on a generalization of this idea to general curved space-times. The early works by \cite{Winter:1986da} and \cite{Calzetta:1987bw} start from an off-shell formulation of two-point functions of the real scalar field and make use of Riemann normal coordinates to obtain a Wigner transformation. Later, reference \cite{Fonarev:1993ht} proposed an off-shell transformation via exponentiated covariant derivatives lifted on the tangent bundle while \cite{Antonsen:1997dc} also proposed to keep the Wigner transform as an operator without taking expectation values.
In this paper, we consider again a covariant Wigner transformation by combining the ideas of the last two references, but this time by using a formulation in terms of canonical quantum field operators that exhibits on-shell closure and that was already proposed within the longitudinal scalar gauge for the metric in \cite{Prokopec:2017ldn}. Thus, we extend our previous work to general curved space-times and derive dynamics for spatially covariant, phase-space operators of the real scalar field, i.e. quadratic operators with a space-time and a spatial momentum dependence. We derive conditions for classical states under which these  phase-space operators describe stochastic distributions of classical particles. The metric is assumed to be derived from the semi-classical Einstein equations, thus fixed through expectation values of the field operators. However, it would not change the formalism within this paper if the metric is assumed to be fixed by an unknown source without any back reaction.
In the hybrid approach in \cite{Prokopec:2017ldn} we assumed a stochastic initial density matrix for these two-point functions to account for stochasticity in cosmological perturbations and we have a similar setting in mind for equations in this paper, although we are focussing primarily on the evolution. 
\par
When studying how classical equations emerge from a quantum field theoretic description, there are two major limits which are a priori of different nature. The first limit concerns the classical stochastic field theory limit of quantum field theory which approximates non-commutating field operators with commuting field operators. At the level of two-point functions, we can also rephrase this limit as having a particle number (the state dependent part) which is much bigger than the quantum contribution originating from the non-commutativity of canonical field operators.\footnote{Note that we are dealing with bosons, a large particle number per momentum for fermions can only be achieved by coarse graining in phase-space.}
It is thus the particle number and not the state independent (vacuum) part of the propagator that dominates loop calculations in this limit.
The classical stochastic field theory limit of quantum field theory should a priori be considered separately from the classical particle limit of quantum field theory which involves an expansion in temporal and spatial gradients with respect to energies and momenta, $\Delta E \Delta t \gg \hbar$ and $\Delta p \Delta x \gg \hbar$. Such a limit is possible after subtracting quantum UV-modes or virtual particles of the Wigner-transformed two-point function which can also be viewed as a special case of coarse-graining. Here, we follow a procedure that is closely related to normal-ordering and involves subtraction of the state-independent part of the two-point function in a normal neighbourhood. The spatial gradient expansion results from assuming that the remaining, state-dependent two-point functions around a collective point on a spatial hypersurface are correlated only in a small neighbourhood relative to spatial gradients taken with respect to that collective point. Thus, the spatial gradient expansion constitutes a separation of spatial scales and it gives rise to an infinite series in the dynamical equations which needs to be truncated. The situation is different for temporal gradients, at least in a one-loop or Gaussian state approximation, since such a state allows for an on-shell closure of the involved two-point functions and takes the form of four first-order differential equations in time. However, enforcing additionally the limit $\Delta E \Delta t \gg \hbar$ for this closed set of equations reduces them to leading order in spatial gradients to the dynamics of non-interacting classical particles, i.e. dynamics described by a collisionless Boltzmann equation or in the context of curved space-time, the Vlasov equation. The effect of self-interactions is analogous to Minkowski space-time where one-loop corrections provide a space-time dependent mass shift \cite{Berges:2015kfa}. Out of the scope of this paper is classical particle scattering as it appears on the right-hand-side of the Boltzmann equation. It can be obtained from the quantum field theory by including two-loop processes which are subsequently approximated with a quasi-particle picture to close on-shell.  
\par
A direct application for the transition from quantum field theory in curved space-time to classical particle physics lies in cosmology and astrophysics. This transition should be studied carefully since in the context of cosmology and astrophysics, physics has many different faces and so it happens that for example dark matter \cite{Ade:2015xua} is -  among many other possible models - a priori believed to be equally well described by a stochastic distribution function of classical non-relativistic, non-interacting, massive particles or a condensate of a stochastic scalar field. The condensate description is easily related to the microscopic scalar field theory whereas the relation between classical particle dark matter to a microscopic theory is less clear and the result of this paper is not only to show that it is indeed well described by a real quantum scalar field but also to systematically keep track of all approximations that lead to the classical particle picture. Maintaining the classical particle picture is a question of scales as we pointed out in the paragraph above and the natural question to ask is whether these scales can be related to other significant scales in the study of large-scale structures (i.e. the scale of non-linearity $k<k_{nl} \sim 0.3~{\rm Mpc}^{-1}$  \cite{Bernardeau:2001qr} or galactic scales $\sim 10 \text{kpc}^{-1}$). Apart from the predictability, fundamental dark matter descriptions may also lead to a transfer of calculational techniques from quantum field theory to make progress on analytical or numerical results in the studies of large-scale structures. \footnote{This point has already been pursued by deriving a Vlasov equation from Wigner transforming the non-relativistic Schr\"odinger-Poisson system for condensates \cite{Widrow:1993qq} \cite{Davies:1996kp}\cite{Uhlemann:2014npa} \cite{Garny:2017xkc} \cite{Mocz:2018ium}. However, the degrees of freedom for one-point functions suffice only to provide an independent mass density and momentum density on the microscopic level. By taking into account coarse-graining, certain momentum distributions may be modelled by exchanging microscopic degrees of freedom below the cut-off for higher moments in phase-space. The connected part of the phase-operators considered in this paper does account for these degrees of freedom without any coarse-graining.} \par

Let us give an overview of the paper.
We will start from an interacting real scalar field theory that is non-minimally coupled to gravity via the Ricci scalar with an arbitrary classical metric $g_{\mu \nu}$ in a 3+1 decomposition and discuss equations of motion and renormalization in the operator formalism. We introduce four composite spatially covariant quantum field operators $\hat{F}_{\phi \phi},\hat{F}_{\Pi \phi},\hat{F}_{\phi \Pi},\hat{F}_{\Pi \Pi} $ by integrating combinations of covariantly translated canonical operators over the tangent space of spatial hypersurfaces.  We rescale them to yield four dimensionally equivalent phase-space density operators $\hat{f}_1^{\pm}(x^{\mu}, p_j)\, , \hat{f}_{2,3}(x^{\mu}, p_j)$ with a dependence on the on-shell momenta. These operators are in fact scalars on the tangent bundle of spatial hypersurfaces for any time. Moreover, we discuss that the state-independent part of these operators should be subtracted in a normal neighbourhood to yield a finite energy-momentum tensor. As a first step towards the classical particle limit, we rewrite hydrodynamic variables like energy density, pressure and fluid velocity in terms of the phase-space operators $\langle \hat{f}_{1,2,3} \rangle$. Afterwards, we derive the dynamics for expectation values $\langle \hat{f}_{1,2,3} \rangle$ in a spatial gradient expansion $\Delta x \Delta p \gg \hbar$ and a one-loop approximation. We then combine two out of these four into the most important phase-space density operator $\hat{f_{1}} = \hat{f}_1^{+} + \hat{f}_1^{-}$ whose expectation value can be related to a classical Boltzmann distribution under certain conditions. Namely, to leading order in the classical particle limit $\Delta x \Delta p \gg \hbar$,  $\Delta t \Delta E \gg \hbar$, the dynamics of the expectation value $\langle \hat{f}_{1}\rangle$ resembles the dynamics of the classical on-shell one-particle phase-space density $f_{\text{cl}} $ for gravitating particles which is given by the truncated BBGKY hierarchy, which is to leading order the Vlasov equation in curved space-time,
\bea
\Big[ p^{\mu} \partial_{\mu} +  p_{\mu} p^{\nu} \Gamma^{\mu}_{\; \nu i} \frac{\partial}{\partial p_i} \Big]  f_{\text{cl}} (x^{\mu}, p_j) &=& 0\, , \\
p^0 (x^{\mu}, p_j) &:=& \sqrt{\big( g^{0j} p_j\big)^2 - g^{00} \big( m^2 + g^{ij} p_i p_j  \big)}\, .
\eea
One could in principle capture higher n-particle distributions by integrating out the gravitational constraint fields. However, here we are mostly interested in giving a field theoretic description of cold dark matter with massive particles where two- and higher n-particle densities $f_{n} (x^0, (x^{i}, p_j)^1,(x^{i}, p_j)^2, ...)$ can be neglected \cite{Bertschinger:1993xt}. 
\par
We work in units where $c=1$ with a mostly plus signature $(-,+,+,+)$.
\pagebreak


\section{Canonical operator formalism in curved space-time}
As opposed to previous approaches for off-shell Wigner transformations in curved space-times \cite{Winter:1986da}\cite{Calzetta:1987bw}\cite{Fonarev:1993ht}\cite{Antonsen:1997dc}, we want to obtain on-shell phase-space densities from the very beginning by working with a Hamiltonian formulation as we have done it for the scalar, linearized longitudinal gauge in \cite{Prokopec:2017ldn}. For convenience we recap the Hamiltonian formulation at the classical level for a massive, real, self-interacting scalar field $\phi$ in curved space-time with metric $g_{\mu \nu}$ in the ADM formalism which is for example discussed in \cite{kiefer2012quantum} or \cite{wald2010general}. We then quantize the matter field and write down the Heisenberg equations for the canonical field operators.
\subsection{ADM decomposition and equations of motion}
We begin by writing down the classical action for the theory 
\be
S_{\text{tot}}\left[ \phi, g_{\mu \nu} \right] = S_{g} \left[ g_{\mu \nu} \right] + S_{m} \left[ \phi, g_{\mu \nu} \right]\, ,
\ee
where the matter action $S_m$ is given by
\be
 S_{m} \left[ \phi, g_{\mu \nu} \right] = -   \int d^4 x \sqrt{-g} \left[ \frac{1}{2} g^{\mu \nu} \partial_{\mu} \phi \partial_{\nu} \phi + \frac{1}{2} \frac{m^2}{\hbar^2} \phi^2 + \frac{1}{2} \xi R \phi^2 + \frac{1}{4!} \frac{\lambda}{\hbar} \phi^4 \right]\, , \label{sMatter}
\ee
and the gravitational action $S_g$ reads
\be
S_{g} \left[ g_{\mu \nu} \right] = \frac{M_P^2}{2 \hbar} \int d^4 x \sqrt{-g} R  \, \label{SemiClassGravityAction} \, . 
\ee
Here, $R$ denotes the four-dimensional Ricci scalar, we have a tree level mass parameter $m^2$ and we allow for a non-minimal coupling as well as a self-interaction given by the tree-level parameters  $\xi$ and $\lambda$, respectively.
We continue by slicing the space-time into spatial hypersurfaces $\Sigma_t$ that are determined by constant values of a four-scalar field $t (x^{\mu})$ whose corresponding vector field  $t^{\mu}$, obeying $t^{\mu} \nabla_{\mu} t = 1$, is given by
\be
t^{\mu} = N n^{\mu}  + N^{\mu} \, , \quad \partial_t = t^{\mu} \partial_{\mu}\, ,
\ee
where $N$ is the lapse function and $N^{\mu}$ is the shift vector such that $n^{\mu}$ is the vector normal to the spatial hypersurface.
We note that $N$ is a four-scalar given by
\be
g (\partial_t , \partial_t ) = - N^2 + N_{\mu} N^{\mu} \,,
\ee
and that
\be
\partial_0 \neq \partial_t \; \text{in general} \, ,
\ee
i.e. we can in principle choose to work with a zero coordinate that is different from our choice of time $t$ used for the slicing into hypersurfaces. 
The next step is to define a projection tensor
\be
\gamma_{\mu \nu} = g_{\mu \nu} + n_{\mu } n_{\nu}\, .
\ee
This allows us the write down the extrinsic curvature associated with our choice of the normal vector field as
\be
K_{\mu \nu} =  - \nabla_{\nu} n_{\mu} - a_{\mu}  n_{\nu} = - \gamma_{\mu}^{\; \alpha} \nabla_{\alpha} n_{\nu} = - \frac{1}{2} \mathcal{L}_n \gamma_{\mu \nu} \, ,
\ee
where $\mathcal{L}_n$ denotes the Lie derivative along $n^{\mu}$ and the acceleration is given by
\be
a_{\mu} = n^{\alpha} \nabla_{\alpha} n_{\mu} = \gamma_{\mu}^{\; \nu} \nabla_{\nu} \log N \, .
\ee
The Ricci scalar can be rewritten as \cite{straumann2012general}
\be
R = {^{(3)} R} + K^2 + K_{\mu \nu} K^{\mu \nu} - \frac{2}{N} \big( \partial_t - N^{\mu} \partial_{\mu} \big) K - \frac{2}{N} {^{(3)}\nabla}_{\mu} {^{(3)}\nabla}^{\mu} N\, ,
\ee
where ${^{(3)}R}$ is the three-dimensional Ricci scalar on spatial hypersurfaces given by
\be 
{^{(3)}R_{\mu \nu \rho}^{\; \quad \sigma}} \big[ \gamma_{\sigma}^{\; \alpha} v_{\alpha} \big] = \big[ {^{(3)}\nabla}_{\mu} {^{(3)}\nabla}_{\nu} - {^{(3)}\nabla}_{\nu} {^{(3)}\nabla}_{\mu} \big] \big[ \gamma_{\rho}^{\; \alpha} v_{\alpha} \big]\, ,\ee
 for some dual vector $v_{\alpha}$ and the covariant derivative on spatial hypersurfaces reads
 \be
 {^{(3)}\nabla}_{\mu} \big[ \gamma_{\nu}^{\; \rho} v_{\rho} \big] =  \gamma_{\mu}^{\; \rho}  \gamma_{\nu}^{\; \sigma}  \nabla_{\rho} \big[ \gamma_{\sigma}^{\; \alpha} v_{\alpha} \big]\, .
 \ee 
 We have already remarked that the zero coordinate $x^0$ and the scalar field $t$ are a priory not related. However, as is often done in the ADM decomposition, we choose the zero coordinate $x^0$ to coincide with $t$,
\be
 x^0 = t \, .
\ee
We then have the following component decomposition of the metric 
\be
g_{\mu \nu} =
\begin{pmatrix}
   - N^2 + N^i N_i & N_i  \\
  N_i & \gamma_{ij} \\
 \end{pmatrix}\, , \; g^{\mu \nu} =
\begin{pmatrix}
    - N^{-2} & N^{-2} N^i  \\
  N^{-2} N^i &  \gamma^{ij} - N^{-2} N^i N^j \\
 \end{pmatrix}\, , \; \sqrt{-g} = N \gamma^{1/2} \, ,
\ee
with
\be
 n^{\mu} = N^{-1}(1, \, - N^i)\,, \quad  n_{\mu} = (-N, \,0)\,  ,
\ee
and $\gamma_{ij}$ being the induced metric on the spatial hypersurface. 
The action for gravity evaluates up to boundary terms\footnote{Since we are only interested in a classical approximation to gravity, these boundary terms can be safely neglected, as they do not influence the dynamics in the semi-classical approximation that we will be using. } to
\be
S_{g} \left[ N, N^k, \gamma_{ij} \right] = \frac{M_P^2}{2 \hbar} \int N dt  \gamma^{1/2} d^3 x \Big[{^{(3)}R} - K^2 + K_{ij } K^{ij}  \Big] \, ,
\ee
where we used
\be
\partial_t \log \gamma^{1/2} = - NK + {^{(3)} \nabla}_i N^i \, .
\ee 
We define the canonical momentum as a classical field configuration by means of the scalar field $t(x^{\mu})$,
\be
\Pi = \frac{\delta S_m}{ \delta \big[ \partial_t \phi \big]} =\sqrt{-g} \frac{n^{\mu} }{N} \partial_{\mu} \phi  = \frac{\gamma^{1/2}}{N} \big[ \partial_t - N^j \partial_j \big] \phi \, ,
\ee
and find for the classical matter action
\be
 S_{m} =    \int N dt  \gamma^{1/2} d^3 x    \left[\frac{1}{2}   \frac{\Pi^2}{\gamma}  - \frac{1}{2}\gamma^{i j } \partial_{i} \phi \partial_{j} \phi  - \frac{1}{2}   \frac{m^2}{\hbar^2} \phi^2  - \frac{1}{2} \xi R \phi^2  -\frac{1}{4!} \frac{\lambda}{\hbar}  \phi^4\right]\, ,
\ee
which is manifestly invariant under spatial coordinate transformations.
 Since we will be dealing mostly with $3+1$ variables in the main parts of the paper, we would like to mention once that it is not the spatial Ricci scalar $^{(3)}R$ but the four-dimensional Ricci scalar $R$ that enters the non-minimal coupling to the matter field $\phi$ and we will sometime refrain from expanding it in a $3+1$ split in order to save space.
\par
We intend to quantize the matter field $\phi$ in a curved space-time with a classical metric $g_{\mu \nu}$ which is an excellent approximation whenever momenta are much smaller than the Planck mass. 
The quantum theory in the operator formalism is formally specified by the time-evolution or the Hamilton operator $\hat{H}$ in \eqref{hamiltonian}, the Heisenberg equations motion \eqref{eqn:phiDot} and  \eqref{eqn:piDot} as well as the equal-time commutation relation \eqref{commRel}. The Hamilton operator $\hat{H}$ is a functional of the canonical (bare) field operators $\hat{\phi}_B$ and $\hat{\Pi}_B$. Moreover, it depends on the bare couplings $m^2_B$, $\xi_B$, ${\lambda_B}$ as well as the classical, possibly stochastic metric $g_{\mu \nu}$ in the $3+1$ split,
\begin{multline}
\hat{H} 
 =\int_{\Sigma_t} N \gamma^{1/2} d^3 x  \Bigg[ \frac{1}{2}  \gamma^{-1} \hat{\Pi}^2_B +  \gamma^{-1/2}{N}^{-1}  \hat{\Pi}_B  {N^j} \partial_{j} \hat{\phi}_B + \frac{1}{2} \gamma^{ij} \partial_{i} \hat{\phi}_B \partial_{j} \hat{\phi}_B +  \frac{1}{2} \frac{m^2_B}{\hbar^2} \hat{\phi}^2_B +  \frac{1}{2}\xi_B R  \hat{\phi}^2_B +\frac{1}{4!} \frac{\lambda_B}{\hbar}  \hat{\phi}^4_B\Bigg] \\ - \hat{\id}\int_{\Sigma_t}N \gamma^{1/2} d^3 x  \Bigg[ \Lambda_B + \kappa_B R + \alpha_{1B} R_{\mu \nu \rho \sigma} R^{\mu \nu \rho \sigma} + \alpha_{2B} R_{\mu \nu} R^{\mu \nu} + \alpha_{3B} R^2  \Bigg]\, , \label{hamiltonian}
\end{multline}
where we refrained from a $3+1$-split of the gravitational counterterms. The counterterms are unavoidable in order to obtain a finite Hamiltonian, once we solve the Heisenberg equations of motion and impose the equal-time commutation relation 
\be
\Big[ \hat{\phi}_B (x^0, x^i)\, , \hat{\Pi}_B(x^0, \widetilde{x}^i) \Big] = i \hbar \delta^{(3)} (x^i - \widetilde{x}^i)\, , \label{commRel}
\ee
where other combinations of canonical fields at equal time commute. 
The Heisenberg equations of motion read
\bea
\mathcal{L}_t  \hat{\phi}_B &=& \partial_t \hat{\phi}_B  =   \frac{N}{\gamma^{1/2} } \hat{\Pi}_B + N^{j} \partial_{j} \hat{\phi}_B \, , \label{eqn:phiDot}\\
\mathcal{L}_t  \hat{\Pi}_B  &=& \partial_t \hat{\Pi}_B + \hat{\Pi}_B  \partial_{\mu} t^{\mu}  = \partial_{j} \Big[ N^{j} \hat{\Pi }_B \Big] +\partial_{i} \Big[ N \gamma^{1/2} \gamma^{ij} \partial_{j} \hat{\phi}_B \Big] \nonumber \\&& \qquad\qquad\qquad\qquad\qquad\qquad\qquad - N \gamma^{1/2} \Big[ \frac{m^2_B}{\hbar^2} \hat{\phi}_B+  \xi_B R \hat{\phi}+ \frac{1}{6} \frac{{\lambda_B}}{\hbar}  \hat{\phi}^3_B \Big]\, . \label{eqn:piDot}
\eea
In covariant notation we find\footnote{Note that
\be
\nabla_{\mu} \Big[ \gamma^{\mu \nu} \nabla_{\nu} \hat{\phi}_B \Big] = {^{(3)}\nabla_{\mu}} {^{(3)}\nabla^{\mu}} \hat{\phi}_B  + a_{\mu}{^{(3)}\nabla^{\mu}} \hat{\phi}_B = {^{(3)}\nabla_{i}} {^{(3)}\nabla^{i} \hat{\phi}_B } + {^{(3)}\nabla_{i}} \log N {^{(3)}\nabla^{i}} \hat{\phi}_B\, .\label{opEqPhi}
\ee} 
\bea
\frac{ n^{\mu}}{N} \nabla_{\mu} \hat{\phi}_B &=&  \frac{\hat{\Pi}_B }{\sqrt{-g}}  \, , \\
\nabla_{\mu} \Big[N  n^{\mu} \frac{\hat{\Pi}_B}{\sqrt{-g}} \Big]  &=& \nabla_{\mu} \Big[ \gamma^{\mu \nu} \nabla_{\nu} \hat{\phi}_B \Big] -  \frac{m^2_B}{\hbar^2} \hat{\phi}_B-  \xi_B R \hat{\phi}_B -\frac{1}{6} \frac{{\lambda_B}}{\hbar}  \hat{\phi}^3_B \, , \label{opEqPi}
\eea
which is equivalent to
\be
\Box \hat{\phi}_B = \nabla_{\mu} \nabla^{\mu} \hat{\phi}_B = \frac{m^2_B}{\hbar^2} \hat{\phi}_B + \xi_B R \hat{\phi}_B +\frac{1}{6} \frac{{\lambda_B}}{\hbar}  \hat{\phi}^3_B \, . \label{opEq}
\ee
In contrast to their classical counterparts, the latter equations exhibit a couple of subtleties of which we have to be aware if we want to formulate phase-space densities that are based on quantum field operators. 
A very important remark we would like to spell out right away is that even the renormalized version of \eqref{opEq} holds strictly speaking only for n-point functions at non-coincident points $x_1, x_2, ... , x_n$ in space-time. The equations of motion do not need to hold for monomials of operators in the coincident limits $x_{i} \rightarrow x_{j} $ due to anomalies that emerge from renormalization (see the summary of section 3.3 in \cite{Hollands:2007zg}) and we will comment more on this anomaly when we discuss the normal ordered energy-momentum tensor entering the semi-classical Einstein equation in the next section.
\par
Renormalizing  n-point functions in the coincident limit or similarly normal ordering the composite operators  appearing within them is unavoidable due to the constraints that have to be imposed on the canonical operators via the commutation relation \eqref{commRel}. These constraints distinguish the quantum field theory from a stochastic field theory whose (commuting) field operators may be given any initial values (formulated in terms of all n-point functions) which are evolved via the analogue of \eqref{opEqPhi} and \eqref{opEqPi} or equivalently of \eqref{opEq} in the stochastic field theory. Apart from the dynamics, which is altered by the non-commutativity, the quantum field theory is thus also different from a stochastic field theory in the sense that their action is constrained to yield a certain value for any n-point function since the commutation relation \eqref{commRel} is independent of any state with respect to which we evaluate these operators. 
We can see this for example  by looking at the two Wightman-functions that solve
\bea
{\Box}_{x} \langle \hat{\phi}_B (x) \hat{\phi}_B (y) \rangle &=& \frac{m^2_B}{\hbar^2} \langle \hat{\phi}_B(x) \hat{\phi}_B (y) \rangle + \xi_B R (x) \langle \hat{\phi}_B (x) \hat{\phi}_B (y) \rangle  +\frac{1}{6} \frac{{\lambda_B}}{\hbar}  \langle \hat{\phi}^3_B (x) \hat{\phi}_B (y) \rangle \, ,\label{eqWightman}
\eea 
where the expectation value refers to an arbitrary state and the same equation holds if the differential operator including the Ricci scalar acts on the other coordinate. No matter which state we choose, the equal-time commutation relation \eqref{commRel} forces us to pick up a bi-solution (solution in both arguments) of equation \eqref{eqWightman} which is singular in the limit $x \rightarrow y$. It also forces us to bestow the Wightman functions with an imaginary part for non-equal times and we note that it is the same singular behavior that yields Greens functions for the Klein-Gordon operator (see e.g. \cite{DeWitt:1960fc}). Still, the state of the system can very well posseses additional non-singular behavior which is exactly the part which is suitable to be described by the kinetic equations we derive below in some approximation. A clear example for this contribution of singular and non-singular behavior to the two-point function is a thermal state in Minkowski space-time which contains the vacuum contribution as well as a finite temperature dependent piece (see e.g. \cite{Quiros:1999jp}).
\par 
Moreover, the distributional nature of quantum fields forces us to renormalize parameters of the theory as soon as composite operators such as $\hat{\phi}^2(x)$ enter physical observables. This becomes apparent if we consider the energy-momentum tensor such that we have to renormalize gravitational couplings as we will review below and it will continue to do so at the level of self-interactions. 
Looking at the operator equations \eqref{opEq}, we can already see that the mass $m_B$, the non-minimal coupling parameter $\xi_B$ and coupling ${\lambda_B}$ will get renormalized since they have to balance the divergent pieces of the composite operator $\hat{\phi}^3$ in the coincident limit which itself may be expressible as a formal series in ${\lambda_B}$ of composite free-field operators.
A better way of writing \eqref{opEq} makes use of a normal ordering procedure that has been developed in the context of algebraic quantum field theory in curved space-time (see \cite{HOLLANDS20151} \cite{Fredenhagen:2014lda} \cite{Brunetti:2015vmh} and references therein, in particular \cite{Hollands:2004yh}). Defining the renormalized field operator $\hat{\phi}$ and the renormalized couplings $m, \xi , \lambda$, equation \eqref{opEq} now reads
\be
\Box \hat{\phi} = \frac{m^2}{\hbar^2} \hat{\phi} + \xi R \hat{\phi} +\frac{1}{6} \frac{{\lambda}}{\hbar} {: \hat{\phi}^3:} \,  \, ,\label{opEqRen}
\ee
where $": (\, . \,) :"$ denotes a normal ordering procedure whose essential ideas are explained in the review \cite{HOLLANDS20151}.
The main observation is that a class of well-defined states - called Hadamard states that cover for example Gaussian states and thermal states - have the same singular behaviour concerning the coincidence limit of their two-point function in the free field limit. The singular behaviour is given in terms of the Hadamard parametrix ${H}(x,y)$ which is a local (normal neighborhood) bi-solution to the free Klein-Gordon equation \eqref{eqWightman} ($\lambda =0$) up to state-dependent terms that remain smooth in the coincident limit, it reads
\be
H(x,y) = \frac{\hbar}{4 \pi^2} \Bigg[ \frac{u(x,y)}{\sigma(x,y) + i 0^+  \tau(x,y)} + v(x,y) \log \Big[ \mu^2 \Big( {\sigma(x,y) + i 0^+ \tau(x,y)} \Big) \Big] \Bigg]\, , \label{hadamardParametrix}
\ee
where the bi-scalar $\sigma(x,y)$ is the signed squared geodesic distance between two points $x,y$ in space-time ($+$ for space-like and $-$ for time-like separations), $\tau(x,y)$ is the difference of some global time function between $y$ and $x$ and $\mu$ is an arbitrary energy scale. Moreover, the bi-scalars $u$ and $v$ are smooth, real valued and depend on the squared mass as well as local  geometric quantities. The bi-scalar $v$ may be written as a formal series in the signed squared geodesic distance $\sigma$ whose coefficients can be determined iteratively \cite{DeWitt:1960fc}. The two-point function $w_2(x,y)$ of any Hadamard state has then locally the form,
\bea
w_2(x,y) &=& \frac{\hbar}{4 \pi^2} \Bigg[ \frac{u(x,y)}{\sigma(x,y) + i 0^+  \tau(x,y)}\nonumber  \\ 
&& \qquad  \qquad + \Big( \sum_{n=0}^N v_n(x,y) \sigma^n(x,y)  \Big) \log \Big[ \mu^2 \Big( {\sigma(x,y) + i 0^+ \tau(x,y)} \Big) \Big] \Bigg] +  R_{N,w} \nonumber \\
&=& H_N (x,y) +R_{N,w} \, , \label{hadamardSplit}
\eea
where $R_{N,w}$ is a smooth, $N+1$-times differentiable remainder that depends on the state.
 Normal ordering of a quadratic monomial of off-shell field operators with $N$ derivatives at the same space-time point is achieved by covariant point-splitting, subtracting the Hadamard parametrix ${H}_N (x,y)$, taking the coincidence limit and fixing a finite number of ambiguities which can be related to the arbitary energy-scale $\mu$. This fixation may be achieved by demanding a certain state or the value of certain renormalized couplings. The deviation from minimal normal ordering (i.e. substracting exclusively terms that diverge in the coincidence limit) is for some monomials necessary to fulfill reasonable requirements as for example stress-energy conservation   \cite{Wald:1978pj}. The latter observation may be understood as a consequence of consistently defining algebraic quantum field theory in curved space-time \cite{Hollands:2004yh}. The procedure of normal ordering quadratic monomials can also be generalized to higher-order monomials and a rigorous definition is given in equation (59) in \cite{Hollands:2004yh}, where it is also discussed that normal ordering obeys the Leibniz rule for off-shell field operators. An anomaly in the free scalar field theory (i.e. failure of the equations of motion to be satisfied for composite operators) is eventually related to the normal ordered operator  $: \hat{\phi} (x) \Big[ \square_x - \frac{m^2}{\hbar^2} - \xi R(x)\Big] \hat{\phi} (x) : \, \propto \hat{\id} Q(x) $ where $Q(x)$ is a classical field constructed from purely geometrical quantities which cannot be set to zero via counterterm ambiguities (a detailed calculation via point-splitting is for example available in \cite{Hack:2010iw}, where, however, the counterterm ambiguities still need to be applied). It is the latter observation that forbids us to enforce the Heisenberg equations for monomials in the coincident limit. It translates to the fact the energy-momentum tensor acquires a trace even if it was classically zero. Moreover, since the above anomaly can be written as an operator identity, it is independent of the state in which one would like to evaluate this operator and thus cannot be argued away for example by choosing a state that has some notion of classicality. It might be negligible but is strictly speaking always present. The whole program of algebraic quantum field theory in curved space-time is then carried forward to include also interactions by defining time-ordered products in order to relate free and interacting field operators via a formal power series in the coupling constant. 
\subsection{Semi-classical Einstein equation and stress-energy renormalization }
The difficulties of quantum field theory in curved space-time are in particular revealed if we ask how to determine the classical metric $g_{\mu \nu}$. The first option is to postulate the metric to have a certain form by means of additional degrees of freedom that couple to the field $\phi$ only indirectly via gravity neglecting any back reaction. 
The second option is to include back-reactions via the renormalized semi-classical Einstein equations,
\be
G_{\mu \nu} \big[ g_{\mu \nu} \big] = \frac{\hbar}{M_P^2}\langle : \hat{T}_{\mu \nu } \big[\hat{\phi}, g_{\mu \nu} \big] : \rangle\, \label{semiClassicalEinstein} ,
\ee
where the normal ordering regularizes the infinite contribution of composite operators such that we are dealing with finite quantities but also with renormalized couplings. The quantum expectation values are taken with respect to some yet unspecified state with possibly stochastic initial conditions (for example in order to account for cosmological setups). Ambiguities in the normal ordering prescription can be interpreted as a change of couplings of a renormalized effective action on the gravitational side. These ambiguities may be fixed by demanding that the left-hand side of the Einstein equation remains in its classical form without a cosmological constant. A standard way to carry out this renormalization is to make use of the effective action that is defined in terms of a path integral which implicitly makes use of a preferred state. This state is unambigious in Minkowski space but fails to be so for general curved space-times. Nonetheless, in the context of slowly-varying space-times one can pick an adiabatic vacuum and calculate the renormalized effective action by methods such as dimensional regularization (this is discussed for example in the standard reference \cite{Birrell:1982ix} as well as the more recent textbook \cite{Parker:2009uva}). 
Thus, we choose the renormalization parameters  for gravity such that they shall neither contain a cosmological constant, nor higher-order geometrical terms other than the four-dimensional Ricci scalar $R$ so that it agrees with the classical action $S_g$ and the renormalized Planck mass is given by $M_P \approx 2.45 \times 10^{18}\, \text{GeV}$ which is another way of phrasing that corrections to the classical Einstein equations without a cosmological constant can safely be neglected at the energy scales that we are preparing experiments at. 
 \par
Keeping in mind that composite operators diverge in the coincidence limit and that we have to be careful evaluating the equations of motion, the energy-momentum operator  in \eqref{semiClassicalEinstein} reads formally
\be
\hat{T}_{\mu \nu} = \partial_{\mu} \hat{\phi}_B \partial_{\nu} \hat{\phi}_B   + \xi_B \big(g_{\mu \nu} \square - \nabla_{\mu} \nabla_{\nu} + R_{\mu \nu} \big) \hat{\phi}^2_B - \frac{g_{\mu \nu}}{2} \Big[ \partial^{\alpha} \hat{\phi}_B \partial_{\alpha} \hat{\phi}_B + \frac{m^2_B}{\hbar^2} \hat{\phi}^2_B +\xi_B R \hat{\phi}^2_B + \frac{1}{12} \frac{\lambda_B}{\hbar}  \hat{\phi}^4_B \Big]  
\, . \label{enMomOp}
\ee
\par
Let us rewrite the energy-momentum tensor given by \eqref{enMomOp} in terms of the canonical field operators and the 3+1 decomposition 
\be
\hat{T}_{\mu \nu} =  \hat{E} n_{\mu} n_{\nu} 
+ \hat{P}_{\mu} n_{\nu}+ \hat{P}_{\nu} n_{\mu} + \hat{S}_{\mu \nu}
  \label{energyMomDecom}\, .
\ee
 One can verify that the non-trivial equation of motion of the canonical field operators is encoded in the bare spatial operator $\hat{S}_{\mu \nu}$. Without going into any details we now take for granted that we have a normal ordering procedure available as we sketched it above and that this procedure  includes perturbative interactions as well. The finite energy, momentum and stress densities with respect to a normal observer,
 \be
 E = \langle : \hat{E} : \rangle\, , \quad P_j = \langle : \hat{P}_j : \rangle\, , \quad S_{jk} = \langle : \hat{S}_{jk} : \rangle\, , 
 \ee
are then according to \eqref{EADM2} to \eqref{SADM2} expressible in terms of the normal ordered equal-time correlators of renormalized fields $\langle :\hat{\Pi}^2 :\rangle$,$\langle  :\big({^{(3)} \nabla_k}\big)^m \hat{\phi} \big({^{(3)} \nabla_j}\big)^{2-m} \hat{\phi} :\rangle$, $\langle {:\hat{\phi}^4 :}\rangle$, ... as well as in terms of the renormalized couplings $m^2$, $\xi$ and $\lambda$. The spatial tensor $S_{jk}$ will always get an anomalous contribution $\gamma_{kj} Q $ after evaluating the normal ordered equation of motion, despite the viewpoint that this anomalous contribution may be safely neglected for a certain choice of a state. We have\footnote{As a cross-check, we verify that also in this $3+1$-split the anomalous trace is indeed given by $Q$ for the configuration $m^2=\lambda=0$ and $\xi=1/6$,
\be
\langle : \hat{T}^{\mu}_{\; \, \mu} : \rangle = S - E  :=  S^{k}_{\;\, k} - E\,, \qquad  \Big(S - E\Big) \Big|_{m^2 = \lambda =0, \, \xi = 1/6} = {Q} \;\Big|_{m^2 = \lambda =0, \, \xi = 1/6}\, .
\ee}
\begin{alignat}{2}
{E} &  =  \frac{1}{2} \Big[\gamma^{-1} \langle : \hat{\Pi}^2  : \rangle+ \langle : {^{(3)}\nabla^{k}} \hat{\phi} {^{(3)}\nabla}_{k} \hat{\phi} : \rangle - 2 \xi {^{(3)}\nabla^{k}}{^{(3)}\nabla}_{k} \langle : \hat{\phi}^2 : \rangle  + \frac{m^2}{\hbar^2} \langle : \hat{\phi}^2 : \rangle \nonumber \\& \qquad - 2 \xi K \gamma^{-1/2}   \Big(   \langle : \hat{\phi}\hat{\Pi} : \rangle +    \langle : \hat{\Pi}\hat{\phi} : \rangle \Big) +\xi \big( {^{(3)} R} + K^2 - K_{ij} K^{ij} \big) \langle :\hat{\phi}^2 : \rangle+ \frac{1}{12} \frac{\lambda}{\hbar} \langle : \hat{\phi}^4 : \rangle \Big] \, ,\label{EADM2} \\
{P}_{j} & =  - \frac{1}{2} \gamma^{-1/2}  \Big[  \langle :\hat{\Pi}  {^{(3)}\nabla}_{j} \hat{\phi}  : \rangle+ \langle : {^{(3)}\nabla}_{j}  \hat{\phi}  \,  \hat{\Pi} : \rangle \Big] + \xi {^{(3)}\nabla}_{j}  \Big[ \gamma^{-1/2} \langle : \hat{\Pi} \,  \hat{\phi} : \rangle +  \gamma^{-1/2} \langle :  \hat{\phi}   \,  \hat{\Pi}: \rangle  \Big]  \nonumber \\ 
 & \qquad \qquad \qquad\qquad \qquad\qquad\qquad   \qquad
 +\xi \Big[  {^{(3)} \nabla^{m}} K_{j m} - {^{(3)} \nabla_{j}} K +    K_{j}^{\; m} {^{(3)} \nabla_{m}}  \Big]\langle : \hat{\phi}^2  : \rangle \, ,\label{PADM2}\\
{S}_{jk} & =      \langle :{^{(3)}\nabla}_{j} \hat{\phi} {^{(3)}\nabla}_{k} \hat{\phi} : \rangle  -  \xi  {^{(3)}\nabla}_{j}{^{(3)}\nabla}_{k} \langle :  \hat{\phi}^2   : \rangle -  \xi   K_{j k} \gamma^{-1/2}  \Big[ \langle : \hat{\Pi} \,  \hat{\phi} : \rangle +    \langle : \hat{\phi}   \, \hat{\Pi} : \rangle \Big] + 2\xi   Q \gamma_{j k}   \nonumber \\
  & \qquad 
   + \xi  \Big[  {^{(3)} R}_{jk} + K K_{jk} - 2  K_{j m}K^{m}_{\; k} - \mathcal{L}_{n} K_{jk} + N^{-1}{^{(3)}\nabla_{j}}{^{(3)}\nabla}_{k}  N  \Big]\langle :\hat{\phi}^2 :\rangle  \nonumber \\
  &   \qquad\qquad 
  - \frac{1}{2}\big(1 - 4\xi \big) \gamma_{j k} \Big[ - \gamma^{-1}  \langle : \hat{\Pi}^2  : \rangle + \langle: {^{(3)}\nabla^{m} }\hat{\phi} {^{(3)}\nabla}_{m} \hat{\phi}: \rangle +  \frac{m^2}{\hbar^2} \langle :  \hat{\phi}^2 :\rangle \nonumber \\ & \qquad\qquad\qquad\qquad \qquad\qquad\qquad\qquad\qquad\qquad\qquad +\xi R \langle :  \hat{\phi}^2 :\rangle + \frac{1}{12}\frac{1  -8 \xi}{1-4\xi} \frac{\lambda}{\hbar} \langle :  \hat{\phi}^4:\rangle \Big]\label{SADM2}
  \, .
 \end{alignat}
It should be clear that the trace ${S}$ appearing in these equations is not to be confused with the total classical action $S_{\text{tot}}$. 
 \par
 As explained below equation \eqref{semiClassicalEinstein} we now choose the renormalized couplings on the left-hand-side of the semi-classical Einstein equation - and thus the normal ordering ambiguities - such, that we are dealing with classical gravity without a cosmological constant.
We then have the expressions found in \cite{rezzolla2013relativistic},
\begin{multline}
\begin{aligned}
 \frac{1}{2} \Big[{^{(3)} R} + K^2 - K_{ij} K^{ij} \Big] & =  \frac{\hbar}{M_P^2} E  \, ,\\
{^{(3)}\nabla_j} K^j_{\; i} -  {^{(3)}\nabla_i} K & = \frac{\hbar}{M_P^2}  {P}_i   \, ,\\
\mathcal{L}_{Nn} K_{ij} + {^{(3)}\nabla_i}  {^{(3)}\nabla_j} N  
-N\Big[  {^{(3)}R_{ij}} +K K_{ij} -2 K_{im}K^{m}_{\; j} \Big]    &=
 \frac{\hbar}{M_P^2} {N} \Big[ \frac{1}{2} \big({S} - {E} \big)\gamma_{ij} -  {S}_{ij} \Big] \, ,
 \end{aligned}
\end{multline}
where we restricted the expressions to spatial indices for tensors in the spatial hypersurface.
As we remarked in the beginning, the split of the two-point function into a part which is singular in the coincident limit (state-independent) and non-singular (state-dependent) part can be read as a split into a manifestly microscopic part inherited from the quantum commutation relation and a part that in principle allows for a macroscopic distribution of particles (among many other possibilities).
Our goal is now to rewrite the quantities $E$, $P_i$ and $S_{ij}$ in terms of integrated phase-space densities which allow for a particle distribution interpretation in certain limits. 
\pagebreak
\section{Wigner operators from canonical fields \label{defWignerSect}}
We have reviewed the Hamiltonian formulation for the real scalar field operator and its conjugate momentum in curved space-time with a classical metric that is given through the semi-classical Einstein equation.
Up to this point we have a description of matter in terms of canonical field operators. We would like to get to a different description by retaining the operator nature and forming a set of transformed objects $\hat{f}_i(x^{\mu}\,, p_i)$ out of the canonical field operators that depend on phase-space variables, where $x^{\mu}$ is a collective space-time point and $p_i$ labels momenta distributed around it. We will construct four such operators $\hat{f}_1^{\pm}$, $\hat{f}_2$ and $\hat{f}_3$. The first two $\hat{f}_{1}^{\pm}$ naturally combine into a single operator $\hat{f}_1$ which may be straightfordwardly interpreted as a fluctuating particle distribution in phase-space under certain conditions. This means that whenever the state, that the operator $\hat{f}_1$ eventually acts on can be characterized as classical, we want to interprete the operator $\hat{f}_1$ as a classical, fluctuating phase-space density in the sense of statistical mechanics. The remaining two phase-space densities $\hat{f}_2$ and $\hat{f}_3$ stem from the relativistic description of the Klein-Gordon equation and do represent degrees of freedom which are absent for classical particle descriptions, so they may be interpreted as giving small backreaction on the operator $\hat{f}_1$ whenever we are in regime where the contributions of $\hat{f}_1$ dominate. We remark, that we see no advantage in reformulating the system in terms of these phase-space operators if the state cannot be characterized as classical. This requirement can be understood by looking at the dynamics of the operators $\hat{f}_{1,2,3}$, which will involve an infinite series of spatial derivatives that needs to be truncated, which is not possible if the state does not allow for a seperation of scales, see also the explanations in \cite{Prokopec:2017ldn}.\par Although we are talking about phase-space \textit{operators} so far, we should note that it is not overly important to retain the operator nature and we will soon drop it by taking expectation values. The reason we mentioned the operator nature in the first place, was to make easier contact to an n-particle distribution for the operator $\hat{f}_1$, such as for example the irreducible two-particle distribution $f_1^{(2)} = \langle \hat{f}_1  \hat{f}_1 \rangle - \langle \hat{f}_1 \rangle \langle \hat{f}_1 \rangle$ which will appear naturally once we switch on interactions or take into account the classical stochastic limit of quantized gravitational perturbations (the role of these higher-order correlators which goes under the name BBGKY hierarchy is discussed for example in \cite{Bertschinger:1993xt} in the context of dark matter). 
However, even if we switch on interactions, we are interested in regimes  where the  higher connected n-point functions are considered to have a small influence on the dynamics  (Gaussian state truncation or resummed 1-loop approximation \cite{Destri:2005qm}),
\be
\langle : \hat{\phi}(x_1)...\hat{\phi}(x_{n+2}) :  \rangle_{\text{connected}} \approx 0 \, , \quad n >2\, .
\ee
This is the case when the self-coupling $\lambda$ multiplied by the number of particles running in the loops is small.
Moreover, we want to consider a state with vanishing one-point functions 
\be
\langle \hat{\phi} \rangle = \langle \hat{\Pi} \rangle =0 \, .
\ee
In principle there is no obstacle in including also one-point functions in the formalism and it is certainly worth studying the influence of condensates. Nonetheless, in order to the keep the scope of the paper focussed we will postpone this discussion.
The reason is that densities which are obtained by Wigner transforming products of one-point functions admit a gradient expansion only after a smoothing procedure \cite{Uhlemann:2014npa} \cite{Garny:2017xkc}, which makes it necessary to deal separately with 
their dynamical equations and the way they react back on the connected part of the two-point function (directly via self-interactions or indirectly via gravity). 
By working with the just mentioned assumptions, we see that the full four-point function entering the energy-momentum tensor becomes
\be
\langle : \hat{\phi}^4  : \rangle \approx 3 \langle : \hat{\phi}^2  : \rangle^2  =   3 \times \includegraphics[width=0.2\linewidth, valign=c]{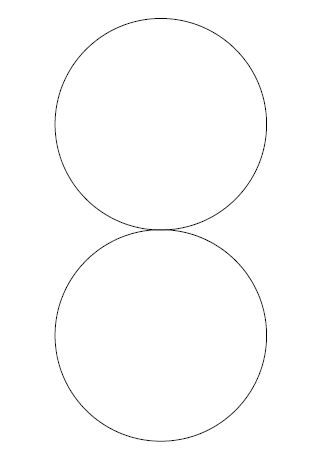}  \, .
\ee
\par
After having discussed our assumptions on the quantum state and the self-interaction, let us now get started by gaining some intuition for the phase-space operators $\hat{f}_i$ that we are after. We know that the energy-momentum tensor for a classical particle distribution in a general relativistic setting is given via second moments for the corresponding classical Boltzmann distribution $f_{\text{class}}$,
\be
T_{\mu \nu}^{\text{class}} (x^{\mu}, p_i) = \int \frac{d^3 p }{\gamma^{1/2}} p_{\mu} p_{\nu} f_{\text{class}}(x^{\mu}, p_i) \, , \quad p_0 = \omega (p_i)\, , \label{classT}
\ee 
where the zero component of the four-momentum is constrained by an on-shell condition. On the other hand we see that, at least in the absence of self-interactions, the energy-momentum tensor of the scalar field theory \eqref{energyMomDecom} is given in terms of quadratic monomials of the canonical operators. It is suggestive to look for some kind of Fourier transform of these quadratic monomials with respect to a shift variable $r^k$ whose conjugate variable $p_k$ may be interpreted as a spatial momentum such that gradient terms in \eqref{energyMomDecom} will contain integrals over these momenta. This spatial Fourier transform is called Wigner transform and it is well known for the special relativistic case with flat metric $\eta_{\mu \nu}$ in terms of a formulation where the zero component of the momentum variable is off-shell (see e.g. \cite{trove.nla.gov.au/work/9783845} for an introduction),
\be
f_{\text{sr}} (x^{\mu},p_{\mu}) \propto \int {d^4r} e^{-\frac{i}{\hbar} r^{\mu} p_{\mu}} \Big\langle \hat{\phi} \Big(x+\frac{r}{2} \Big)\hat{\phi} \Big(x-\frac{r}{2} \Big) \Big\rangle \, . \label{WignerMink}
\ee
It is important to note that taking moments of these densities in $k_0$ will yield more than one density as it was worked out in \cite{Garbrecht:2002pd} for FLRW space-times and the relation to the energy density in \eqref{classT} has been provided there.  \par Furthermore, fully general relativistic Wigner transforms have been proposed at the level of two-point functions for scalar fields using local expansions \cite{Winter:1986da}  \cite{Calzetta:1987bw} as well as non-perturbative expression based on an operator formulation \cite{Fonarev:1993ht} \cite{Antonsen:1997dc}. However, all of the latter four fully covariant proposals are based descriptions that leave the zero component of the momentum off-shell which does not make a closed set of differential equations for all involved degrees of freedom  manifest if we assume a Gaussian state truncation.\footnote{The Gaussian state truncation implies that the effect of interaction via connected higher n-point functions is neglected, if the latter had substantial effect and were taken into account, the system would not close anyway unless a quasi-particle approximation is applied.} In \cite{Prokopec:2017ldn} we were working with linearized gravitational fields in longitudinal gauge and defined equal-time densities via two-point functions of canonical field operator which did not depend on off-shell momenta. 
We concluded that in a large mass, non-relativistic limit only a combination of two out of four two-point functions may be regarded as classical Boltzmann distribution whereas the other two two-point functions are to leading order highly oscillatory and otherwise suppressed. Although these oscillatory densities need not to be observable themselves, they still can have the potential to influence the classical particle density.
\par
In addition to the work of \cite{Winter:1986da}  \cite{Calzetta:1987bw} \cite{Fonarev:1993ht} \cite{Antonsen:1997dc}, it is our goal the make the application of the kinetic representation of quantum field theory for generic curved space-times more feasible by generalizing the description in terms of canonical fields in linearized longitudinal gauge in \cite{Prokopec:2017ldn}  to arbitrary metrices.
This framework would firstly allow us once more to identify the quantity that comes closest to a classical Boltzmann phase-space density (which is non-trivial for a real scalar field in inhomogeneous setups, i.e. it is not simply the integrated version of the various off-shell densities discussed in \cite{Winter:1986da}  \cite{Calzetta:1987bw} \cite{Fonarev:1993ht} \cite{Antonsen:1997dc}) and secondly, to systematically study the effect of highly oscillatory state contributions on the dynamics of the slowly varying part of the state that comes closest to a classical particle description.\footnote{The last viewpoint is similar to the analysis in \cite{Namjoo:2017nia} in which  relativistic correction on the non-relativistic amplitude of an interacting real scalar field in flat space-time have been worked out to yield.}
 
Looking at \eqref{WignerMink}, we see that the main ingredient will be a covariant shift of the canonical fields  which has been worked out by \cite{Fonarev:1993ht} by treating space and time on equal footing. We apply the idea here to our canonical formulation on spatial hypersurfaces.
\par
Before we can present our definition, we introduce the following differential operator on the spatial hypersurfaces $\Sigma_t$,
\be
{^{(3)}{\nabla}_k^{\text{H}}}  := {^{(3)}\nabla}_k - r^l {^{(3)}\Gamma^{n}_{\; k l }} \frac{\partial}{\partial r^n}  \, , \label{horizontalLiftCovDer}
\ee
where ${^{(3)}\nabla}$ is the time-dependent covariant derivative on the spacelike hypersurfaces $\Sigma_t$ and ${^{(3)}\Gamma^{n}_{\; k l }}$ are the associated connection coefficients. The differential operator \eqref{horizontalLiftCovDer} is in fact the horizontal lift of the covariant derivative ${^{(3)}\nabla}$ (induced on $\Sigma_t$ via the 3+1 decomposition) to the tangent bundle $\text{T}\Sigma_t$ (see for example \cite{de2011methods} or \cite{Sarbach:2013uba} for an introduction to induced covariant derivatives on tangent bundles). This covariant derivative satisfies ${^{(3)}{\nabla}_k^{\text{H}}} r^j = 0$ . \par
Let $\hat{X} $ and $\hat{Y}$ denote canonical operators $\hat{\phi}$ and $\gamma^{-1/2} \hat{\Pi}$. 
If we combine a pair of canonical operators $\lbrace \hat{X}_x,\,\hat{Y}_y \rbrace$  into a single operator  $\hat{X}_x \hat{Y}_y$  it will yield a state-independent and moreover UV-divergent part that can be defined in a normal neighbourhood around a collective point. In order to capture this region for operators at equal times, let $\Theta\big[r^k,l_N(x^{\mu}) \big]$ be a cut-off function for the spatial tangent space at $x^{\mu}$ that vanishes for values of $r^k$ which yield a spatial geodesic $s (x^{\mu} , r^k)$ with initial tangent vector $r^k$ emanating from $x^{\mu}$  whose associated distance $||s||(x^{\mu} , r^k)$  is bigger than the radius of a spatial, normal neighbourhood specified by the scalar $l_N (x^{\mu})$ around the point $x^{\mu}$ which much smaller than the scale provided by the curvature but much bigger than a typical momentum scale, $ {(\Delta p)}^{-1}  \ll l_N \ll R^{-1/2}$. Now we define for each pair of spatially separated canonical operators a function that removes the state-independent part of the operator $\hat{X} \big[s (x^{\mu} , r^k/2) \big] \hat{Y} \big[s (x^{\mu} , -r^k/2) \big]$ within a normal region around it, 
\bea
H_{\phi \phi}\big[ x^{\mu}, r^k\big] &:=& H _{\lambda} \big[ y, z \big] \Big|_{y =s(x^{\mu}, r^k/2), \, z = s(x^{\mu}, -r^k/2)} \, ,\\
H_{\phi \Pi}\big[ x^{\mu}, r^k\big] &:=& \big( n_{\nu}  {\nabla^{\nu}} \big)_z H_{\lambda} \big[ y, z \big] \Big|_{y =s(x^{\mu}, r^k/2),\,  z = s(x^{\mu}, -r^k/2)} 
\, ,\\
H_{\Pi \phi}\big[ x^{\mu}, r^k\big] &:=&  \big( n_{\nu}  {\nabla^{\nu}} \big)_y H_{\lambda} \big[ y, z \big] \Big|_{y =s(x^{\mu}, r^k/2),\, z = s(x^{\mu}, -r^k/2)}
\, ,\\
H_{\Pi \Pi}\big[ x^{\mu}, r^k\big] &:=&  \big( n_{\nu}  {\nabla^{\nu}} \big)_y \big( n_{\rho}  {\nabla^{\rho}} \big)_z H_{\lambda} \big[ y, z \big]\Big|_{y =s(x^{\mu}, r^k/2), \, z = s(x^{\mu}, -r^k/2)}\, ,
\eea
where $H_{\lambda}$ has to be computed perturbatively in the self-coupling $\lambda$ in a normal neighbourhood. The free-field limit  $H_{\lambda=0}$ is given by \eqref{hadamardParametrix}.
 We define the associated, spatially covariant, Wigner operator as
\begin{alignat}{2}
\hat{F}_{X Y}(x^{\mu}, p_k) &:=  \gamma^{1/2}(x^{\mu}) \int_{T\Sigma_t}  dr^{3}  e^{-\frac{i }{\hbar}r^k p_k}   \Bigg[ \exp \Bigg( {\frac{r^k}{2}{^{(3)}{\nabla}_k^{\text{H}}}(x^{\mu})} \Bigg) \hat{X} (x^{\mu}) \Bigg]  \Bigg[ \exp \Bigg({-\frac{r^k}{2}{^{(3)}{\nabla}_k^{\text{H}}}(x^{\mu})} \Bigg)  \hat{Y} (x^{\mu}) \Bigg] \nonumber\\ & \qquad\qquad\qquad\qquad
- \hat{1} \gamma^{1/2}(x^{\mu}) \int_{T\Sigma_t}  dr^{3} e^{-\frac{i }{\hbar}r^k p_k}   \Theta\big[ r^j ,  l_{\text{N}}(x^{\mu}) \big] H_{XY}\big[ x^{\mu}, r^k\big] \\ \nonumber
&=:  \gamma^{1/2}(x^{\mu}) \int_{T\Sigma_t}  dr^{3}  e^{-\frac{i }{\hbar}r^k p_k}  \\ & \qquad\qquad\qquad \times  {:\Bigg[ \exp \Bigg( {\frac{r^k}{2}{^{(3)}{\nabla}_k^{\text{H}}}(x^{\mu})} \Bigg) \hat{X} (x^{\mu}) \Bigg]  \Bigg[ \exp \Bigg({-\frac{r^k}{2}{^{(3)}{\nabla}_k^{\text{H}}}(x^{\mu})} \Bigg)  \hat{Y} (x^{\mu}) \Bigg]:}\, ,  \label{genDefWignerOp}
\end{alignat}
with   $X,Y \in \lbrace\phi, \gamma^{-1/2} \Pi \rbrace$ 
and the corresponding expectation values 
\be
{F}_{\phi \phi} = \langle\hat{F}_{\phi \phi} \rangle\,, \quad {F}_{\phi \Pi} = \langle\hat{F}_{\phi \Pi} \rangle\,,\quad {F}_{\Pi \phi} = \langle\hat{F}_{\Pi \phi} \rangle\,,\quad {F}_{\Pi \Pi} = \langle\hat{F}_{\Pi \Pi} \rangle\, .
\ee 
In the definition \eqref{genDefWignerOp}, $H_{XY}$ subtracts the state-independent part in a normal neighbourhood around $x^{\mu}$ and may be viewed as a off-coincident normal ordering.\footnote{See also the inclusion of normal ordering for Minkowski space Wigner transformations in \cite{trove.nla.gov.au/work/9783845} which is not restricted to a normal neighbourhood due to vanishing curvature.} We can similarly view the subtraction as a coarse-graining with respect to quantum UV-modes and it is interesting to note that boundary terms arising from the normal neighbourhood can give rise to noise terms as they appear in stochastic inflation \cite{Starobinsky:1986fx}. Including these type of terms in kinetic equations is however beyond the scope of this paper.
Although a state-independent, off-coincident normal ordering operation is discussed in the context of algebraic quantum field theory for free fields by means of the Hadamard parametrix \cite{HOLLANDS20151}, we are not aware that this definition is extended to interacting fields at the rigorous level that algebraic quantum field theory operates on. However, we think it is important to signal that such a procedure is necessary to obtain operators that describe real particle fluctuation and exclude virtual particles. Moreover, normal ordering is clearly demanded if we take moments in the momenta $p_k$ of \eqref{genDefWignerOp} which would otherwise result in coincident limit two-point functions that are divergent, instead we have for example
\be
\frac{1}{(2 \pi \hbar)^3}\int \frac{d^3 p}{\gamma^{1/2} } \hat{F}_{\phi \phi} (x^{\mu}, p_k) = {: \hat{\phi}^2:} \,(x^{\mu})\, , 
\ee
which is crucial if we want to rewrite the energy-momentum tensor in terms of $\hat{F}_{\phi \phi}$, $\hat{F}_{\Pi \phi}$, $\hat{F}_{\phi \Pi}$, $\hat{F}_{\Pi \Pi}$.

On the other hand, the details of the normal ordering procedure should not affect the effective description that we are after whenever infrared and ultraviolet physics decouple which concretely amounts in this paper to neglect firstly, anomalous contributions of the matter two-point functions, secondly, boundary terms of the state-independent part due to the normal region and thirdly, 2-loop corrections that will again pick up quantum contributions running in the loop.
\par
Let us discuss the other ingredients appearing \eqref{genDefWignerOp}. We verify in appendix \ref{proofExpCovDer} that powers of the spatially covariant shift operator ${\frac{r^k}{2}{^{(3)}{\nabla}_k^{\text{H}}}}$ yield the following when acting on scalar densities  $f(x^{\mu})$ with weight zero, 
\be
\Big[ { {r^k}{^{(3)}{\nabla}_k^{\text{H}}}} \Big]^n  f =  {r^{i_1}...r^{i_n} } {^{(3)}{\nabla}_{i_1}}... {^{(3)}{\nabla}_{i_n}} f = \Bigg[{ {r^k} \Bigg( \partial_k -  {^{(3)}\Gamma^m_{\; kl}} r^l \frac{\partial}{\partial r^m} }  \Bigg)\Bigg]^n f  \, , \label{horLiftOnfunc}
\ee
which allows us to consider the definition \eqref{genDefWignerOp} even without the introduction of geometrical objects on the tangent bundles  $\text{T}\Sigma_t$. We also realize that any change of spatial coordinates in \eqref{genDefWignerOp}  can be absorbed into the integration variable $r^k$ that then transforms as a 3-vector and leaves the measure invariant thanks to the spatial determinant factor. The conjugate momentum $p_k$ then transforms as a covariant 3-vector. \par Equation \eqref{horLiftOnfunc} also reveals, that the covariant shift operators reduce to spatial translations when working with Riemann normal coordinates since only symmetrized covariant derivatives enter. It shows that the exponentials acting in \eqref{genDefWignerOp} translate the operators $\hat{X},\hat{Y}$ from point $x^{\mu}$ to the point specified by the spatial geodesic emanating at $x^{\mu}$ with tangent vector $r^k$ in opposite directions. Taking expectation values of the operator $\hat{f}_{XY}$ will then contain the information in the momentum space representation on how $\hat{X}$ and $\hat{Y}$ are correlated around the collective point $x^{\mu}$ where the normal ordering takes care that UV correlations that are purely quantum are removed up to boundary terms. However, the cut-off related to the normal neighbourhood around $x^{\mu}$ should have small effect as long as the state-dependent field correlations are restricted to regions much smaller than the inverse curvature. The gradients with respect to the  center coordinate $x^{\mu}$ then quantify how this correlation changes in space-time.
\par
Given the canonical operators of the real scalar field theory, we have four different Wigner operators $\hat{F}_{\phi \phi}$, $\hat{F}_{\Pi \phi}$, $\hat{F}_{\phi \Pi}$, $\hat{F}_{\Pi \Pi}$ whose dynamics are determined by the dynamics of the operators $\hat{\phi}$ and $\hat{\Pi}$. Unfortunately, the calculation is tedious and some techniques to perform it have to be introduced. Let us therefore make some easier observation before that and save the difficulties for later. 
\par
We observe that the operators $\hat{F}_{\phi \phi}$, $\hat{F}_{\Pi \phi}$, $\hat{F}_{\phi \Pi}$, $\hat{F}_{\Pi \Pi}$ are dimensionally inequivalent and not all of them are real. In order to rescale the Wigner operators in units of energy, we consider the free particle energy via the 3+1 decomposition
\be
\omega_p = N p^0 = \Big( m^2 + \gamma^{kl} p_k p_l \Big)^{1/2}\, ,
\ee and define the following dimensionally equivalent phase-space density operators and the corresponding expectation values
\begin{alignat}{2}
\label{deff+} {f}_{1}^{+} &= \langle \hat{f}_{1}^{+} \rangle  &:=& \; \frac{1}{(2 \pi \hbar)^3}\frac{1}{2 \hbar} \Big[ \frac{\omega_p}{\hbar} \langle \hat{F}_{\phi \phi} \rangle  +  \frac{\hbar}{\omega_p} \langle  \hat{F}_{\Pi \Pi} \rangle\Big] \, ,\\
\label{deff-} {f}_{1}^{-} &=\langle  \hat{f}_{1}^{-} \rangle &:= &\; \frac{1}{(2 \pi \hbar)^3}\frac{i}{2\hbar} \Big[ \langle\hat{F}_{ \Pi \phi }\rangle -  \langle\hat{F}_{ \phi \Pi } \rangle \Big]\,, \\
\label{deff1}{f}_{2} &= \langle\hat{f}_{2} \rangle &:=&\; \frac{1}{(2 \pi \hbar)^3}\frac{1}{2 \hbar} \Big[ \frac{\omega_p}{\hbar} \langle  \hat{F}_{\phi \phi}\rangle - \frac{\hbar}{\omega_p} \langle  \hat{F}_{\Pi \Pi} \rangle\Big] \, , \\
\label{deff2}{f}_{3} &=  \langle\hat{f}_{3}\rangle &:=&\; \frac{1}{(2 \pi \hbar)^3} \frac{1}{2\hbar} \Big[\langle  \hat{F}_{ \Pi \phi }\rangle + \langle  \hat{F}_{ \phi \Pi }\rangle \Big]\, .
\end{alignat}
We note that the phase-space density operators $\hat{f}_1^{+}$, $\hat{f}_2$ and $\hat{f}_3$ are even functions of the momentum $p_k$ whereas $\hat{f}_1^-$ is an odd function of the momentum. From here on we will work mostly with expectation values of operators which is clarified by omitting the hats.
\par
Making use of delta functions, setting the connected four-point functions to zero, dropping the anomalous contribution and boundary terms, we can express the energy-momentum tensor in terms of the phase-space densities \eqref{deff+} to \eqref{deff2} as follows,
\begin{alignat}{2}
{E} \;  &=  \int \frac{d^3 p}{\gamma^{1/2}} \omega_p \, {f}_1^+ - 2 \xi  \hbar K \int \frac{d^3 p}{\gamma^{1/2}} f_3  
+\frac{\hbar^2}{8} \Big[1  -8 \xi \Big] {^{(3)}\nabla_k}{^{(3)}\nabla^k}\int \frac{d^3 p}{\gamma^{1/2}} \frac{ {f}_1^+ + {f}_2 }{\omega_p} \nonumber \\   &\qquad
+\xi \frac{\hbar^2 }{2} \Big( {^{(3)}R} +K^2 - K_{ij} K^{ij}  \Big)\int \frac{d^3 p}{\gamma^{1/2}} \frac{ {f}_1^+ + {f}_2 }{\omega_p}
+  \lambda \frac{\hbar^3}{8} \Bigg[ \int \frac{d^3 p}{\gamma^{1/2}} \frac{ {f}_1^+ + {f}_2 }{\omega_p} \Bigg]^2 \, ,\label{eMomDecom1} \\
\nonumber {P}_{k} \;   &=  \int \frac{d^3 p}{ \gamma^{ 1/2}} {p_k} {f}_1^{-} - 
 \frac{\hbar}{2} \big[1-4 \xi  \big] {^{(3)}\nabla_k} \int \frac{d^3 p}{ \gamma^{ 1/2}} {f}_3 
\\&\qquad + \xi \hbar^2 \Big[  {^{(3)} \nabla^{m}} K_{j m} - {^{(3)} \nabla_{j}} K +    K_{j}^{\; m} {^{(3)} \nabla_{m}}  \Big] \int \frac{d^3 p}{\gamma^{1/2}} \frac{ {f}_1^+ + {f}_2 }{\omega_p}  \, ,\label{eMomDecom2}\\ \nonumber 
\qquad{S}_{km} \;   &=    \int  \frac{d^3 p}{\gamma^{ 1/2}}  \frac{p_k p_m}{\omega_p}\big({f}_1^+  + {f}_2  \big) - 2 \xi \hbar K_{km}\int \frac{d^3 p}{ \gamma^{ 1/2}} {f}_3  +   \frac{\hbar^2}{4} \Big[1-4 \xi \Big] {^{(3)}\nabla_k}{^{(3)}\nabla_m} \int \frac{d^3 p}{\gamma^{1/2}} \frac{ {f}_1^+  + {f}_2  }{\omega_p}  \\ \nonumber
& \quad -\gamma_{km} \big[1- 4 \xi \big]\Bigg[\int \frac{d^3 p}{\gamma^{1/2}} \omega_p \,  {f}_2   + \frac{\hbar^2}{8} {^{(3)}\nabla^j}{^{(3)}\nabla_j}\int \frac{d^3 p}{\gamma^{1/2}} \frac{ {f}_1^+  + {f}_2  }{\omega_p}  \Bigg]  \\ 
& \quad  \nonumber 
+ 2 \xi \hbar  \Big[ R \gamma_{km} + {^{(3)} R}_{km} + K K_{km} - 2  K_{k j}K^{j}_{\; m} - \mathcal{L}_{n} K_{km} + N^{-1}{^{(3)}\nabla_{j}}{^{(3)}\nabla}_{k}  N  \Big]\int \frac{d^3 p}{\gamma^{1/2}} \frac{ {f}_1^+  + {f}_2  }{\omega_p} 
\\ 
& \quad  
- \gamma_{km}\lambda \frac{\hbar^3}{2} \big[1-8 \xi \big] \Bigg[ \int \frac{d^3 p}{\gamma^{1/2}} \frac{ {f}_1^+ + {f}_2 }{\omega_p} \Bigg]^2
\, , \label{eMomDecom3}
\end{alignat}
where we used relations of the type 
\be
{:\partial_i \hat{\phi}  \partial_j \hat{\phi}:} = \frac{1}{4} {{^{(3)}\nabla_i}{^{(3)}\nabla_j} {:\hat{\phi}^2}:}  + \int \frac{d^3 p}{(2 \pi \hbar)^3} \gamma^{- 1/2} \frac{p_i p_j}{\hbar^2} \hat{F}_{\phi \phi}\, .
\ee
We can also write down energy-momentum conservation $\nabla_{\mu} \langle : \hat{T}^{\mu \nu}: \rangle  = 0$ in terms of this 3+1 decomposition (see for example \cite{rezzolla2013relativistic}),
\be
\partial_t \big(\gamma^{1/2} {E} \big) + \partial_i \big[\gamma^{1/2}\big(N {P}^i - N^i \hat{E} \big) \big] = N \gamma^{1/2} \big( S_{ij} {K}^{ij} - {P}^i \partial_i \ln N \big) \, , 
\ee
\be
\partial_t \big(\gamma^{1/2} {P}_j \big) + \partial_i \big[\gamma^{1/2}\big(N {P}^i_{\, j} - N^i {P}_j \big) \big] = N \gamma^{1/2} \big( \frac{1}{2} {S}^{ik} \partial_j \gamma_{ik} + N^{-1}{P}_i \partial_j N^i  - {E} \partial_j \ln N \big) \, ,
\ee
which however does not help very much since the involved quantities are still constrained via the phase-space densities $f_{i}$. We thus have to know their dynamics which we will postpone to a later section as promised.
\par 
Let us see what we can read off from the decomposition \eqref{eMomDecom1} to \eqref{eMomDecom3}. By writing $\hat{f}_{1} =\hat{f}_{1}^{+} + \hat{f}_{1}^{-}$ and using the parity properties of these operators , we arrive at the form
\begin{alignat}{2}
E &= \int \frac{d^3 p}{\gamma^{1/2}} \omega_p f_1 + \hbar^2  \mathcal{O} \big( f_1 \label{EClass}
\, \big)+  \hbar^2  \mathcal{O} \big( f_2 \big)+ \mathcal{O} \big(  \xi, {\lambda} \big)\, , \\
P_k &=   \int \frac{d^3 p}{ \gamma^{ 1/2}} {p_k} f_1 +  \hbar \,  \mathcal{O} \big( f_3 \big)  + \mathcal{O} \big( \xi \big) \, ,\label{PClass} \\
{S}_{k m} &=    \int \frac{d^3 p}{\gamma^{ 1/2}}  \frac{p_k p_m}{\omega_p} f_1
+ \mathcal{O} \big(f_2  \big) 
+ \hbar^2  \mathcal{O} \big(f_1  \big)+ \mathcal{O} \big( \xi, {\lambda} \big)\, .\label{SClass}
\end{alignat}
We can do the same 3+1 projection with the classical energy-momentum tensor \eqref{classT} whose building blocks can be fluctuating phase-space densities that still need to be averaged over in statistical context. Since we could have written down the equautions \eqref{EClass} to \eqref{SClass} also at the level of renormalized operators, we can tentatively identify the operator $\hat{f}_{1}$ as a fluctuating phase-space density at the level of the normal projected energy-momentum tensor, up to certain correction terms. The average $f_1$ is viewed as the one-particle distribution in phase-space.
Let us discuss under which conditions this identification is justified.
 \par The first conditions concerns a spatial gradient expansions  proportional to the Planck constant $\hbar$ where spatial gradients with respect to the variable $x^i$ are compared to either the energy $\omega_p$ or the momentum $p_k$ within spatial momentum integrals of a phase-space density $\langle \hat{f}_i(x^{\mu}, p_j) \rangle$. If we picture a non-relativistic setting where $m \gg |p_k|$ for any $f_i(x^{\mu}, p_j)$, we see that the gradient expansion is applicable if the energy-scales satisfy $m \gg \Delta p \approx \frac{\hbar}{\Delta r} \gg \frac{\hbar}{\Delta x}$.\footnote{We commented more on this expansion in the context of dark matter in \cite{Prokopec:2017ldn}.}
The relation between the short distance difference scale $\Delta r$ and the long distance, center coordiate scale  $\Delta x$ lies at the heart of the Wigner transformation. In the context of general relativity, it corresponds to locally homogeneous two-point functions depending only on the (covariantly generalized) difference coordinate of the involved operators subsequently yielding only a momentum dependence around $\Delta p$ which is then corrected on larger scales $\Delta x$ via gravitational inhomogeneities (plus additional effects due to self-interactions). We underpin once more, that it depends on the state in Hilbert space whether or not these corrections are small since the higher-order spatial gradient corrections are strictly speaking always present. 
Typical correction terms in the dynamics of phase-space operators will include
\be
\mathcal{O} \big( \hbar \big) {f}_{i} (t,x, p) \sim \Big\lbrace \hbar  \partial_k \frac{\partial}{\partial p_k}, \,\frac{\hbar \partial_k}{\sqrt{m^2 + \gamma^{ij} p_i p_j }}  , \,  \frac{\hbar}{{p^k \partial_k}}{{^{(3)}\square}}, \, ... \Big\rbrace {f}_{i} (t,x, p) \, .
\label{spatialGradientApprox}
\ee 
Another obvious condition that should be satisfied in order to treat the operator $\hat{f}_{1}=\hat{f}_{1}^{+} + \hat{f}_{1}^{-}$ similarly to a classical, particle-associated fluctuating phase-space density concerns the expectation values of the two operators $\hat{f}_{2,3}$ appearing in \eqref{EClass} to \eqref{SClass}. If we wanted an averaged energy-momentum tensor from the field theoretic description that looks almost identical to the one obtained from an averaged classical particle description, the densities $f_{2,3}$ would have to be chosen small initially. 
Note that the assumption that $f_1^{+}$ should be regarded as the dominant density in comparison with $f_2$ is supported by observation that
\be 
\int \frac{d^3 p}{\gamma^{1/2}} \omega_p f_1 \geq \left| \int \frac{d^3 p}{\gamma^{1/2}} \omega_p f_2\right| \, , 
\ee
which follows from their very definition. The bound can be hit for example for homogeneous condensates $\langle \hat{\phi} \rangle (t) \propto \sin (mt/\hbar)$. On the other, only the density $f_1^{-}$ can fulfil the job as a classical particle phase-space density since it is the only odd density and thus clearly favoured in comparison to $f_3$ in \eqref{PClass}. 
However, even if we make the identification \eqref{EClass} to \eqref{SClass} initially, we have to be sure that the dynamics keeps the influence of the fluctations ${f}_{2,3}$ small over time which relates to requirements on the parameters of the theory ($m,\xi, \lambda$). It is clear that we expect for example from a strongly interacting regime ${\lambda} \gg 1$ many more effects than a mass renormalization and pressure correction, since the Gaussian state approximation breaks down and higher n-point functions enter the dynamics.
\par
Let us summarize what we have found so far. We have provided a spatially covariant set of three even and one odd  quadratic equal-time  operators and their expectation values \eqref{deff+} to \eqref{deff2} that have units of phase-space densities and that depend on a space-time point $x^{\mu}$ and spatial three-momentum $p_i$. There is no dependence on a off-shell zero-momentum component. By looking at the 3+1 decomposition of the energy-momentum tensor \eqref{EClass} to \eqref{SClass}, we identified a distinguished combination of one even and the odd operator $\hat{f}_{1} = \hat{f}_{1}^{+}+\hat{f}_{1}^{-}$ which appears to mimic a fluctuating, classical phase-space density in the sense of statistical mechanics whenever it is acting on a state that is classical enough such that it admits a spatial gradient expansion. The remaining two even operators $\hat{f}_{2,3} $ represent degrees of freedom that stem from the fundamental relativistic, field theoretic description and we have argued that the expectation values ${f}_{2,3} $  should taken to be small for a purely particle-like interpretation. However, they can in principle have a significant role for the evolution of the system and it is worth studying how such additional components from the field theoretic description correct the classical particle picture. 
\section{Hydrodynamic cold dark matter limit: from normal observers to fluid rest-frame}
Our original motivation to identify the phase-space densities $f_{1,2,3}$ was to study cold dark matter from a field theoretic description  that allows for a systematic inclusion of relativistic effects \cite{Prokopec:2017ldn}. This section is devoted to making a closer contact to a cold dark matter description that is formulated in terms of hydrodynamic variables. We want to show as a proof of concept how the hydrodynamical variables, that are used for the classical particle description, can be derived from the theory of scalar quantum field in a certain classical limit. It turns out that this map is already non-trivial even at the level of vanishing self-interaction and minimal coupling to gravity which is why we stick to the simplified case $\lambda = \xi =0$ in this section.
We were discussing the energy-momentum tensor in a 3+1 decomposition. The projected quantities 
\be
\langle :\hat{E}: \rangle = E \,,  \quad \langle: \hat{P}_k: \rangle = P_k \,,  \quad \langle: \hat{S}_{kl}: \rangle = S_{kl}  \, ,
\ee
that appear in the energy-momentum tensor \eqref{eMomDecom1} to \eqref{eMomDecom3} are related to the observer specified from any other frame by the normal vector $n^{\mu}$. This normal (also referred to as Eulerian) observer measure an energy density $E$, a momentum $P_i$ and a stress tensor $S_{ij}$. 
However, especially in the context of cosmology it is standard to work with a different decomposition that assumes a hydrodynamic representation of energy-momentum which relates to an observer co-moving with the fluid.  The fluid is specified from any other frame by the four-velocity $u^{\mu}$ that corresponds to an observer moving with a fluid element and the energy-momentum tensor for a hydrodynamic representation is usually written as 
\be
\langle : \hat{T}_{\mu \nu}: \rangle \equiv T_{\mu \nu} = \big( e + P  \big) u_{\mu} u_{\nu} +  P g_{\mu \nu} + \pi_{\mu \nu}\, , \quad u^{\mu} \pi_{\mu \nu} = 0 \, , \quad \pi^{\mu}_{\; \mu} = 0\, .
\ee
In this formulation one assumes that energy and momentum are expressible in terms of a fluid with rest-frame energy density $e$, pressure $P$ and non-isotropic stresses $\pi_{ij}$. The energy density $e$ that is measured by the observer moving with the fluid is in general different from the energy density $E$ measured by the normal observer. 
We remark that
\be
u^{\mu} = -n^{\nu}u_{\nu} \big( n^{\mu} + v^{\mu} \big) = W \big( n^{\mu} + v^{\mu} \big) \, , \quad W = \frac{1}{\sqrt{1-v^i v^j \gamma_{ij}}}\, ,
\ee
where $v^{\mu}$ is the spatial part of the four-velocity with respect to the normal vector and $W$ is the Lorentz factor \cite{rezzolla2013relativistic}.
We then have the following relations
\bea
\label{S}
E &=&  W^2 \big( e  + \gamma^{ij} v_i v_j P \big) + \pi_{ij}{v^i v^j} \, ,\\
\label{S_i} {P}^i  &=& \big( e + P \big) W^2 v^i + \pi_{kl} v^k \gamma^{li}\, , \\
\label{S_ij} {S}^{ij}  &=& (e+P)W^2 v^i v^j  + P \gamma^{ij} + \gamma^{ik} \gamma^{jl} \pi_{kl} \, .
\eea
We would now like to invert the relations \eqref{S} to \eqref{S_ij}, which is in principle a complicated task. However, it can be done in principle exactly and we will do it for the case of the real scalar field stress-energy tensor, where we set the self-coupling ${\lambda}$ and the non-minimal coupling to gravity $\xi$ for simplicity to zero,
\be
\langle : \hat{T}_{\mu \nu} : \rangle_{{\lambda}= \xi =0} = \langle : \partial_{\mu} \hat{\phi} \partial_{\nu} \hat{\phi} : \rangle - \frac{g_{\mu \nu}}{2} \Big[ \langle : \partial^{\alpha} \hat{\phi} \partial_{\alpha} \hat{\phi}: \rangle + \frac{m^2}{\hbar^2} \langle : \hat{\phi}^2 : \rangle \Big] \, . \label{TMuNuScalar}
\ee
It turns out, that it is more convenient at this point to first work  without any time-slicing and define the object
\be
 \chi^{\mu}_{\; \nu} := \langle : \partial^{\mu} \hat{\phi} \partial_{\nu} \hat{\phi} : \rangle \, ,
\ee
which is the key ingredient, if we want to consider non-perfect fluids.
The energy density is the negative eigenvector of the fluid four-velocity, whereas the pressure is one third of the sum of the principal stresses, which are eigenvalues belonging to the spatial part of the energy-momentum tensor.
The task is thus to find the eigenvalues of energy-momentum \eqref{TMuNuScalar}, which amounts to finding the eigenvalues of $\chi^{\mu}_{\; \nu}$, which is initially for every point in space an arbitrary matrix that obeys the Cayley-Hamilton equation
\begin{multline}
\big[\chi^4 \big]^{\mu}_{\; \nu} - \tr \big[ \chi \big]  \big[\chi^3 \big]^{\mu}_{\; \nu} + \frac{1}{2} \Big[\big( \tr \big[ \chi \big] \big)^2 -  \tr \big[\chi^2 \big]   \Big]\big[\chi^2 \big]^{\mu}_{\; \nu} \\ - \frac{1}{6} \Big[ \big(\tr  \big[ \chi\big] \big)^3 - 3  \tr \big[ \chi^2 \big]  \tr \big[ \chi \big] + 2 \tr \big[ \chi^3 \big] \Big] \chi^{\mu}_{\; \nu}  +  \delta^{\mu}_{\; \nu}  \det \chi = 0\, .
\end{multline}
The eigenvalues ${\sigma}$ of this matrix are then subject to the quartic equation
\be
{\sigma}^4 + \widetilde{b}  {\sigma}^3 + \widetilde{c}  {\sigma}^2 + \widetilde{d}  {\sigma} + \widetilde{e}  = 0\, ,
\ee 
where
\bea
\widetilde{b} &=& - \tr \big[ \chi \big]   \, , \\
\widetilde{c} &=& \frac{1}{2} \Big[\big( \tr \big[ \chi \big] \big)^2 -  \tr \big[\chi^2 \big]   \Big] \, , \\
\widetilde{d} &=& - \frac{1}{6} \Big[ \big(\tr  \big[ \chi\big] \big)^3 - 3  \tr \big[ \chi^2 \big]  \tr \big[ \chi \big] + 2 \tr \big[ \chi^3 \big] \Big] \, , \\
\widetilde{e} &=&   \det \chi  \, .
\eea
We express these traces in terms of two-point functions of canonical field operators in appendix \ref{traces}.
The solutions of the quartic eigenvalue equation may be written as
\bea
{{\sigma}}_0 &=& -\frac{\widetilde{b}}{4}  - \big| \widetilde{S} \big| - \frac{1}{2} \Bigg[-4 \widetilde{S}^2 - 2 \widetilde{p} + \frac{\widetilde{q}}{\big|\widetilde{S}\big|} \Bigg]^{1/2}
\, , \\
{{\sigma}}_1 &=& -\frac{\widetilde{b}}{4}  - \big|\widetilde{S}\big| + \frac{1}{2} \Bigg[-4 \widetilde{S}^2 - 2 \widetilde{p} + \frac{\widetilde{q}}{\big|\widetilde{S}\big|} \Bigg]^{1/2}
\, , \\
{{\sigma}}_{2,3} &=& -\frac{\widetilde{b}}{4}  + \big|\widetilde{S}\big| \pm \frac{1}{2} \Bigg[-4 \widetilde{S}^2 - 2 \widetilde{p} - \frac{\widetilde{q}}{\big|\widetilde{S}\big|} \Bigg]^{1/2}
\,  ,
\eea
in terms of the following quantities
\bea
\widetilde{p} &:=& \frac{8 \widetilde{c}-3 \widetilde{b}^2}{8} = \frac{1}{8}  \big(\tr \big[ \chi \big] \big)^2 - \frac{1}{2} \tr \big[ \chi^2 \big] \, , \\
\widetilde{q} &:=& \frac{ \widetilde{b}^3 - 4 \widetilde{b} \widetilde{c} + 8 \widetilde{d}}{8} = - \frac{1}{24} \big(\tr \big[ \chi \big] \big)^3 + \frac{1}{4} \tr \big[ \chi^2 \big] \tr \big[ \chi \big] - \frac{1}{3}  \tr \big[ \chi^3 \big] \, , \\
\widetilde{S} &:=& \frac{1}{2} \Bigg[ -\frac{2}{3} \widetilde{p} + \frac{1}{3} \Bigg( \widetilde{Q} + \frac{\Delta_0}{\widetilde{Q}} \Bigg) \Bigg]^{1/2} \, , \\
\widetilde{Q} &:=& \Bigg[\frac{\Delta_1}{2} + \frac{1}{2} \Big(\Delta_1^2 - 4\Delta_0^3 \Big)^{1/2} \Bigg]^{1/3}\, , \\
\Delta_0 &:=& \widetilde{c}^2 - 3 \widetilde{b} \widetilde{d} + 12 \widetilde{e}\, ,\\
\Delta_1 &:=& 2 \widetilde{c}^3 - 9 \widetilde{b}\widetilde{c}\widetilde{d} +27 \widetilde{b}^2 \widetilde{e} + 27 \widetilde{d}^2 - 72 \widetilde{c} \widetilde{e}\, .
\eea
We can identify the eigenvalue that will be related to the energy density $e$ by looking at the limiting case where the full scalar field two-point function reduces into products of classical fields and thus yields a perfect fluid energy-momentum tensor
\be
\big( \chi_{\text{cl}}\big)^{\mu}_{\; \nu}= \partial^{\mu} \langle \hat{\phi} \rangle  \partial_{\nu} \langle \hat{\phi} \rangle = \partial^{\mu} \phi_{\text{cl}}  \partial_{\nu} \phi_{\text{cl}}\, .
\ee
In this case, all coefficients of the quartic eigenvalue equation vanish except for $\widetilde{b}$ and we find
\be
{{\sigma}}_0^{\text{cl}} = \partial^{\mu} \phi_{\text{cl}}  \partial_{\mu} \phi_{\text{cl}}\, , \quad
{{\sigma}}_{1,2,3}^{\text{cl}}  = 0\, .
\ee
Setting
\be
 \langle \partial_{\mu} \hat{\phi} \partial_{\nu} \hat{\phi} \rangle - \frac{g_{\mu \nu}}{2} \Big[ \langle \partial^{\alpha} \hat{\phi} \partial_{\alpha} \hat{\phi}\rangle + \frac{m^2}{\hbar^2} \langle\hat{\phi}^2\rangle \Big] = \big( e + P  \big) u_{\mu} u_{\nu} +  P g_{\mu \nu} + \pi_{\mu \nu}\, ,
 \ee
and taking $-{{\sigma}}_0$ as the eigenvalue corresponding to the eigenvector of the four-velocity $u^{\mu}$, we have
\bea
 e &=& - {{\sigma}}_0 + \frac{1}{2} \Big[ \langle \partial^{\alpha} \hat{\phi} \partial_{\alpha} \hat{\phi}\rangle + \frac{m^2}{\hbar^2} \langle\hat{\phi}^2\rangle \Big]\, ,\\
 P &=& \frac{1}{3} \Big[- {{\sigma}}_0 - \frac{1}{2}\langle \partial^{\alpha} \hat{\phi} \partial_{\alpha} \hat{\phi}\rangle  - \frac{3}{2} \frac{m^2}{\hbar^2} \langle\hat{\phi}^2\rangle  \Big]\,.
 \eea
 We still have to identify the four-velocity itself, which can be done by rewriting the Cayley-Hamilton equation as
 \be
  \prod_{\mu=0}^3 \big( \chi - \id {{\sigma}}_{\mu} \big) = 0\, ,
 \ee
 which tells us that we have four potential eigenvectors for the eigenvalue ${{\sigma}}_0$ and we label them by the letter $\kappa$, 
 \begin{multline}
 \big(u_{\kappa} \big)^{\mu} :=  \langle  \partial^{\mu} \hat{\phi} \partial_{\nu} \hat{\phi} \rangle \langle \partial^{\nu} \hat{\phi}\partial_{\rho} \hat{\phi} \rangle \langle \partial^{\rho} \hat{\phi}\partial_{\kappa} \hat{\phi} \rangle - 
 \big( {{\sigma}}_1 + {{\sigma}}_2 + {{\sigma}}_3 \big) \langle  \partial^{\mu} \hat{\phi} \partial_{\nu} \hat{\phi} \rangle \langle \partial^{\nu} \hat{\phi}\partial_{\kappa} \hat{\phi} \rangle \\ 
 + \big( {{\sigma}}_1 {{\sigma}}_2  + {{\sigma}}_2 {{\sigma}}_3  + {{\sigma}}_1 {{\sigma}}_3 \big) \langle  \partial^{\mu} \hat{\phi} \partial_{\kappa} \hat{\phi} \rangle
 - {{\sigma}}_1 {{\sigma}}_2 {{\sigma}}_3 \delta^{\mu}_{\; \kappa}\, .
 \end{multline}
 However, by considering the homogeneous case for classical fields we see that the only reasonable choice is $\kappa=0$.
Taking into account a normalisation factor $\alpha$, we are left with
 \begin{multline}
  u^{\mu} = \alpha \Big[  \langle  \partial^{\mu} \hat{\phi} \partial_{\nu} \hat{\phi} \rangle \langle \partial^{\nu} \hat{\phi}\partial_{\rho} \hat{\phi} \rangle \langle \partial^{\rho} \hat{\phi}\partial_{0} \hat{\phi} \rangle - 
 \big( {\sigma}_1 + {\sigma}_2 + {\sigma}_3 \big) \langle  \partial^{\mu} \hat{\phi} \partial_{\nu} \hat{\phi} \rangle \langle \partial^{\nu} \hat{\phi}\partial_{0} \hat{\phi} \rangle \\ 
 + \big( {\sigma}_1 {\sigma}_2  + {\sigma}_2 {\sigma}_3  + {\sigma}_1 {\sigma}_3 \big) \langle  \partial^{\mu} \hat{\phi} \partial_{0} \hat{\phi} \rangle
 - {\sigma}_1 {\sigma}_2 {\sigma}_3 \delta^{\mu}_{\; 0} \Big]\, , \quad u^{\mu} u_{\mu } = -1\, .
\end{multline}
 Note that the above reproduces the classical field identification (${\sigma}_i =0$),
  \bea
 e_{\text{cl}} &=&  - \frac{1}{2} \partial^{\mu} \phi_{\text{cl}}  \partial_{\mu} \phi_{\text{cl}}  + \frac{1}{2} \frac{m^2}{\hbar^2}  \phi_{\text{cl}}^2  \, ,\\
 P_{\text{cl}} &=& - \frac{1}{2} \partial^{\mu} \phi_{\text{cl}}  \partial_{\mu} \phi_{\text{cl}}  - \frac{1}{2} \frac{m^2}{\hbar^2}  \phi_{\text{cl}}^2\, ,\\
 u_{\text{cl}}^{\mu} &=&  \alpha \big(\chi_{\text{cl}}^3 \big)^{\mu}_{\; 0} =  \frac{\partial^{\mu}  \phi_{\text{cl}} }{\big(-\partial^{\nu}  \phi_{\text{cl}}\partial_{\nu}  \phi_{\text{cl}} \big)^{1/2}}\,, \quad \alpha =  \Big[- \big(\chi_{\text{cl}}^6 \big)^0_{\; 0} \Big]^{1/2} \,.
 \eea
 We would like to check whether our identification of energy density and pressure yields meaningful expressions beyond the special case where the two-point function reduces to classical fields.
 We consider the limiting case where the mass $m$ constitutes the biggest energy scale and we can perturbatively expand with respect to this scale. This non-relativistic expansion with parameter $\varepsilon_p = p^2/m^2$ is another approximation on top of the gradient approximation that we have explained in the previous chapter and which is denoted by $\varepsilon_{\hbar} \propto \hbar^2 m^{-2} {^{(3)}\nabla_x^2},...$.
We find
\begin{multline}
 \widetilde{b} = - \langle \partial^{\mu} \hat{\phi} \partial_{\mu} \hat{\phi} \rangle = \int \frac{d^3 p}{\gamma^{1/2}}   \omega_p \big( {f}_1^{+}  -  f_2  \big)  - \gamma^{ij} \int \frac{d^3 p}{\gamma^{1/2}}   p_i p_j\frac{{f}_1^{+}  + {f}_2  }{\omega_p} \\-
\frac{\hbar^2}{4} {^{(3)} \nabla_i}  {^{(3)} \nabla^i}   \int \frac{d^3 p}{\gamma^{1/2}}  \frac{{f}_1^{+} +  {f}_2 }{\omega_p} \\ 
= \int \frac{d^3 p}{\gamma^{1/2}}   \omega_p \big( {f}_1^{+}  -  {f}_2  \big) \Big[ 1 + \mathcal{O}\big( \varepsilon_{p} \big) + \mathcal{O}\big( \varepsilon_{\hbar} \big) \Big] \, .
\end{multline}
However, this expansion is only meaningful if ${f}_1^{+}  \neq {f}_2 $ which needs not to be satisfied for arbitrary times and initial conditions. Just for illustration one can consider  the classical field case in a perturbed FLRW-universe. The solution will be oscillatory 
\be
\phi_{\text{cl}} (x) \propto \sqrt{\rho_{\text{cl}}(x)} \cos \big[ m \int^{x^0} a d\tilde{x}^0  - v(x) - \theta \big] \,, \label{1PIPHi}
\ee 
and thus the correlator $\langle : \hat{\Pi}^2 : \rangle$ is periodically and for short times not determined by the scale $m$ but by a smaller energy scale
\be 
 {\Pi}_{\text{cl}} \propto m \sqrt{\rho_{\text{cl}}(x)} \sin \big[ ... \big] + \dot{v}_{\text{cl}}(x)\sqrt{\rho_{\text{cl}}(x)} \sin \big[... \big]    - \dot{\sqrt{\rho_{\text{cl}}}(x)} \cos \big[ ...\big] \, .
\ee
However, the case of classical fields is itself not problematic since we already have the exact answer for $e,P$ and $u^{\mu}$. We only wanted to make the reader aware that an expansion with respect to the scale $m$ might be more subtle than one would naively expect. Still, in order to make progress with the non-relativistic limit we will assume that 
\be
 {f}_2  \propto \mathcal{O} \big( \varepsilon_{\lbrace p,\hbar \rbrace}  \big) {f}_1^{+}   \, ,
 \ee
which matches one of the conditions for a pure particle limit, that we formulated in the end of the last section. The symbol $\varepsilon_{\lbrace p,\hbar \rbrace}$ denotes a correction in either   $\varepsilon_{p}$ or  $\varepsilon_{\hbar}$. It is clear from the one-point function analysis in \eqref{1PIPHi} that this condition requires a description that goes beyond coherent states (unless an averaging procedure is employed).
Once this condition is satisfied, it makes sense to continue the expansion with respect to the scale $m$ and find
\be
\frac{\widetilde{c}}{\widetilde{b}^2} = \mathcal{O} \big( \varepsilon_{\lbrace p,\hbar \rbrace}  \big) \, , \quad
\frac{\widetilde{d}}{\widetilde{b}^3} = \mathcal{O} \big( \varepsilon_{\lbrace p,\hbar \rbrace }^2  \big) \, , \quad
\frac{\widetilde{e}}{\widetilde{b}^4} = \mathcal{O} \big( \varepsilon_{ \lbrace p,\hbar \rbrace }^{5/2}  \big) \, ,
\ee
where we used
\begin{multline}
\langle  : \partial^{\mu} \hat{\phi} \partial_{\nu} \hat{\phi} :\rangle \langle : \partial^{\nu} \hat{\phi}\partial_{\mu} \hat{\phi} : \rangle = \Bigg[ \int \frac{d^3 p}{\gamma^{1/2}}   \omega_p \Big({f}_1^{+}   - {f}_2   \Big) \Bigg]^2 \\ - 2 \gamma^{ij}  \Bigg[\frac{\hbar}{2}{^{(3)} \nabla_i} \int \frac{d^3 p}{\gamma^{1/2}}{f}_3  + \int \frac{d^3 p}{\gamma^{1/2}} p_i {f}_1^{-}  \Bigg]  \Bigg[\frac{\hbar}{2}{^{(3)} \nabla_j} \int \frac{d^3 p}{\gamma^{1/2}} {f}_3  - \int \frac{d^3 p}{\gamma^{1/2}} p_j {f}_1^{-}  \Bigg] \\   +\gamma^{jk} \gamma^{il}\Bigg[ \int \frac{d^3 p}{\gamma^{1/2}}   p_i p_j\frac{{f}_1^{+}  +{f}_2  }{\omega_p} +
\frac{\hbar^2}{4} {^{(3)} \nabla_i}  {^{(3)} \nabla_j}   \int \frac{d^3 p}{\gamma^{1/2}}  \frac{{f}_1^{+}  + {f}_2  }{\omega_p} \Bigg] \\ \times \Bigg[ \int \frac{d^3 p}{\gamma^{1/2}}   p_k p_l \frac{{f}_1^{+} + {f}_2  }{\omega_p} +
\frac{\hbar^2}{4} {^{(3)} \nabla_k}  {^{(3)} \nabla_l}   \int \frac{d^3 p}{\gamma^{1/2}}  \frac{{f}_1^{+}  + {f}_2  }{\omega_p} \Bigg] \, ,
\end{multline}
and similar expression for the cubic trace and the determinant.
We can now perturb the quartic equation for the eigenvalue ${\sigma}_0$ around its zero order solution $\overline{{\sigma}_0} = - \widetilde{b}$ and  find
\begin{alignat}{2}
{\sigma}_0 &= - \widetilde{b} + \frac{\widetilde{c}}{\widetilde{b}}+ \mathcal{O} \big( \varepsilon_{\lbrace p, \hbar \rbrace}^2 \big)= \langle : \partial^{\mu} \hat{\phi} \partial_{\mu} \hat{\phi} : \rangle - \frac{1}{2} \frac{\langle : \partial^{\mu} \hat{\phi} \partial_{\mu} \hat{\phi} : \rangle^2 - \langle : \partial^{\mu} \hat{\phi} \partial_{\nu} \hat{\phi} : \rangle \langle : \partial^{\nu} \hat{\phi}\partial_{\mu} \hat{\phi} : \rangle}{\langle : \partial^{\mu} \hat{\phi} \partial_{\mu} \hat{\phi} : \rangle} + \mathcal{O} \big( \varepsilon_{\lbrace p, \hbar \rbrace}^2 \big) \\
&=-\gamma^{-1} \langle : \hat{\Pi} \hat{\Pi} : \rangle + \langle : \hat{\Pi} \hat{\Pi} : \rangle^{-1} \langle :\hat{\Pi} \partial_i \hat{\phi} : \rangle \gamma^{ij} \langle : \partial_j \hat{\phi} \hat{\Pi} :  \rangle + \mathcal{O} \big( \varepsilon_{\lbrace p, \hbar \rbrace}^2 \big) \\
&= -\int \frac{d^3 p}{\gamma^{1/2}}   \omega_p \Big({f}_1^{+}   - {f}_2   \Big) + \Bigg[ \int \frac{d^3 p}{\gamma^{1/2}}   \omega_p \Big({f}_1^{+}    - {f}_2   \Big)\Bigg]^{-1 } \\ &  \times\gamma^{ij}  \Bigg[\frac{\hbar^2}{4}{^{(3)} \nabla_i} \int \frac{d^3 p}{\gamma^{1/2}} {f}_3 {^{(3)} \nabla_j} \int \frac{d^3 p}{\gamma^{1/2}} {f}_3  - \int \frac{d^3 p}{\gamma^{1/2}} p_i {f}_1^{-} \int \frac{d^3 p}{\gamma^{1/2}} p_j {f}_1^{-} \Bigg]   + \mathcal{O} \big( \varepsilon_{\lbrace p, \hbar \rbrace}^2 \big)\, .
\end{alignat}
When we now calculate the energy density up to this order, we find that the leading order  contribution containing ${f}_2 $ drops out
\begin{multline}
 e = \int \frac{d^3 p}{\gamma^{1/2}}   \omega_p {f}_1^{+}      - \Bigg[ \int \frac{d^3 p}{\gamma^{1/2}}   \omega_p \Big({f}_1^{+}    - {f}_2  \Big)\Bigg]^{-1}\gamma^{ij}  \Bigg[\frac{\hbar^2}{4}{^{(3)} \nabla_i} \int \frac{d^3 p}{\gamma^{1/2}} {f}_3 {^{(3)} \nabla_j} \int \frac{d^3 p}{\gamma^{1/2}} {f}_3  \\ - \int \frac{d^3 p}{\gamma^{1/2}} p_i {f}_1^{-} \int \frac{d^3 p}{\gamma^{1/2}} p_j {f}_1^{-} \Bigg]  + \frac{\hbar^2}{8} {^{(3)} \nabla_j}{^{(3)} \nabla^j} \int \frac{d^3 p}{\gamma^{1/2}}    \frac{{f}_1^{+}    + {f}_2  }{\omega_p} + \mathcal{O} \big( \varepsilon_{\lbrace p, \hbar \rbrace}^2 \big)\, .
 \end{multline}
 Considering the pressure, we find that the dependence on the density ${f}_2 $ is still present to leading order,
 \begin{multline}
 P =  \frac{1}{3}\int \frac{d^3 p}{\gamma^{1/2}}  \gamma^{ij} \frac{p_i p_j}{\omega_p} \Big({f}_1^{+}   + {f}_2   \Big) -\int \frac{d^3 p}{\gamma^{1/2}}   \omega_p  {f}_2   \\+\frac{1}{3}
\Bigg[ \int \frac{d^3 p}{\gamma^{1/2}}   \omega_p \Big({f}_1^{+}    - {f}_2   \Big)\Bigg]^{-1}\gamma^{ij}  \Bigg[\frac{\hbar^2}{4}{^{(3)} \nabla_i} \int \frac{d^3 p}{\gamma^{1/2}} {f}_3 {^{(3)} \nabla_j} \int \frac{d^3 p}{\gamma^{1/2}} {f}_3  \\ - \int \frac{d^3 p}{\gamma^{1/2}} p_i {f}_1^{-} \int \frac{d^3 p}{\gamma^{1/2}} p_j {f}_1^{-} \Bigg]
 - \frac{\hbar^2}{24} {^{(3)} \nabla_j}{^{(3)} \nabla^j} \int \frac{d^3 p}{\gamma^{1/2}}    \frac{{f}_1^{+}    + {f}_2  }{\omega_p}
 + \mathcal{O} \big( \varepsilon_{\lbrace p, \hbar \rbrace}^2 \big) \,.
 \end{multline}
Similarly, let us compute the four-velocity to next-to-leading order.
This can done by considering 
 \begin{multline}
  u^{\mu} = \alpha \Big[  \langle : \partial^{\mu} \hat{\phi} \partial_{\nu} \hat{\phi} :\rangle \langle : \partial^{\nu} \hat{\phi}\partial_{\rho} \hat{\phi} :\rangle \langle : \partial^{\rho} \hat{\phi}\partial_{0} \hat{\phi} :\rangle \\ - 
 \big( \langle :  \partial^{\nu} \hat{\phi} \partial_{\nu} \hat{\phi} :\rangle  - {\sigma}_0 \big) \langle :  \partial^{\mu} \hat{\phi} \partial_{\nu} \hat{\phi} :\rangle \langle : \partial^{\nu} \hat{\phi}\partial_{0} \hat{\phi} :\rangle  + \mathcal{O}\big( \varepsilon_{\lbrace p, \hbar \rbrace}^{2} \big) \Big]\, , \quad u^{\mu} u_{\mu } = -1\, .
 \end{multline}
 We  compute the non-normalized Lorentz factor first
 \begin{multline}
 \frac{W}{\alpha} = - \frac{n_{\nu}u^{\nu}}{\alpha}  = 
 -N \gamma^{-3} \langle : \hat{\Pi} \hat{\Pi} :\rangle^3 -N^k \gamma^{-5/2}\langle \hat{\Pi} \partial_k \hat{\phi} :\rangle  \langle : \hat{\Pi} \hat{\Pi} :\rangle^2  \\ + N \gamma^{-2} \langle : \hat{\Pi} \hat{\Pi} :\rangle \langle : \hat{\Pi} \partial_i \hat{\phi} :\rangle \gamma^{ij}  \langle : \partial_j \hat{\phi}  \hat{\Pi} :\rangle + N\gamma^{-2} \gamma^{ij} \langle : \partial_i \hat{\phi} \partial_j \hat{\phi} :\rangle  \langle : \hat{\Pi} \hat{\Pi} :\rangle^2+ \mathcal{O}\big( \varepsilon_{\lbrace p, \hbar \rbrace}^{3/2} \big)\, .
 \end{multline}
 Next, we compute
 \begin{multline}
 \frac{u^k}{\alpha} =  N^k \gamma^{-3} \langle : \hat{\Pi} \hat{\Pi} :\rangle^3+ \frac{N^k N^s}{N} \gamma^{-5/2} \langle : \hat{\Pi} \hat{\Pi} :\rangle^2 \langle : \hat{\Pi} \partial_s \hat{\phi} :\rangle + \gamma^{kl}N\langle : \partial_l \hat{\phi}  \hat{\Pi} :\rangle \gamma^{-2} \langle : \hat{\Pi} \hat{\Pi} :\rangle^2\\
 + \gamma^{kl} N^s \gamma^{-2} \langle :\partial_l \hat{\phi}  \hat{\Pi} :\rangle\langle : \hat{\Pi} \hat{\Pi} :\rangle \langle : \hat{\Pi}   \partial_s \hat{\phi} :\rangle 
 - N^k\gamma^{-2} \langle : \hat{\Pi} \hat{\Pi} :\rangle\langle : \hat{\Pi} \partial_i \hat{\phi} :\rangle  \gamma^{ij}\langle: \partial_j \hat{\phi}  \hat{\Pi} :\rangle
  \\ - N^k\gamma^{-2} \langle : \hat{\Pi} \hat{\Pi} :\rangle^2 \langle :  \partial_i \hat{\phi}  \partial_j \hat{\phi} :\rangle  \gamma^{ij}+ \mathcal{O}\big( \varepsilon_{\lbrace p, \hbar \rbrace}^{3/2} \big)\, .
 \end{multline}
 Finally, we can calculate the spatial part of the four-velocity without the need to explicitly calculate the normalization factor $\alpha$,
 \begin{multline}
 v^k  =  \frac{\alpha^{-1}u^{k} }{\alpha^{-1} W} + \frac{N^k}{N}
 = - \gamma^{kl} \frac{\langle : \partial_l \hat{\phi}  \hat{\Pi}  : \rangle}{\langle : \hat{\Pi}   \hat{\Pi} : \rangle} + \mathcal{O}\big( \varepsilon_{\lbrace p, \hbar \rbrace}^{3/2} \big) \\
 = - \gamma^{kl} \Bigg[ \int \frac{d^3 p}{\gamma^{1/2}}   \omega_p \Big({f}_1^{+}    -{f}_2  \Big)\Bigg]^{-1} \Bigg[\frac{\hbar}{2}{^{(3)} \nabla_i} \int \frac{d^3 p}{\gamma^{1/2}} {f}_3  - \int \frac{d^3 p}{\gamma^{1/2}} p_i {f}_1^{-}  \Bigg] + \mathcal{O}\big( \varepsilon_{\lbrace p, \hbar \rbrace}^{3/2} \big) \, ,
 \end{multline}
 and the normalized Lorentz factor is read-off in the standard way
 \be
 W = 1 + \frac{1}{2} v^i v^j \gamma_{ij} + \mathcal{O}\big( v^4 \big) = 1 + \frac{1}{2} \gamma^{kl} \frac{\langle : \partial_k \hat{\phi}  \hat{\Pi} : \rangle\langle: \partial_l \hat{\phi}  \hat{\Pi} : \rangle}{\langle : \hat{\Pi}   \hat{\Pi} : \rangle^2} + \mathcal{O}\big( \varepsilon_{\lbrace p, \hbar \rbrace}^{2} \big)\, .
 \ee
In order to recover expressions in terms of a purely classical particle distribution, we need to assume that the $\varepsilon_{\hbar}$ corrections are negligible with respect to the $\varepsilon_{p}$ corrections and that our initial state is allowing for a hierarchy
$m \gg \Delta p \gg \frac{\hbar}{\Delta x} $.  
Once we put forward the identification of even ($ {f}_1^{+} $) and odd (${f}_1^{-} $) phase-space densities that we discussed in the previous section, we end up with the following expressions,
\begin{multline}
 e = \int \frac{d^3 p}{\gamma^{1/2}}   \omega_p {f}_1^{+}   + \Bigg[ \int \frac{d^3 p}{\gamma^{1/2}}   \omega_p {f}_1^{+}   \Bigg]^{-1}\gamma^{ij}  \Bigg[  \int \frac{d^3 p}{\gamma^{1/2}} p_i {f}_1^{-} \int \frac{d^3 p}{\gamma^{1/2}} p_j {f}_1^{-} \Bigg]  +\mathcal{O} \big( \varepsilon_\hbar \big) + \mathcal{O} \big( \varepsilon_p^2\big) \\
 = \int \frac{d^3 p}{\gamma^{1/2}}   \omega_p {f}_{1}   
  + v^k \int \frac{d^3 p}{\gamma^{1/2}} p_k    {f}_{1} +\mathcal{O} \big( \varepsilon_\hbar \big) + \mathcal{O} \big( \varepsilon_p^2\big)   \, ,
 \end{multline}
 \begin{multline}
 P =  \frac{1}{3}\int \frac{d^3 p}{\gamma^{1/2}}  \gamma^{ij} \frac{p_i p_j}{\omega_p} {f}_1^{+}   \\ -\frac{1}{3}
\Bigg[ \int \frac{d^3 p}{\gamma^{1/2}}   \omega_p {f}_1^{+}  \Bigg]^{-1}\gamma^{ij}  \Bigg[ \int \frac{d^3 p}{\gamma^{1/2}} p_i {f}_1^{-} \int \frac{d^3 p}{\gamma^{1/2}} p_j {f}_1^{-} \Bigg]
+\mathcal{O} \big( \varepsilon_\hbar \big) + \mathcal{O} \big( \varepsilon_p^2\big) 
\\ 
=  \frac{1}{3}\int \frac{d^3 p}{\gamma^{1/2}}  \gamma^{ij} \frac{p_i p_j}{m}  {f}_{1}    -\frac{1}{3}
v^k\int \frac{d^3 p}{\gamma^{1/2}} p_k {f}_{1}   
+\mathcal{O} \big( \varepsilon_\hbar \big) + \mathcal{O} \big( \varepsilon_p^2\big)  \,,
 \end{multline}
  \begin{multline}
 v_k  
 =  \Bigg[ \int \frac{d^3 p}{\gamma^{1/2}}   \omega_p {f}_1^{+}    \Bigg]^{-1} \Bigg[  \int \frac{d^3 p}{\gamma^{1/2}} p_k {f}_1^{-}  \Bigg] +\mathcal{O} \big( \varepsilon_\hbar \big)  + \mathcal{O} \big( \varepsilon_p^{3/2}\big)
 \\ =  \Bigg[ \int \frac{d^3 p}{\gamma^{1/2}}    {f}_{1}   \Bigg]^{-1}  \int \frac{d^3 p}{\gamma^{1/2}} \frac{p_k}{m}  {f}_{1}   +\mathcal{O} \big( \varepsilon_\hbar \big)  + \mathcal{O} \big( \varepsilon_p^{3/2}\big)
   \, .
  \end{multline}
These expressions are identical to the expressions one would obtain for a distribution of classical non-relativistic particles in curved space-time.
The above identification shows once more that such classical distributions may be represented by two-point functions of real scalar field operators via the Wigner transformation \eqref{genDefWignerOp} and the subsequent recombination \eqref{deff+} to \eqref{deff2}, always provided we are given a state that behaves classical enough. An example for such a state was given in \cite{Pirk:1989bs} for an FRLW-universe with a particular vacuum choice. We identify the correlators $f_{2,3}$ in our paper with combinations of the squeezing contributions $\langle \hat{a}_k \hat{a}_{-k} \rangle$ and $\langle \hat{a}_k^{\dagger} \hat{a}_{-k}^{\dagger} \rangle$ in their paper ($\hat{a}_k$ and $\hat{a}^{\dagger}_k$ denote creation and annihilation operators, respectively), which they eventually dropped. The density $f_1$ in our paper, that approximates a classical particle phase-space density, is expressible in terms of $\langle \hat{a}_k^{\dagger} \hat{a}_{k}\rangle$  in their paper and gives an intuitive interpretation of $\hat{f}_1$ as a counting operator. The state-independent (or in this setting vacuum) contributions in \cite{Pirk:1989bs} were removed by hand, which corresponds to the normal ordering prescription. Let us remark that the starting point in \cite{Pirk:1989bs} is a phase-space description that makes use of an off-shell momentum variable which makes it in our opinion difficult to take the other degrees of freedom encoded in  $f_{2,3}$ into account (they were dropped in \cite{Pirk:1989bs} as they are in the review literature \cite{trove.nla.gov.au/work/9783845} for Minkowski space-time). 
\section{Dynamics of phase-space densities}
In the previous sections, we have interpreted the averaged phase-space densities \eqref{deff+} to \eqref{deff2} always with respect to the energy-momentum tensor, without self-interactions and without non-minimal couplings to the geometry.  The goal of this section is to work out their dynamics in a spatial gradient approximation including the non-minimal coupling to the curvature and even including self-interaction in a one-loop approximation where we assume, for simplicity, that one-point functions $\langle \hat{\phi} \rangle$, $\langle \hat{\Pi} \rangle$ are absent.\footnote{ One-point functions are straightforwardly included by shifting the canonical operators. This shift is necessary since the gradient approximation cannot simply be applied for Wigner transforms of products of classical fields without a smoothing procedure. We discuss this also in \cite{Prokopec:2017ldn} and list the references where such a procedure is pursued.} 
Another point we have to stress again, is that we will not include anomalous contributions in the following kinetic equations. This means  that we assume those contributions  to be negligible, which remain after the equations of motion have been applied on the terms that normal order the phase-space operators in \eqref{genDefWignerOp}, which is well justified for the energy scales we are interested in, since such anomalous contributions are of order $R M_P^{-2}$ at the level of the energy-momentum tensor.
Moreover, we assume contributions, that result from the boundary of the normal neighbourhood, to be negligible which is a requirement on the state that goes hand in hand with the spatial gradient expansion.
\par
The dynamics for the averaged phase-space densities 
$f_{1}^{\pm}$, $f_2$, $f_3$ given in \eqref{deff+} to \eqref{deff2} can be derived by first considering the expectation values  ${F}_{\phi \phi}$, ${F}_{\phi \Pi}$, ${F}_{\Pi \phi}$, ${F}_{\Pi \Pi}$ given via \eqref{genDefWignerOp} and acting with a time-derivative, commuting it with the exponential shift operators, using the equations of motions for the canonical fields, commuting the resulting operators back and rewriting them in such a way that they act on the expectation values of ${F}_{\phi \phi}$, ${F}_{\phi \Pi}$, ${F}_{\Pi \phi}$, ${F}_{\Pi \Pi}$ which is the most difficult part of the calculation. Finally, we rewrite  everything in terms of the dimensionally rescaled quantities \eqref{deff+} to \eqref{deff2}. The spatial gradient approximation truncates the infinite series of spatial derivatives, that results from commuting various differential operators. We keep on the other hand all time derivatives and thus all degrees of freedom. Since the calculation involves a number of lengthy expressions, it is unavoidable to introduce some notation. A lot of technical details of this procedure are deferred to Appendix \ref{defAndId} to  \ref{dynWig}.
\par
First, we define
\bea
\hat{u}^{\pm} &:=& \exp \Bigg[{\pm \frac{r^k}{2}{^{(3)}{\nabla}^H_k}}  \Bigg] \hat{\phi} \, ,  \quad \; \qquad
\hat{v}^{\pm} := \exp \Bigg[{\pm \frac{r^k}{2}{^{(3)}{\nabla}^H_k}}  \Bigg] \Big[\gamma^{-1/2} \hat{\Pi} \Big] \, , \\
N^{\pm} &:= & \exp \Bigg[{\pm \frac{r^k}{2}{^{(3)}{\nabla}^H_k}}  \Bigg] N \, , \quad 
(NK)^{\pm} :=  \exp \Bigg[{\pm \frac{r^k}{2}{^{(3)}{\nabla}^H_k}}  \Bigg] (NK) \,, \\
\big[  N {:\hat{\phi}^2:} \big]^{\pm}  &:= &\exp \Bigg[{\pm \frac{r^k}{2}{^{(3)}{\nabla}^H_k}}  \Bigg]\big[  N {:\hat{\phi}^2:} \big] \, , \quad 
(NR)^{\pm} :=  \exp \Bigg[{\pm \frac{r^k}{2}{^{(3)}{\nabla}^H_k}}  \Bigg] (NR) \,,
\eea
where $R$ is the four-dimensional Ricci scalar.
Moreover, we will need a couple of differential operators denoted by $\mathcal{T}^{\pm}_*$, $\mathcal{M}^{\pm}_*$, $\big({^{(3)} \square}\big)^{\pm}_*$ and $({^{(3)} \nabla} N)^{\pm}_*$, which are calculated in a gradient approximation in appendix \ref{commis} based on the general identities in \ref{defAndId}.
We find the following expressions, up to anomalous contributions, boundary terms and higher-order correlators which are all assumed to be small,
\begin{multline}
\gamma^{1/2} \partial_t \Big[ \gamma^{-1/2} \langle \hat{F}_{\phi \phi} \rangle \Big]
= \frac{1}{2}\int_{T\Sigma_t} dr^{3} \gamma^{1/2} e^{-\frac{i }{\hbar}r^k p_k} \big( N^{+} +N^{-}\big) \langle{: \hat{v}^+ \hat{u}^- 
+\hat{u}^+ \hat{v}^-  :}\rangle \\
+ \frac{1}{2} \int_{T\Sigma_t} dr^{3} \gamma^{1/2} e^{-\frac{i }{\hbar}r^k p_k} \big( N^{+} -N^{-}\big) \langle
{:\hat{v}^+ \hat{u}^-  - \hat{u}^+ \hat{v}^- :} \rangle \\
+\int_{T\Sigma_t} dr^{3} \gamma^{1/2} e^{-\frac{i }{\hbar}r^k p_k} \Big(\mathcal{T}^+_* +\mathcal{T}^-_* +\mathcal{M}^+_* +\mathcal{M}^-_* \Big) \langle {: \hat{u}^+ \hat{u}^- :} \rangle \, , \label{F00BeforeInt}
\end{multline}
\begin{multline}
\frac{1}{2}\gamma^{1/2} \partial_t \Big[ \gamma^{-1/2}  \langle \hat{F}_{\Pi \phi}   + \hat{F}_{\phi \Pi} \rangle  \Big]
=\frac{1}{2} \int_{T\Sigma_t} dr^{3}_{T\Sigma_t} \gamma^{1/2} e^{-\frac{i }{\hbar}r^k p_k}  \big(N^+ + N^- \big) \langle : \hat{v}^+ \hat{v}^-  : \rangle \\
+\frac{1}{2}\int dr^{3}_{T\Sigma_t} \gamma^{1/2} e^{-\frac{i }{\hbar}r^k p_k} \Big(\mathcal{T}^+_* + \mathcal{T}^-_* +\mathcal{M}^+_* + \mathcal{M}^-_* + \frac{1}{2}(NK)^{+} +\frac{1}{2}(NK)^{-}\Big) \langle  : \hat{v}^+ \hat{u}^-  + \hat{u}^+ \hat{v}^-  : \rangle  \\
+\frac{1}{2}\int_{T\Sigma_t} dr^{3} \gamma^{1/2} e^{-\frac{i }{\hbar}r^k p_k} \Big(\frac{1}{2} (NK)^{+} -\frac{1}{2}(NK)^{-}\Big) \langle { : \hat{v}^+ \hat{u}^-  - \hat{u}^+ \hat{v}^-   :} \rangle \\
+\frac{1}{2} \int_{T\Sigma_t} dr^{3} \gamma^{1/2} e^{-\frac{i }{\hbar}r^k p_k} \Big( N^+ \big({^{(3)} \square}\big)^+_* + N^- \big({^{(3)} \square}\big)^-_* -   \frac{m^2}{\hbar^2}\big( N^{+} +N^{-}\big)- \xi \big[ (NR)^{+} +(NR)^{-}\big]  \\-   \frac{1}{2 } \frac{{\lambda}}{\hbar}\big( \big[  N  \langle :\hat{\phi}^2:  \rangle \big] ^{+} +\big[  N  \langle: \hat{\phi}^2  :\rangle\big] ^{-}\big)  
+({^{(3)} \nabla} N)^{+}_*+({^{(3)} \nabla} N)^{-}_*   \Big)\langle : \hat{u}^+ \hat{u}^- : \rangle \, ,\label{F+BeforeInt}
\end{multline}
\begin{multline}
\frac{1}{2}\gamma^{1/2} \partial_t \Big[ \gamma^{-1/2}  \langle \hat{F}_{\Pi \phi} -\hat{F}_{\phi \Pi} \rangle  \Big]
= -\frac{1}{2} \int_{T\Sigma_t} dr^{3} \gamma^{1/2} e^{-\frac{i }{\hbar}r^k p_k}  \big(N^+ - N^- \big) \langle: \hat{v}^+ \hat{v}^- : \rangle \\
+\frac{1}{2}\int_{T\Sigma_t} dr^{3} \gamma^{1/2} e^{-\frac{i }{\hbar}r^k p_k} \Big(\mathcal{T}^+_* + \mathcal{T}^-_* +\mathcal{M}^+_* + \mathcal{M}^-_* + \frac{1}{2}(NK)^{+} +\frac{1}{2}(NK)^{-} \Big) \langle: \hat{v}^+ \hat{u}^-  - \hat{u}^+ \hat{v}^-  : \rangle \\
+\frac{1}{2}\int_{T\Sigma_t} dr^{3} \gamma^{1/2} e^{-\frac{i }{\hbar}r^k p_k} \Big( \frac{1}{2}(NK)^{+} -\frac{1}{2}(NK)^{-}\Big)\langle: \hat{v}^+ \hat{u}^-  + \hat{u}^+ \hat{v}^- :  \rangle \\
+\frac{1}{2} \int_{T\Sigma_t} dr^{3} \gamma^{1/2} e^{-\frac{i }{\hbar}r^k p_k} \Big( N^+ \big({^{(3)} \square}\big)^+_* - N^- \big({^{(3)} \square}\big)^-_* -   \frac{m^2}{\hbar^2}\big( N^{+} - N^{-}\big) -\xi \big[ (NR)^{+} -(NR)^{-}\big]  \\ - \frac{1}{2 } \frac{{\lambda}}{\hbar}\big( \big[  N \langle : \hat{\phi}^2: \rangle \big] ^{+} -\big[  N \langle :\hat{\phi}^2 : \rangle\big] ^{-}\big)  +
({^{(3)} \nabla} N)^{+}_* - ({^{(3)} \nabla} N)^{-}_*   \Big)\langle: \hat{u}^+ \hat{u}^- :\rangle \, ,\label{F-BeforeInt}
\end{multline}
\begin{multline}
\gamma^{1/2} \partial_t \Big[ \gamma^{-1/2} \langle \hat{F}_{\Pi \Pi}  \rangle  \Big]
= +\int_{T\Sigma_t} dr^{3} \gamma^{1/2} e^{-\frac{i }{\hbar}r^k p_k} \Big(\mathcal{T}^+_* +\mathcal{T}^-_* +\mathcal{M}^+_* +\mathcal{M}^-_* +  (NK)^{+} +(NK)^{-} \Big) \langle:  \hat{v}^+ \hat{v}^- : \rangle \\
-\frac{1}{2}\int_{T\Sigma_t} dr^{3} \gamma^{1/2} e^{-\frac{i }{\hbar}r^k p_k} \Big(N^+ \big({^{(3)} \square}\big)^+_* - N^- \big({^{(3)} \square}\big)^-_*  -  \frac{m^2}{\hbar^2}\big( N^{+} - N^{-}\big) - \xi \big[ (NR)^{+} -(NR)^{-}\big]\\ -\frac{1}{2 } \frac{{\lambda}}{\hbar}\big( \big[  N \langle:\hat{\phi}^2: \rangle \big] ^{+} -\big[  N  \langle : \hat{\phi}^2 : \rangle \big] ^{-}\big)  +({^{(3)} \nabla} N)^{+}_* - ({^{(3)} \nabla} N)^{-}_*    \Big) \langle: \hat{v}^+ \hat{u}^-  - \hat{u}^+ \hat{v}^-  :\rangle  \\
+\frac{1}{2}\int_{T\Sigma_t} dr^{3} \gamma^{1/2} e^{-\frac{i }{\hbar}r^k p_k}  \Big(N^+ \big({^{(3)} \square}\big)^+_* + N^- \big({^{(3)} \square}\big)^-_*  -  \frac{m^2}{\hbar^2}\big( N^{+} + N^{-}\big) -\xi \big[ (NR)^{+} +(NR)^{-}\big] \\ -\frac{1}{2 } \frac{{\lambda}}{\hbar}\big( \big[  N \langle : \hat{\phi}^2  : \rangle \big] ^{+} +\big[  N \langle : \hat{\phi}^2 : \rangle \big] ^{-}\big)   + ({^{(3)} \nabla} N)^{+}_* + ({^{(3)} \nabla} N)^{-}_* \Big)\langle:  \hat{v}^+ \hat{u}^-  + \hat{u}^+ \hat{v}^- :  \rangle  \, .\label{F11BeforeInt}
\end{multline}
The dynamical equations for $\hat{F}_{\phi \phi}$, $\hat{F}_{\Pi \phi}$, $\hat{F}_{\phi \Pi}$, $\hat{F}_{\Pi \Pi}$ take a convenient form in terms of the horizontal lift of the covariant derivative \cite{de2011methods} on the cotangent bundle of spatial hypersurfaces
\be
{D}_k := {^{(3)} \nabla}_{k} + p_l {^{(3)} \Gamma^l_{\; k j}} \frac{\partial}{ \partial p_j}\,, \quad D_k p_j = 0\, .
\ee
The latter derivative transforms covariantly under a change of spatial coordinates.
For brevity and to illustrate the structure, we write down the dynamics for the Wigner transformed expectation values ${F}_{\phi \phi}$, ${F}_{\Pi \phi}$, ${F}_{\phi \Pi}$, ${F}_{\Pi \Pi}$ only to leading order in the spatial gradient expansion and the next-to-leading order expressions may be found in appendix \ref{dynWig},
\begin{multline}
 \partial_t {F}_{\phi \phi} 
=   \Big[ N + \mathcal{O}\big(\hbar^2\big) \Big]   \Big[ {F}_{\Pi \phi} + {F}_{\phi \Pi } \Big] 
+ \frac{i }{2}  \hbar \Big[ N_{;k} \frac{\partial}{\partial p_k}  + \mathcal{O}\big(\hbar^2\big) \Big]  \Big[ {F}_{\Pi \phi} - {F}_{\phi \Pi } \Big]
 \\ + \Bigg[N^k  D_k    -
  p_k N^k_{\; ; m} \frac{\partial}{\partial p_m}     - NK 
 + \mathcal{O}\big(\hbar^2\big) \Bigg]  {F}_{\phi \phi}  \, ,
\end{multline}
\begin{multline}
\frac{1}{2}\partial_t \big( {F}_{\Pi \phi} +{F}_{\phi \Pi} \big)
=   \Big[ N + \mathcal{O}\big(\hbar^2\big) \Big]  {F}_{\Pi \Pi } 
+\frac{1}{2} \Bigg[N^k  D_k    -
  p_k N^k_{\; ; m} \frac{\partial}{\partial p_m}     
  + \mathcal{O}\big(\hbar^2\big)\Bigg] \big( {F}_{\Pi \phi} +{F}_{\phi \Pi} \big)
 \\ +\hbar \frac{i}{4}  \Big[\big(NK\big)_{;j} \frac{\partial}{\partial p_j} + \mathcal{O}\big(\hbar^2\big)\Big] \big( {F}_{\Pi \phi}-{F}_{\phi \Pi} \big) \\
- \frac{1}{\hbar^2}\Bigg[ N {m^2}+N \gamma^{kj} {p_k p_j} +\frac{1}{2} \hbar{{\lambda}}  \Big[N+ \mathcal{O}\big(\hbar^2\big)\Big] \int \frac{d^3 q }{\gamma^{1/2}} {F}_{\phi \phi} (q)   + \mathcal{O}\big(\hbar^2\big)\Bigg]{F}_{\phi \phi}  \, ,
\end{multline}
\begin{multline}
\frac{i}{2} \partial_t \big( {F}_{\Pi \phi} -{F}_{\phi \Pi} \big) 
= 
\frac{ \hbar}{2} \Big[  N_{;k} \frac{\partial}{\partial p_k}  + \mathcal{O}\big(\hbar^2\big)  \Big] {F}_{\Pi \Pi } \\
+\frac{i}{2} \Bigg[N^k  D_k    -
  p_k N^k_{\; ; m} \frac{\partial}{\partial p_m}     
  + \mathcal{O}\big(\hbar^2\big)\Bigg] \big( {F}_{\Pi \phi} -{F}_{\phi \Pi} \big)
-\frac{\hbar}{4}\Big[  \big(NK\big)_{;j} \frac{\partial}{\partial p_j}  + \mathcal{O}\big(\hbar^2\big) \Big]\big( {F}_{\Pi \phi}+{F}_{\phi \Pi} \big)\\
-\frac{1}{2\hbar}\Bigg[
 2N  p_j D^j -  \omega_p^2 N_{;k} \frac{\partial}{\partial p_k}   + \mathcal{O}\big(\hbar^2\big)  
-\frac{1}{2} {\hbar} {{\lambda}} \Big[  \big[N \int \frac{d^3 q }{\gamma^{1/2}} {F}_{\phi \phi} (q)\big]_{;k} \frac{\partial}{\partial p_k}  + \mathcal{O}\big(\hbar^2\big)   \Big]  \Bigg] {F}_{\phi \phi}  \, ,
\end{multline}
\begin{multline}
 \partial_t  {F}_{\Pi \Pi}
=\Bigg[N^k  D_k +NK   -
  p_k N^k_{\; ; m} \frac{\partial}{\partial p_m}     
  + \mathcal{O}\big(\hbar^2\big)\Bigg] {F}_{\Pi \Pi}\\
-\frac{i}{2\hbar}\Bigg[
 2N  p_j D^j   + \mathcal{O}\big(\hbar^2\big)  
-\frac{1}{2} {\hbar} {{\lambda}} \Big[  \big[N \int \frac{d^3 q }{\gamma^{1/2}} {F}_{\phi \phi} (q)\big]_{;k} \frac{\partial}{\partial p_k}  + \mathcal{O}\big(\hbar^2\big)   \Big]  \Bigg]\Big[ {F}_{\Pi \phi} -  {F}_{\phi \Pi}\Big]  \\
- \frac{1}{\hbar^2}\Bigg[ N {m^2}+N\gamma^{kj} {p_k p_j}  \Big] +  \mathcal{O}\big(\hbar^2\big)  
+\frac{1}{2} \hbar {{\lambda}} \Big[  N \int \frac{d^3 q }{\gamma^{1/2}} {F}_{\phi \phi} (q) + \mathcal{O}\big(\hbar^2\big)  \Big] \Bigg] \Big[ {F}_{\Pi \phi} +  {F}_{\phi \Pi}\Big]  \, .
\end{multline}
The next step is to convert the dynamical equation for the dimensionally unequal expectation values $F_{\phi \phi}$, $F_{\Pi \phi}$, $F_{\phi \Pi}$ and $F_{\Pi \Pi}$ into dynamical equations for the dimensionally equal phase-space densities $f^{\pm}_1$, $f_2$ and $f_3$, that we defined in \eqref{deff+} to \eqref{deff2}. It turns out that several leading order terms cancel in this dimensional rescaling, such that some next-to-leading order terms of the previous equations turn into leading order terms for the equations of the rescaled quantities. We would have to include even higher order terms in the previous calculation for $F_{\phi \phi}$, $F_{\Pi \phi}$, $F_{\phi \Pi}$ and $F_{\Pi \Pi}$ in order to get to next-to-leading order terms for rescaled quantities $f^{\pm}_1$, $f_2$ and $f_3$. However, we see that even certain leading order corrections are of order $\hbar$ and thus first order terms concerning the spatial gradient expansion. We find
\begin{multline}
 \partial_t {f}_1^{+}  
=     \Bigg[ N^k  D_k    -
  p_k N^k_{\; ; m} \frac{\partial}{\partial p_m}   \Bigg]{f}_1^{+}  
-\Bigg[ NK +  \frac{ p_m p_k }{\omega_p^2}  N^{k\, m}_{\;\; ; }   \Bigg] {f}_2 
- \frac{1}{\omega_p}\Big[
  N  p_j      D^j   
-  \omega_p^2N_{;m}  \frac{\partial}{\partial p_m}  \Big]{f}_1^{-}   \\
+\frac{\hbar}{\omega_p}\Bigg[   \frac{1}{2} p_j  N_{;k}  \frac{\partial}{\partial p_k }   D^j   + \frac{1}{4}N D_j D^j    
-\frac{1}{3} N p_i  p_j {^{(3)}R^{i \; \; \; \; j}_{\; q m }}  \frac{\partial^2}{\partial p_q \partial p_m} \\ -\frac{1}{12}N p_i {^{(3)}R^i_{\; k}} \frac{\partial}{\partial p_k} + \frac{1}{6} N {^{(3)}R}-\xi N R \Bigg] {f}_3  \\
+\frac{{\lambda}}{2}  {\omega_p}\Big[ N\frac{\hbar^3}{\omega_p^2} \int \frac{d^3 q }{\gamma^{1/2}} \frac{ {f}_1^{+}(q) + {f}_2(q)}{\omega_q} \Big]_{;k} \frac{\partial}{\partial p_k}{f}_1^{-} \\
-\frac{\lambda}{2}\frac{\omega_p}{\hbar }  \Bigg[  N \frac{\hbar^3}{\omega_p^2} \int \frac{d^3 q}{\gamma^{1/2}} \frac{ {f}_1^{+}(q) + {f}_2(q)}{\omega_q} -  \frac{\hbar^2}{8}   \frac{\hbar^3}{\omega_p^2}  \big[N \int \frac{d^3 q }{\gamma^{1/2}} \frac{ {f}_1^{+}(q) + {f}_2(q)}{\omega_q} \big]_{; k s} \frac{\partial^2}{\partial p_k \partial p_s}  \Bigg] {f}_3  \, , \label{finf1+}
\end{multline}
\begin{multline}
  \partial_t  {f}_1^{-}  
= \Bigg[N^k  D_k    -
  p_k N^k_{\; ; m} \frac{\partial}{\partial p_m}     
  \Bigg] {f}_1^{-}   
-\frac{\hbar}{2} \big(NK\big)_{;j} \frac{\partial}{\partial p_j}   {f}_3 
-\frac{1}{ \omega_p}\Big[ N  p_j      D^j   
-  \omega_p^2N_{;m}  \frac{\partial}{\partial p_m}   \Big] {f}_1^{+}  \\
-\frac{1}{ \omega_p}\Big[ N  p_j      D^j   
+  N_{;m}  {p_l \gamma^{lm}}  \Big] {f}_2 
 +\frac{{\lambda}}{2} \omega_p \Big[ N \frac{ \hbar^3}{\omega_p^2} \int \frac{d^3 q }{\gamma^{1/2}} \frac{ {f}_1^{+}(q) + {f}_2(q)}{\omega_q} \Big]_{;k}
\frac{\partial}{\partial p_k}  \Big[ {f}_1^{+} + {f}_2 \Big]  \, ,\label{finf1-}
\end{multline}
\begin{multline}
 \partial_t {f}_2   
= 2 \frac{\omega_p}{\hbar}  N    {f}_3
+ {\omega_p}   N_{;k} \frac{\partial}{\partial p_k} {f}_1^{-} +
   \Bigg[ N^k  D_k    -
  p_k N^k_{\; ; m} \frac{\partial}{\partial p_m} -\frac{p_i p_k}{\omega_p^2} \big(N K^{ij} - N^{i \; j}_{\; ;} \big) \Bigg]{f}_2  \\
- \Bigg[ NK + 
\frac{ p_m p_k }{\omega_p^2}  N^{k\, m}_{\;\; ; }      \Bigg] {f}_1^{+} 
+ N\frac{p_j}{\omega_p}   D^j    {f}_1^{-}   \\
-\frac{\hbar}{\omega_p} \Bigg[   \frac{1}{2} p_j  N_{;k}  \frac{\partial}{\partial p_k }   D^j   + \frac{1}{4}N D_j D^j    
-\frac{1}{3} N p_i  p_j {^{(3)}R^{i \; \; \; \; j}_{\; q m }}  \frac{\partial^2}{\partial p_q \partial p_m} \\ -\frac{1}{12}N p_i {^{(3)}R^i_{\; k}} \frac{\partial}{\partial p_k} + \frac{1}{6} N {^{(3)}R} -\xi N R \Bigg] {f}_3  \\
+\frac{{\lambda}}{8}  {\omega_p}\Big[ N\frac{\hbar^3}{\omega_p^2} \int \frac{d^3 q }{\gamma^{1/2}} \frac{ {f}_1^{+}(q) + {f}_2(q)}{\omega_q} \Big]_{;k} {f}_1^{-} \\
+ \frac{\lambda}{2} \frac{\omega_p}{\hbar } \Bigg[  N \frac{\hbar^3}{\omega_p^2} \int \frac{d^3 q }{\gamma^{1/2}} \frac{ {f}_1^{+}(q) + {f}_2(q)}{\omega_q} -  \frac{\hbar^2}{8}    \frac{\hbar^3}{\omega_p^2}\big[ N \int \frac{d^3 q }{\gamma^{1/2}} \frac{ {f}_1^{+}(q) + {f}_2(q)}{\omega_q} \big]_{; k s} \frac{\partial^2}{\partial p_k \partial p_s}  \Bigg] {f}_3    \, ,\label{finf2}
\end{multline}
\begin{multline}
\partial_t {f}_3 
=   \Bigg[N^k  D_k    -
  p_k N^k_{\; ; m} \frac{\partial}{\partial p_m}     \Bigg] {f}_3 
+\frac{\hbar}{2}  \big(NK\big)_{;j} \frac{\partial}{\partial p_j} {f}_1^{-}
\\
- \frac{\hbar}{\omega_p}\Bigg[ 2\frac{\omega_p^2}{\hbar^2} N - \frac{\omega_p^2}{4}  N_{; qm}  \frac{\partial^2}{\partial p_q \partial p_m}  - \frac{1}{4} \frac{p_i p_j}{\omega_p^2} N_{;}^{\; ij}   - \frac{1}{2} p_j  N_{;m}  \frac{\partial}{\partial p_m}  D^j +\frac{1}{2} \frac{p_i p_j}{\omega_p^2}   N_{;}^{\,i} D^j   -  \frac{1}{4}N D_j D^j   
   \\
+\frac{1}{3} N p_i  p_j {^{(3)}R^{i \; \; \; \; j}_{\; q m }}  \frac{\partial^2}{\partial p_q \partial p_m}   +\frac{1}{12}N p_i {^{(3)}R^i_{\; m}}\frac{\partial}{\partial p_m}     
  +\frac{1}{4}N  \frac{p_i p_j}{\omega_p^2} {^{(3)}R^{ ij}}    - \frac{1}{6} N {^{(3)}R}   + \xi N R \Bigg]{f}_2  \\
  - \frac{\hbar}{\omega_p}\Bigg[  \frac{1}{2} p_m N_{; \; \, q}^{\,m}  \frac{\partial}{\partial p_q} 
 +
\frac{1}{4}  N_{; \; j}^{\;j } 
  -  \frac{1}{2}   \frac{p_i p_j}{\omega_p^2}   N_{;}^{\; ij}
  - \frac{1}{2} p_j  N_{;m}  \frac{\partial}{\partial p_m}  D^j +\frac{1}{2} \frac{p_i p_j}{\omega_p^2}   N_{;}^{\,i} D^j  -  \frac{1}{4}N D_j D^j   
   \\
+\frac{1}{3} N p_i  p_j {^{(3)}R^{i \; \; \; \; j}_{\; q m }}  \frac{\partial^2}{\partial p_q \partial p_m}   +\frac{1}{12}N p_i {^{(3)}R^i_{\; m}}\frac{\partial}{\partial p_m}     
  +\frac{1}{4}N  \frac{p_i p_j}{\omega_p^2} {^{(3)}R^{ ij}}    - \frac{1}{6} N {^{(3)}R} + \xi N R \Bigg] {f}_1^{+}  \\
  -\frac{\lambda}{2} \frac{\omega_p}{\hbar }  \Bigg[  N \frac{\hbar^3}{\omega_p^2} \int \frac{d^3 q }{\gamma^{1/2}} \frac{ {f}_1^{+}(q) + {f}_2(q)}{\omega_q} -  \frac{\hbar^2}{8} \frac{\hbar^3}{\omega_p^2} \big[ N  \int \frac{d^3 q }{\gamma^{1/2}} \frac{ {f}_1^{+}(q) + {f}_2(q)}{\omega_q} \big]_{; k s} \frac{\partial^2}{\partial p_k \partial p_s}  \Bigg]  \Big[ {f}_1^{+} + {f}_2 \Big] \, .\label{finf3}
\end{multline}
Equations \eqref{finf1+} to \eqref{finf3} are the main result of this paper.\footnote{Although we excluded states containing one-point functions for simplicity, they will give rise to similar equations subject to a constraint equation due to the lack of degrees of freedom - such equations have been derived for example in \cite{Widrow:1993qq} where higher time derivatives and thus degrees of freedom were dropped. However, the conditions to obtain a leading order classical particle Vlasov equation \eqref{Vlasov} can only be satisfied on time-averages over the expectation values. This can be understood for example by considering the Minkowski space-time limit where the two solutions of $\langle \hat{f}_2 \rangle $ that are determined by condensates are given by $\langle \hat{f}_2 \rangle_{\text{cond}}^{\text{flat}} = \alpha \cos(2 \omega_p t) + \beta \sin (2 \omega_pt) $. Fixing the proportionality constants of these two solutions to be zero,  as we were able to do it for the general case,  would also set $\langle \hat{f}_1^{+} \rangle_{\text{cond}}^{\text{flat}} = (\alpha^2 + \beta^2)^{1/2}$ to zero and yields only a trivial solution of the system. The resolution is thus to keep all the degrees of freedom and perform a time-averaging in this case.} These equations are an effective description of the state-dependent (normal ordered) part  of the dynamics of a real scalar field quantum state in curved space-time in the language of phase-space variables $(x^{\mu},p_k)$, under the assumption that the state admits a  gradient and loop expansion. For macroscopic observables of systems, that have some notion of classicality, the quantities $f_1^{\pm}(x^{\mu},p_k)$ and $f_{2,3}(x^{\mu},p_k)$ should dominate over the state-independent part coming from the quantum commutation relation. They can be given any initial value that is compatible with the spatial gradient approximation and their symmetry properties. 
\par
We first note that all equations for the operators  $f_1^{\pm}$ and $f_{2,3}$ are spatially covariant and provide in principle candidates for phase-space density operators. However, we also need to realize that $f_{1}^+$ and ${f}_{2,3}$ are even functions in $p_k$ whereas ${f}_1^{-}$ is an odd function in the momentum. Thus, only some combination these two-point functions can account for the degrees of freedom of a classical particle phase-space density.
A promising candidate is read off from the first pair of equations \eqref{finf1+} and \eqref{finf1-} in the non-interacting limit,
\begin{multline}
 \Bigg[ \partial_{t} -  N^k {^{(3)}D}_k          + \big( {^{(3)} \nabla}_j N^k \big) p_k  \frac{\partial}{\partial p_j} + N  \frac{p^k}{\omega_p} {^{(3)}D}_k - \omega_p \big[ \partial_j  N \big] \frac{\partial}{\partial p_j} \\ +\frac{{\lambda}}{2}  {\omega_p}\Big[ N\frac{\hbar^3}{\omega_p^2} \int \frac{d^3 q }{\gamma^{1/2}} \frac{ {f}_1^{+}(q)}{\omega_q} \Big]_{;k} \frac{\partial}{\partial p_k} + \mathcal{O}\big( \hbar^2 \big)\Bigg] \Big[ {f}_1^{+} + {f}_1^{-}   \Big] 
 =      \mathcal{O} \big({f}_{2,3} \big)  \, . \label{almostVlasov}
\end{multline}
By rewriting the above equation for ${f}_{1} = f_1^+ + f_1^-$, we find the  Vlasov equation with a one loop correction, that can be interpreted as a mass shift, as well as source terms that are due to the additional correlators in the scalar field description and higher-order spatial gradient corrections. Undoing the ADM-decomposition, the equation reads
\bea
\Big[ p^{\mu} \partial_{\mu} +  p_{\mu} p^{\nu} \Gamma^{\mu}_{\; \nu i} \frac{\partial}{\partial p_i} +\frac{{\lambda}}{2}  {\omega_p}\Big[ N\frac{\hbar^3}{\omega_p^2} \int \frac{d^3 q }{\gamma^{1/2}} \frac{ {f}_1(q)}{\omega_q} \Big]_{;k} \frac{\partial}{\partial p_k} + \mathcal{O}\big( \hbar^2 \big) \Big] {f}_{1} (x^{\mu}, p_j) &=&   \mathcal{O} \big({f}_{2,3} \big)  \, , \label{Vlasov} \\
p^0 (x^{\mu}, p_j) := \sqrt{\big( g^{0j} p_j\big)^2 - g^{00} \big( m^2 + g^{ij} p_i p_j  \big)} = \sqrt{\big( g^{0j} p_j\big)^2 - g^{00} \omega_p^2}  && \, .
\eea
We remark that within the one loop approximation we do not find $2 \rightarrow 2$ particle scattering processes which come from self-energy diagrams whose first contribution is proportional to $\lambda^2$.\footnote{It is the one-loop approximation that allows the system to close on-shell since two-loop contributions will integrate off-shell energies which are not only supported on the mass-shell. However, one can employ a quasi-particle approximation for the 2-loop contributions which eventually leads also to a $2 \rightarrow 2$ particle scattering contribution as it appears on the right hand side of the classical particle Boltzmann equation. These properties for the $\lambda \phi^4$ theory are known in Minkowsky space \cite{trove.nla.gov.au/work/9783845}\cite{Berges:2015kfa}, but a general curved space-time discussion is still lacking.}  However, the self-masses $\propto \lambda$ are included and - depending on the problem - may already give significant corrections to the dynamics of the Vlasov equation. \par
Combining the other pair of equations \eqref{finf2} and \eqref{finf3} shows that ${f}_2$ and ${f}_3$ are to leading order oscillators with frequency of the particle energy $\omega_p$. Thus, equations \eqref{finf1+} to \eqref{finf3} generalize the Vlasov equation for relativistic particles in curved space-time by including the additional densities $f_{2,3}$. The latter densities can be rewritten as higher-order time derivatives acting on $f_1^{\pm}$. 
We conclude that if we wanted to recover the limit of a classical particle density, we would have to impose a state such that $ {f}_2 $ is initially of higher order in $\hbar$ and also remains of higher order in $\hbar$, which then translates into a condition for ${f}_3 $ and finally into ${f}_2  \sim \mathcal{O}(\hbar^2)  \big( {f}_1^{+}  \, , {f}_1^{-} \big) $ (these are rough estimates and it remains to be studied whether such conditions can be maintained by the dynamics). First-order corrections to \eqref{Vlasov}, that are contained in \eqref{finf1+} to \eqref{finf3}, may be obtained by expanding the phase-space densities into harmonics and see how the oscillatory terms back-react on the non-oscillatory part of the density $f_1$ via the self-interaction terms or via non-linear terms that are obtained by making use of the Einstein equations.
Also keeping in mind a generalization in terms of higher-loop effects, we think that the advantage of our formalism lies in an end-to-end link between quantum field theory and particle kinetics in curved space-time, that allows one to systemically include field theoretic corrections, while being able to refer to a (in some sense modified) particle interpretation.
\section{Generalized cold dark matter kinetics in linearized gravity}
In the last section we have dealt with a set of fairly general but lengthy equations. The idea of this section is to see how they reduce to more feasible sets of equations once we apply them to the concrete cosmological set up of cold dark matter perturbations between galactic scales and the Hubble horizon. The main result is a generalization of the kinetic description of classical particle cold dark matter as it discussed for example in \cite{Bernardeau:2001qr}. \par
Let us fix a linearly perturbed metric in FLRW background in the generalized Newtonian gauge, that includes vector and tensor perturbations and which is also referred to as Poisson gauge \cite{Bertschinger:1993xt} \cite{Bruni:1996im}. We label equal-time hypersurfaces by the variable $\eta$  and denote spatial coordinates by $x$,
\be
\partial_t \rightarrow \partial_{\eta} = \big(\,.\,\big)^{\prime} \, .
\ee Indices for the linear quantities are raised and lowered by the comoving background spatial metric $\delta_{ij}$ as in \cite{Malik:2008im}. The (3+1)-dimensional metric takes the form
\be
g_{\mu \nu}  = a^2
\begin{pmatrix}
   - (1 + 2 \Phi_N )& - s_i  \\
  - s_i & \delta_{ij}(1 - 2 \Psi_N) + h_{ij}  \\
 \end{pmatrix}\, , \; \delta^{kj} \partial_k s_j = 0\, , \; \delta^{kj} \partial_k h_{ji} = 0\, , \;\delta^{ij} h_{ij} = 0\, ,
\ee
such that the spatial metric, its inverse and its determinant are given to linear order by
\be
\gamma_{ij} = a^2 \big[ \delta_{ij}(1 - 2 \Psi_N) + h_{ij}\big]\, , \; \gamma^{ij} = a^{-2}\big[\delta^{ij}(1 + 2 \Psi_N) - h^{ij} \big]\, , \; {\gamma}^{1/2}= a^3 \big( 1 - 3 \Psi_N \big) \, ,
\ee
and the lapse function and shift vector read
\be
  N = a(1 + \Phi_N) \,, \quad N^i = -s^i = - \delta^{ij} s_j =  a^{-2} \delta^{ij} N_j\, .
\ee
We define a gravitational perturbation parameter related to the metric perturbations by
\be
\varepsilon_g \sim \Phi_N \, , s_i \, ,\Psi_N \, , h_{ij} \ll 1\, .
\ee
We have
\be
{^{(3)}\Gamma^l_{\; km}} = \delta^{ls} \delta_{mk}  \partial_s \Psi_N -\delta^l_{\;m} \partial_k \Psi_N - \delta^{l}_{\; k} \partial_m \Psi_N 
+\frac{1}{2}\big( \partial_k h^l_{\; m} + \partial_m h^l_{\;k} - \delta^{sl} \partial_s h_{km} \big)\,.
\ee
Let us collect further geometrical quantities, that appear in the Einstein equation in ADM decomposition.
\be
{^{(3)}R}_{ij} = \delta_{ij} \Delta \Psi_N + \partial_i \partial_j \Psi_N - \frac{1}{2} \Delta h_{ij} \, , \quad
{^{(3)}R} =  \frac{4}{a^2} \Delta \Psi_N\, , 
\ee
\bea
K_{ij} &=& - a \mathcal{H} \big[  \delta_{ij}(1- \Phi_N - \mathcal{H}^{-1}\Psi_N^{\prime} - 2 \Psi_N) + h_{ij}\big] - \frac{a}{2} \big[ s_{i,j} +s_{j,i} + h_{ij}^{\prime} \big] \, ,\\
K&=& - 3 a^{-1} \mathcal{H} (1- \Phi_N - \mathcal{H}^{-1}\Psi_N^{\prime})\, , \quad K^2 = 3 K_{ij} K^{ij} + \mathcal{O}\big( \varepsilon_g^2 \big)  \, .
\eea
We split the normal observer momentum vector and stress tensor in scalar, vector and tensor components,
\bea
{P}_i  &=& a \partial_i P_L + a {P}_i^{T}\, , \quad  \partial^j P^T_j = 0\, \label{momdecom},\\
{S}_{ij} &=& \frac{a^2\delta_{ij}}{3} S + a^2 \big(  \partial_i \partial_j S^A - \frac{\Delta}{3} \delta_{ij} S^A + \partial_i S_j + \partial_j S_i + S_{ij}^{TT} \big) ,   \partial^j S_j =   \partial^j S_{ij}^{TT}   = \delta^{ij} S_{ij}^{TT} = 0 .\label{stressdecom}
\eea
Indices for the quantities on the right-hand-side of \eqref{momdecom} and \eqref{stressdecom} will be raised and lowered with the flat three-dimensional metric.
Let us write down the Einstein equations in terms of the perturbed metric
\bea
  3 \mathcal{H}^2  + 2 \Delta \Psi_N  & =&  \frac{\hbar}{M_P^2} a^2\Big[  E -3 \mathcal{H} P_L  \Big] \, ,\label{E1}\\
\frac{1}{2} \Delta s_i  & = & - \frac{\hbar}{M_P^2} a^2 P_i^T \, , \label{E2}
\\
 \Phi_N - \Psi_N  &=& -
   \frac{\hbar}{M_P^2} a^2  S_A  \, , \label{E3}\\
    h_{ij}^{\prime \prime}  +2 \mathcal{H}  h_{ij}^{\prime}   
-  \Delta h_{ij}    & =&
 \frac{\hbar}{M_P^2}  2a^2{S}^{\text{TT}}_{ij}  \label{E4} \, ,
 \eea
 Energy-momentum conservation reads in linearized gravity
 \begin{multline}
\partial_{\eta} \big(a^3 \big[ 1 - 3 \Psi_N \big]E \big)  + a^3 \partial_i \Big(\big[\delta^{ij}(1 + \Phi_N-  \Psi_N) - h^{ij} \big]a^{-1} {P}_j + \delta^{ij} s_j  {E} \Big) \\+ a^3 \Big(  \mathcal{H} \big[ 1 - \mathcal{H}^{-1}\Psi_N^{\prime} - 3 \Psi_N\big]S  +  \big[ s_{i ,j} + \frac{1}{2} h^{\prime}_{ij} \big] {S}^{ij}  + a^{-1}\delta^{ij}{P}_j \partial_i \Phi_N \Big) = 0 \, ,  \label{econsLin}
\end{multline}
\begin{multline}
\partial_{\eta} \big(a^3 \big[ 1 - 3 \Psi_N \big] {P}_j \big) + a^3  \partial_i \big(a\big[ 1 + \Phi_N- 3 \Psi_N \big] {S}^i_{\, j} +  s^i {P}_j \big) \\ = a^4 \big(a^2 \frac{1}{2} {S}^{ik} \partial_j h_{ik} - S \partial_j \Psi_N - a^{-1} {P}_i \partial_j s^i  - {E} \partial_j \Phi_N \big) \, ,
\end{multline}
which does not help much unless we know how $S_{ij}$ depends on $E$ and $P_i$.
Note that it is suggestive to approximate these equations further with the linearized Einstein equations and rewrite $E, P_i, S_{ij}$ in terms of the gravitational perturbations. However, the resulting non-linear terms are not necessarily small since they involve gradient terms of the type $\mathcal{H}^{-2} \Delta $. Some of them may become important around the scale where the density contrast in Fourier space defined via $E(\eta,k) = \bar{E}(\eta) + \delta E(\eta, k)$ is of order one, $\bar{E}^{-1} \delta E (k_{NL}) \propto{\mathcal{H}^{-2}} k^2_{NL} \Psi_N(k_{NL}) \approx 1$. This scale is on the order of  roughly $k^{-1}_{\text{NL}} \approx 5 \, \text{Mpc}$. We emphasize that linearization in the gravitational perturbations can still be valid  on these scales, although the density contrast has to be treated non-linearly.
In the context of cosmological large-scale structures one is typically interested in the evolution on sub-Hubble scales ($ k^{-1}_{H} \lesssim 10^{4} \text{Mpc} $). We capture the corrections, that result from separations with respect to this scale, by introducing a perturbation parameter $\varepsilon_{H}$,
\be
\mathcal{O}\big( \varepsilon_{H}^{-1} \big) \delta g_{\mu \nu} \sim \mathcal{H}^{-2} \Delta\delta g_{\mu \nu} \gg   \delta g_{\mu \nu} \,.
\ee
This expansion allows us for example to drop several corrections in $E$, $P_i$ and $S_{ij}$, that are related to the perturbation of the determinant of the spatial metric $\delta \gamma^{1/2}$. On the other hand, the smallest large-scale structures we are interested in are related to galactic scales $k^{-1}_{\text{g}} \sim 10 \, \text{kpc}$. In order to be consistent with our perturbative schemes, we have to contrast the scale $k^{-1}_{\text{g}}$ with the de Broglie wave length $k^{-1}_{dB} f_i (\vec{k},\vec{p}) \sim \hbar \parallel \frac{\partial}{\partial p_j} f_i (\vec{k},\vec{p}) \parallel$ which was related to the spatial gradient expansion that we have used to derive the kinetic equations \eqref{finf1+} to \eqref{finf3}. By using typical galaxy velocities of $v_{\text{g}} \approx 10^{-3} c$ we can express the de Broglie wavelength in terms of the Compton wavelength $k_{\text{C}}^{-1} \propto \hbar m^{-1}$ as $k^{-1}_{\text{dB}} \sim 10^3 k^{-1}_{\text{C}}$. For dark matter, the mass of which is at the electroweak scale ($\sim 10^{2} \, \text{GeV}$), we find that de Broglie wavelength is of order $k^{-1}_{\text{dB,EW}} \sim 10^{-33} \, \text{kpc} $ and thus spatial gradient corrections can be safely neglected, whereas for ultralight dark matter with mass $\sim 10^{-31} \, \text{GeV}$ we find $k^{-1}_{\text{dB,UL}} \sim 10^{6} \, \text{kpc}$ such that gradient corrections can play a role at galactic scales.
However, let us focus here on the less exotic case where $k^{-1}_g \gg k^{-1}_{\text{dB}}$. Moreover, we are given a non-relativistic expansion by means of the galactic velocities 
\be
\varepsilon_p    \sim \frac{p_i p_j \gamma^{ij}}{m^2}     \sim 10^{-6}   \, ,
\ee
such that the particle energy is dominated by the mass. 
This relation justifies at least for certain mass ranges the inclusion of a self-coupling term in the kinetic equations, as we will see shortly.
We also want to consider small corrections to the classical particle density picture and demand
\be
f_{1} \gg \left| f_{2,3} \right|\, .
\ee
In order to stick close to the cold dark matter scenario, we also want a first bound on the dark matter interactions such that it does not source the Hubble rate too much 
\be
\lambda \frac{\hbar^3}{m^3} \int \frac{d^3 p}{\gamma^{1/2}} {f}_1^+ \ll 1 \, . \label{interactionConst}
\ee
Moreover, we want to keep the influence of the non-minimal coupling fairly small such that it cannot spoil the smallness of gradients or gravitational perturbations,
\be
 {\hbar^2}|  \xi R | \lesssim  {m^2} \, .
\ee
We now express the leading order terms of the right-hand-side of \eqref{E1} to \eqref{E4} in terms of the phase-space densities such that we can plug the constraint equations back in the kinetic equations for $f_{1}^{\pm}$, $f_2$ and $f_3$ and solve them together with the gravitational wave equation. In accordance with the slightly more general discussion around \eqref{EClass} to \eqref{SClass}, we find that the gravitational perturbations get their leading order contributions between galactic scales and the Hubble scale from the two phase-space densities $f_{1}^{\pm}$ (as is the case for the classical particle cold dark matter scenario if we split the classical density into even and odd parts). The Poisson equation reads
\be
  3 \mathcal{H}^2  + 2 \Delta \Psi_N  \approx  \frac{\hbar}{M_P^2} \frac{m}{a} \int {d^3 p} \, {f}_1^+ = \frac{\hbar}{M_P^2} \frac{m}{a} \int {d^3 p} \, {f}_1\, ,
\ee
and we note that the constraint \eqref{interactionConst} relates the mass and the coupling via
\be
\lambda \frac{\hbar^3}{m^3} \int \frac{d^3 p}{\gamma^{1/2}} {f}_1^+ \ll 1 \quad \longrightarrow \, \quad\lambda \Big( \frac{\hbar^2 \mathcal{H}^2}{a^2 m^2} \Big) \frac{M_P^2}{m^2} \sim  \lambda \frac{10^{-8} \big(\text{eV}\big)^4}{m^4} \ll 1. \label{constSelfHub}
\ee
It is now clear that for masses around the electroweak scale the interaction energy does not influence the Hubble rate whereas it can become important for ultralight particles already for very small couplings.
Moreover, vector perturbations and the gravitational slip are given by
\bea
\frac{1}{2} \Delta^2 s_i  & \approx  & - \frac{\hbar}{M_P^2} a^{-1} \Big[ \Delta \int {d^3 p} \, {p_k} {f}_1^{-} - {\partial_i\partial^k} \int {d^3 p} \, {p_k} {f}_1^{-}\Big] \, , 
\\
 \Delta^2 \big( \Phi_N - \Psi_N \big) &\approx & 
   \frac{\hbar}{M_P^2} \frac{3}{2} a^{-3} \Big[\frac{\Delta}{3}\delta^{kj} - {\partial^k\partial^j}\Big]   \int {d^3 p} \, {p_k}p_j  {f}_1^{+} \, .
   \eea
Note that we can replace $f_1^{\pm}$ with $f_1 = f_1^+ + f_1^-$ in these equations due to their symmetry properties.
The only dynamical gravitational perturbation are the traceless, transverse tensor perturbations which obey
 \begin{multline}
     h_{ij}^{\prime \prime}  +2 \mathcal{H}  h_{ij}^{\prime}   
-  \Delta h_{ij}     \approx \\
 \frac{\hbar}{M_P^2}  \frac{2}{m a^3} \int {d^3 p} \Bigg[  \, {p_i}p_j   - \frac{\delta_{ij}}{3} p_k p_m \delta^{km} +   p_k p_m \frac{\Delta^{-2}}{2} \Big(\partial_i \partial_j + \Delta \delta_{ij} \Big)\Big(\partial^k \partial^m - \frac{ \Delta}{3} \delta^{km} \Big) \\ + p_k p_m \delta^{km} \frac{2}{3} \Delta^{-1} \partial_i \partial_j  - \Delta^{-1} \Big( p_k p_j \partial_i \partial^k  + p_k p_i  \partial_j \partial^k \Big)   \Bigg]{f}_1^{+} \, .
 \end{multline}
Thus, the Einstein equations look to leading order in our perturbation parameters the same, whether we use a classical particle phase-space density or the density derived from the scalar quantum field, $f_1 = f_1^+ + f_1^-$. The densities $f_{2,3}$ enter at higher order. However, the dynamics for this source in the Einstein-equations is generalized by the following set of differential equations including the densities $f_{2,3}$.
We find for the phase-space density $f_1^+$,
\begin{multline}
\big({f}_1^{+}  \big)^{\prime} + \Big[   s^k  \partial_k   -
  \big(\partial_m s^k \big) p_k \frac{\partial}{\partial p_m}   \Big]{f}_1^{+}  
\approx     \\
-  \Bigg[   \delta^{jk} \frac{p_j}{m a}    \partial_k    
-  m a \partial_k \Bigg[  \Phi_N - \frac{\lambda}{2} \frac{\hbar^3}{m^3 a^3} \int d^3 q f_1^{+}(q) \Bigg] \frac{\partial}{\partial p_k}  \Bigg]{f}_1^{-}   \\
+3 \Big[   \mathcal{H} - \Psi_N^{\prime}   \Big] {f}_2 
-\frac{\hbar}{ma} \Bigg[   \frac{1}{4} \delta^{ij} \partial_i \partial_j  +  \frac{\lambda}{2} \frac{\hbar^2}{m^2 a^2} \int d^3 q f_1^{+}(q) \Bigg] {f}_3  \, , \label{f1+LinGrav}
\end{multline}
where we drop higher-order terms involving relativistic corrections or gradients that are small compared to the mass scale. Note that the last term in \eqref{f1+LinGrav} may be important for certain combinations of masses and self-couplings, which is still consistent with the constraint \eqref{constSelfHub},
\be
|| {\partial_{\eta} } ||^{-1} \lambda \frac{\hbar^2}{m^2 a^2} \int d^3 q f_1^{+}(q) \sim \lambda \frac{\hbar \mathcal{H}}{a m}  \frac{M_P^2}{m^2} \sim  \lambda \frac{10^{-3} \big(\text{GeV}\big)^3}{m^3}.
\ee
Maybe more importantly, the self-interaction term multiplying $f_1^{-}$ in \eqref{f1+LinGrav},
\be
\partial_k \Bigg[  \Phi_N - \frac{\lambda}{2} \frac{\hbar^3}{m^3 a^3} \int d^3 q f_1^{+}(q) \Bigg] \sim \partial_k \Bigg[  \Phi_N - \frac{\lambda}{2} \frac{ \hbar^2 \mathcal{H}^2}{m^2 a^2 } \frac{M_P^2}{m^2} \frac{\Delta \Psi_N  }{\mathcal{H}^2} \Bigg]
\ee
can compete with the Newtonian potential at the non-linear scale where $\Delta \Psi_N \sim \mathcal{H}^2$ and still obey the constraint \eqref{constSelfHub} for certain combinations of mass and self-coupling,
\be
\Phi_N (k_{\text{NL}}) \sim  \frac{\lambda}{2} \frac{ \hbar^2 \mathcal{H}^2}{m^2 a^2 }\frac{M_P^2}{m^2} \frac{ k_{\text{NL}}^2  }{\mathcal{H}^2} \Psi_N (k_{\text{NL}}) \, \quad \text{for} \quad   \frac{\lambda}{2} \frac{ \hbar^2 \mathcal{H}^2}{m^2 a^2 }\frac{M_P^2}{m^2} \sim 10^{-5} \ll 1\, .
\ee

We also note that the gravitational vector perturbations enter at this order as a corrective for the time derivative, which is true for all four densities as we will see in a moment. Tensor perturbations enter in equations like \eqref{f1+LinGrav} in various places, however, such terms are all of higher-order in the spatial gradient expansion. The same is again true for the dynamical equations of the other densities. Also, terms involving the non-minimal coupling $\xi$ are of higher-order in all equations.
For the odd density $f_1^-$ we find,
\begin{multline}
 \big({f}_1^{-}  \big)^{\prime} + \Big[   s^k  \partial_k   -
  \big(\partial_m s^k \big) p_k \frac{\partial}{\partial p_m}   \Big]{f}_1^{-} \approx   
\\ -  \Bigg[   \delta^{jk} \frac{p_j}{m a}    \partial_k     -  m a \partial_k \Bigg[  \Phi_N - \frac{\lambda}{2} \frac{\hbar^3}{m^3 a^3} \int d^3 q f_1^{+}(q) \Bigg] \frac{\partial}{\partial p_k}   \Bigg]{f}_1^{+} \\   
-\frac{\hbar}{2} \big(\partial_j \Psi_N^{\prime} ) \frac{\partial}{\partial p_j}   {f}_3 
- \delta^{jk} \frac{p_j}{m a}    \partial_k  
 {f}_2   \, . \label{f1-Lin}
\end{multline}
The term involving the density $f_3$ is probably negligible for $f_1 \gg | f_{2,3} |$, however we kept it to see the type of the leading order term for $f_3$. 
The differential equations for $f_2$ and $f_3$ read
\begin{multline}
\big({f}_2  \big)^{\prime} + \Big[   s^k  \partial_k   -
  \big(\partial_m s^k \big) p_k \frac{\partial}{\partial p_m}   \Big]{f}_2  
\approx   2 \frac{\omega_p}{\hbar}  a \big( 1 + \Phi_N \big)    {f}_3\\
+ \Bigg[   \delta^{jk} \frac{p_j}{m a}    \partial_k     -  m a \partial_k \Bigg[  \Phi_N - \frac{\lambda}{2} \frac{\hbar^3}{m^3 a^3} \int d^3 q f_1^{+}(q) \Bigg] \frac{\partial}{\partial p_k}   \Bigg]  {f}_1^{-} \\
+3 \big( \mathcal{H} -  \Psi_N^{\prime} \big) {f}_1^{+}   
-\frac{\hbar}{ma} \Bigg[   \frac{1}{4} \delta^{ij} \partial_i \partial_j - \frac{\lambda}{2} \frac{\hbar^2}{m^2 a^2} \int d^3 q f_1^{+}(q)  \Bigg] {f}_3  \, , \label{f2LinGrav}
\end{multline}
\begin{multline}
\label{f3LinGrav} \big({f}_3  \big)^{\prime} + \Big[   s^k  \partial_k   -
  \big(\partial_m s^k \big) p_k \frac{\partial}{\partial p_m}   \Big]{f}_3  
\approx   - 2 \frac{\omega_p}{\hbar}  a \big( 1 + \Phi_N \big)    {f}_2 +
\hbar \frac{3}{2}  \big( \partial_j \Psi_N^{\prime} \big) \frac{\partial}{\partial p_j} {f}_1^{-}
\\
+ \frac{\hbar}{ma} \Bigg[   \frac{1}{4} \delta^{ij} \partial_i \partial_j -  \frac{\lambda}{2} \frac{\hbar^2}{m^2 a^2} \int d^3 q f_1^{+}(q)   \Bigg] \big(  f_1^+  + {f}_2 \big)    \, ,
\end{multline}
where we also included higher-order gradient terms acting on $f_3$ and the self-coupling, as they might play a role in determining the non-oscillatory behavior of $f_2$ and $f_3$. 
Also, the correction to the rest-mass energy may be included for the first term on the right-hand-side of \eqref{f2LinGrav} and \eqref{f3LinGrav},
\be
\omega_p \approx m \Big(1  +  \frac{1}{2} \frac{p_i p_j \delta^{ij}}{ m^2 a^2}  \Big) \, .
\ee
We remark once more, that the equations \eqref{f1+LinGrav} to \eqref{f3LinGrav} reduce to the classical particle, cold dark matter phase-space dynamics if we can approximate $f_{2,3} \approx 0$ and set $\lambda=0$. However, we think that the additional densities  $f_{2,3}$ have the potential to significantly alter the evolution of $f_1^{\pm}$ for certain combinations of parameters. 
As a first step, we are currently investigating the effect of the oscillatory densities $f_{2,3}$ on the density $f_1$ and thus the Hubble rate $\mathcal{H}$ in the homogeneous limit. 
\section{Conclusion and outlook}
Motivated by dark matter models for large-scale structures we introduced a spatially covariant framework based on canonical field operators $\hat{\phi}$, $\hat{\Pi}$ to study the transition from the quantum theory of a self-interacting real scalar field on curved space-time to the kinetic theory of classical particles  by using a spatial gradient expansion. We also included a non-minimal coupling to the Ricci scalar, since it is required at the level of bare parameters and non-renormalized interaction terms.
We used a c-number metric that is determined through the semi-classical Einstein equations, although in principle we could have taken any classical metric for our deviation. It was in this sense unrestricted. The metric is a c-number with respect to quantum expectation values but might be taken to be stochastic as a one-point function to account for stochastic features of cosmological settings. 
 Moreover, we considered a Gaussian state or one-loop truncation and neglected the effect of connected higher-order n-point functions related to the self-coupling, anomalous contributions, that result from the renormalization procedure. These effects can in principle be included and it depends on the scales and couplings of the underlying problem whether they become relevant.
 \par
In \eqref{deff+} to \eqref{deff2}, we identified four  phase-space operators $\hat{f}_1^{\pm},\hat{f}_{2,3}$ which depend on a space-time point $x^{\mu}$ and a three-momentum $p_k$. Two of them can be combined and interpreted as a fluctuating phase-space density $\hat{f}_1 = \hat{f}_1^+ + \hat{f}_1^-$, the average of which, $f_1 = \langle \hat{f}_1 \rangle$,  describes a classical statistically-distributed one-particle density, whenever the quantum state of the system is such that the other two phase-space operators are on average small $f_1 = \langle \hat{f}_1 \rangle \gg | \langle \hat{f}_{2,3} \rangle|$ (expectation values of n factors of $\hat{f}_1$ are after subtraction of their disconnected piece similarly interpreted as n-particle phase-space densities) .
This picture is consistent when we rewrite the hydrodynamic energy density, pressure and velocity in the non-interacting limit in terms of momentum integrals over $f_1$. However, the main result of this paper are the dynamical equations \eqref{finf1+} to \eqref{finf3} for the phase-space densities $f_{1}^{\pm}$, $f_{2,3}$ which describe up to one-point functions all degrees of freedom of a Gaussian state. We are not aware that equations \eqref{finf1+} to  \eqref{finf3}  have been derived elsewhere for general curved space-times. Moreover, these equations support the interpretation that the density $f_1$ has a limit as a classical one-particle density since the equations \eqref{finf1+} to  \eqref{finf3} reduce to the Vlasov equation \eqref{Vlasov} to lowest order in the gradient expansion and upon neglecting the densities $f_2$ and $f_3$ and the self-interaction which amounts to a mass correction in the one-loop approximation. 
\par
In the derivation of the kinetic description of the real scalar quantum field, we argue that it is necessary to normal order the involved quadratic field operators \eqref{genDefWignerOp}  also in the off-coincident limit, since only then one is able to extract quantities, that yield a well-defined renormalized energy-momentum tensor  and whose dynamics can be approximated with a finite number of spatial derivatives.
As far as we know, this problem has not been addressed in detail in the context of quantum kinetic theory in curved space-time and it should be further investigated whether the boundary terms related to the local subtraction can be given a quantum noise interpretation.
\par
Eventually, we have used the general kinetic equations \eqref{finf1+} to \eqref{finf3} to extend our earlier results on scalar field dark matter with linearized gravity \cite{Prokopec:2017ldn} to include vector and tensor perturbations as well as self-interaction terms. The resulting equations generalize previous cold dark matter descriptions and have, as far as we know, never been studied before. Note, that we did not include condensates or one-point functions, a popular description of dark matter that goes under the name fuzzy, wave or axion dark matter, which has been around for a long time \cite{Turner:1983he} \cite{Sin:1992bg} \cite{Lee:1995af} \cite{Hu:2000ke} \cite{Goodman:2000tg} \cite{Peebles:2000yy} \cite{Marsh:2015daa} \cite{Hui:2016ltb}. Equipped with a very small mass, the real scalar field condensate leads to different behaviour on small scales. Recently strong bounds on the mass of fuzzy dark matter have been obtained \cite{Irsic:2017yje} and more elaborate models combining fuzzy and cold dark matter were suggested \cite{Armengaud:2017nkf}. Such a condensate component of the state is easily incorporated into our formalism by adding source terms for the Einstein equations \eqref{eMomDecom1} to \eqref{eMomDecom3} via the shift $\hat{f}_{XY} \rightarrow \hat{f}_{(X-\langle X \rangle)(Y-\langle Y \rangle)}$ in \eqref{genDefWignerOp}. The dynamics of the condensates can quickly be derived by taking expectation values of the dynamical equations for the canonical field operators \eqref{opEqPhi} and \eqref{opEqPi}  whose non-linear terms have to be expressed in terms of one-point functions and the one-particle phase-space densities (which are related to the connected part of the two-point function). The coupling between one-point functions and the connected two-point functions happens then directly via one-loop self-interactions or indirectly via the gravitational fields and it is promising to study whether and on which scales the particle or the condensate nature dominates (such a dark matter model, that differentiates between different matter phases depending on the scales has been proposed by \cite{Berezhiani:2015bqa}). Moreover, our formalism can be useful in studying how a Quintessence field \cite{Tsujikawa:2013fta} that goes beyond a condensate, can play role in large-scale structure dynamics. In this case an additional degree of freedom has to be added to play the role of dark matter itself. Another application we have in mind for our formalism is to study the interplay between gravitational waves and the real scalar field on space and time scales where the other gravitational potentials give negligible effects.
 \par
We think our results are important to systematically include special and general relativistic corrections to dark matter models and study their range of applicability. By using non-equilibrium quantum field theory techniques like the Schwinger-Keldysh formalism and the classical limits we have established in this paper, we also hope to give soon alternative routes in analytical and numerical studies of dark matter beyond fluid approximations, particularly concerning the small scale behavior of large-scale structures.

\paragraph{Acknowledgments.}
This work is part of the research programme of the Foundation for Fundamental Research on Matter (FOM), which is part of the Netherlands Organisation for Scientific Research (NWO). This work is in part supported by the D-ITP consortium, a program of the Netherlands Organization for Scientific Research (NWO) that is funded by the Dutch Ministry of Education, Culture and Science (OCW). 
\pagebreak
\appendix
\section{Appendix}
\subsection{Exponentiating covariant derivatives \label{proofExpCovDer}}
Assume that 
\be
\Big[ { {r^k}{^{(3)}{\nabla}_k^H}} \Big]^n  \hat{\phi} =  {r^{i_1}...r^{i_n} } {^{(3)}{\nabla}_{i_1}}... {^{(3)}{\nabla}_{i_n}} f(x^{\mu})  = \Bigg[{ {r^k} \Big[ \partial_k -  {^{(3)}\Gamma^n_{\; kl}} r^l \frac{\partial}{\partial r^n} }  \Big]\Bigg]^n f(x^{\mu}) \,,
\ee
holds for a certain $n>2$. The case $n=2$ is satisfied as can be verified by a quick calculation. We now show that the relation holds also for $n+1$ provided it holds for $n$ and we are done
\begin{multline}
\Big[ { {r^k}{^{(3)}{\nabla}_k^H}} \Big]^{n+1} f(x^{\mu}) =  {r^{i_1}...r^{i_{n+1}} } {^{(3)}{\nabla}_{i_1}}... {^{(3)}{\nabla}_{i_{n+1}}} f(x^{\mu}) \\
= r^k {r^{i_1}...r^{i_{n}} }  \Bigg[\partial_k\Big[ {^{(3)}{\nabla}_{i_1}}... {^{(3)}{\nabla}_{i_{n}}} f(x^{\mu}) \Big] \\ -{^{(3)} \Gamma^{j_1}_{\; k i_1}}\Big[ {^{(3)}{\nabla}_{j_1}}... {^{(3)}{\nabla}_{i_{n}}} f(x^{\mu}) \Big] - ... - {^{(3)}\Gamma^{j_n}_{\; k i_n}  } \Big[ {^{(3)}{\nabla}_{i_1}}... {^{(3)}{\nabla}_{j_{n}}} f(x^{\mu}) \Big]\Bigg] \\
= r^k  \Bigg[r^{i_1}...r^{i_{n}}   \partial_k\Big[ {^{(3)}{\nabla}_{i_1}}... {^{(3)}{\nabla}_{i_{n}}} f(x^{\mu}) \Big] \\ -  {^{(3)} \Gamma^{j_1}_{\; k l}}  r^{l}  
\frac{\partial}{\partial r^{j_1}} \Big[r^{i_1}...r^{i_{n}} {^{(3)}{\nabla}_{i_1}}... {^{(3)}{\nabla}_{i_{n}}} f(x^{\mu}) \Big]+ r^{i_1} {^{(3)} \Gamma^{j_1}_{\; k l}}  r^{l}  
\frac{\partial}{\partial r^{j_1}} \Big[r^{i_2} ...r^{i_{n}} {^{(3)}{\nabla}_{i_1}}... {^{(3)}{\nabla}_{i_{n}}} f(x^{\mu}) \Big] \\ - r^{i_1}...r^{i_{n}} {^{(3)}\Gamma^{j_2}_{\; k i_2}  } \Big[ {^{(3)}{\nabla}_{i_1}}{^{(3)}{\nabla}_{j_2}}... {^{(3)}{\nabla}_{i_{n}}} f(x^{\mu}) \Big] - ... - r^{i_1}...r^{i_{n}} {^{(3)}\Gamma^{j_n}_{\; k i_n}  } \Big[ {^{(3)}{\nabla}_{i_1}}... {^{(3)}{\nabla}_{j_{n}}} f(x^{\mu}) \Big]\Bigg] \\
=
r^k  \Bigg[r^{i_1}...r^{i_{n}}   \partial_k\Big[ {^{(3)}{\nabla}_{i_1}}... {^{(3)}{\nabla}_{i_{n}}}f(x^{\mu}) \Big]  -  {^{(3)} \Gamma^{n}_{\; k l}}  r^{l}  
\frac{\partial}{\partial r^{n}} \Big[r^{i_1}...r^{i_{n}} {^{(3)}{\nabla}_{i_1}}... {^{(3)}{\nabla}_{i_{n}}} f(x^{\mu}) \Big] \Bigg] \\
={ {r^k}\Big[ \partial_k -  {^{(3)}\Gamma^n_{\; kl}} r^l \frac{\partial}{\partial r^n} }  \Big]\Bigg[r^{i_1}...r^{i_{n}} {^{(3)}{\nabla}_{i_1}}... {^{(3)}{\nabla}_{i_{n}}} f(x^{\mu}) \Bigg] = \Bigg[{ {r^k} \Big[ \partial_k -  {^{(3)}\Gamma^n_{\; kl}} r^l \frac{\partial}{\partial r^n} }  \Big]\Bigg]^{n+1} f(x^{\mu}) \, .
\end{multline}
\subsection{Traces in terms of two-point functions of canonical fields \label{traces}}
The hydrodynamic representation of the energy-momentum tensor \eqref{TMuNuScalar},
\be
\langle : \hat{T}_{\mu \nu} : \rangle_{\lambda=\xi=0} = \langle : \partial_{\mu} \hat{\phi} \partial_{\nu} \hat{\phi} : \rangle - \frac{g_{\mu \nu}}{2} \Big[ \langle : \partial^{\alpha} \hat{\phi} \partial_{\alpha} \hat{\phi} : \rangle + \frac{m^2}{\hbar^2} \langle : \hat{\phi}^2 :  \rangle \Big] \, ,
\ee 
is related to its diagonalization which can be written in terms of traces of
\be
 \chi^{\mu}_{\; \nu} := \langle : \partial^{\mu} \hat{\phi} \partial_{\nu} \hat{\phi} :  \rangle \, .
\ee
These three traces read in terms of the two-point functions of canonical field operators as follows,
\begin{multline}
\begin{aligned}
\tr \big[\chi \big] = \langle : \partial^{\mu} \hat{\phi} \partial_{\mu} \hat{\phi} : \rangle &= - \gamma^{-1} \langle : \hat{\Pi} \hat{\Pi} : \rangle + \gamma^{ij} \langle :  \partial_i \hat{\phi} \partial_j \hat{\phi} :  \rangle \, ,\\
\tr \big[\chi^2 \big] = \langle : \partial^{\mu} \hat{\phi} \partial_{\nu} \hat{\phi} :  \rangle \langle : \partial^{\nu} \hat{\phi}\partial_{\mu} \hat{\phi} :  \rangle &=\gamma^{-2} \langle :  \hat{\Pi} \hat{\Pi} : \rangle^2- 2 \gamma^{-1} \langle : \hat{\Pi} \partial_i \hat{\phi} : \rangle \gamma^{ij} \langle : \partial_j \hat{\phi} \hat{\Pi}  : \rangle \\ &\quad +  \langle : \partial_i \hat{\phi} \partial_j \hat{\phi} :  \rangle\gamma^{jk} \gamma^{il} \langle : \partial_k \hat{\phi} \partial_l \hat{\phi} : \rangle \, , \\
\tr \big[\chi^3 \big] =\langle : \partial^{\mu} \hat{\phi} \partial_{\nu} \hat{\phi}: \rangle  \langle : \partial^{\nu} \hat{\phi}\partial_{\rho} \hat{\phi}: \rangle  \langle : \partial^{\rho} \hat{\phi}\partial_{\mu} \hat{\phi}: \rangle &=- \gamma^{-3} \langle : \hat{\Pi} \hat{\Pi}: \rangle ^3 + 3 \gamma^{-2}\langle : \hat{\Pi} \hat{\Pi}: \rangle  \langle : \hat{\Pi} \partial_i \hat{\phi}: \rangle  \gamma^{ij} \langle : \partial_j \hat{\phi} \hat{\Pi} : \rangle   \\  &\quad
- 3 \gamma^{-1} \langle : \hat{\Pi} \partial_i \hat{\phi}: \rangle  \gamma^{ij}\langle : \partial_j \hat{\phi} \partial_l \hat{\phi} : \rangle  \gamma^{kl} \langle : \partial_k \hat{\phi} \hat{\Pi} : \rangle  \\  &\quad +   \gamma^{mn}\langle : \partial_n \hat{\phi} \partial_j \hat{\phi}: \rangle \gamma^{jk} \langle : \partial_k \hat{\phi} \partial_l \hat{\phi}: \rangle  \gamma^{il}\langle : \partial_i \hat{\phi} \partial_m \hat{\phi}: \rangle  \, ,
\end{aligned}
\end{multline}
\begin{gather}
\begin{aligned}
\widetilde{b} &=\gamma^{-1} \langle : \hat{\Pi} \hat{\Pi}: \rangle  - \gamma^{ij} \langle : \partial_i \hat{\phi} \partial_j \hat{\phi}: \rangle     \, , \\
\widetilde{c} &= -\gamma^{-1} \langle : \hat{\Pi} \hat{\Pi}: \rangle \gamma^{ij} \langle : \partial_i \hat{\phi} \partial_j \hat{\phi}: \rangle + \gamma^{-1} \langle : \hat{\Pi} \partial_i \hat{\phi}: \rangle  \gamma^{ij} \langle : \partial_j \hat{\phi} \hat{\Pi} : \rangle   - \frac{1}{2} \Big[ \gamma^{ij} \langle : \partial_i \hat{\phi} \partial_j \hat{\phi}: \rangle    \Big]^2  \\ & \qquad + \frac{1}{2}  \langle : \partial_i \hat{\phi} \partial_j \hat{\phi}: \rangle \gamma^{jk} \gamma^{il} \langle : \partial_k \hat{\phi} \partial_l \hat{\phi}: \rangle   \, , \\
\widetilde{d} &=  \frac{1}{2} \gamma^{-1} \langle : \hat{\Pi} \hat{\Pi}: \rangle 
   \Bigg[ \Big[ \gamma^{ij} \langle : \partial_i \hat{\phi} \partial_j \hat{\phi}: \rangle    \Big]^2  -  \langle : \partial_i \hat{\phi} \partial_j \hat{\phi}: \rangle \gamma^{jk} \gamma^{il} \langle : \partial_k \hat{\phi} \partial_l \hat{\phi}: \rangle  \Bigg] \\  &\quad
+ \gamma^{-1} \langle : \hat{\Pi} \partial_i \hat{\phi}: \rangle  \gamma^{ij}\langle : \partial_j \hat{\phi} \partial_l \hat{\phi} : \rangle  \gamma^{kl} \langle : \partial_k \hat{\phi} \hat{\Pi} : \rangle  \\  &\quad - \frac{1}{3}   \gamma^{mn}\langle : \partial_n \hat{\phi} \partial_j \hat{\phi}: \rangle \gamma^{jk} \langle : \partial_k \hat{\phi} \partial_l \hat{\phi}: \rangle  \gamma^{il}\langle : \partial_i \hat{\phi} \partial_m \hat{\phi}: \rangle  \, , \\ 
& \quad  - \gamma^{ij} \langle : \partial_i \hat{\phi} \partial_j \hat{\phi}: \rangle   \ \Bigg[ \gamma^{-1} \langle : \hat{\Pi} \partial_i \hat{\phi}: \rangle  \gamma^{ij} \langle : \partial_j \hat{\phi} \hat{\Pi} : \rangle  \\ & \qquad\qquad\qquad\qquad\qquad\qquad - \frac{1}{2}\langle : \partial_i \hat{\phi} \partial_j \hat{\phi}: \rangle \gamma^{jk} \gamma^{il} \langle : \partial_k \hat{\phi} \partial_l \hat{\phi}: \rangle  +\frac{1}{6} \Big[\gamma^{ij} \langle : \partial_i \hat{\phi} \partial_j \hat{\phi}: \rangle   \Big]^2 \Bigg]
\\
\widetilde{e} &=  - \frac{1}{6} \tilde{\epsilon}^{ijk}\Bigg[\gamma^{-1} \langle : \hat{\Pi} \hat{\Pi}: \rangle  + \frac{N^k}{N} \gamma^{-1/2}\langle : \hat{\Pi} \partial_k \hat{\phi}: \rangle   \Bigg] \Bigg[\gamma^{-1} \tilde{\epsilon}^{lmn} \Big[ \langle : \partial_i \hat{\phi} \partial_l \hat{\phi}: \rangle \langle : \partial_j \hat{\phi} \partial_m \hat{\phi}: \rangle \langle : \partial_k \hat{\phi} \partial_n \hat{\phi}: \rangle \Big] \\ & \qquad \qquad + \frac{1}{2}  \gamma^{-1/2}\tilde{\epsilon}_{lmn} \gamma^{la}\gamma^{mb} \frac{N^n}{N}\langle : \hat{\Pi} \partial_i \hat{\phi}: \rangle  \langle : \partial_a \hat{\phi} \partial_j \hat{\phi}: \rangle \langle : \partial_b \hat{\phi} \partial_k \hat{\phi}: \rangle    \Bigg] \, .
\end{aligned}
\end{gather}
The determinant is given by
\begin{multline}
\det \chi =  - \frac{1}{6} \tilde{\epsilon}^{ijk}\Big[\gamma^{-1} \langle : \hat{\Pi} \hat{\Pi}: \rangle  + \frac{N^k}{N} \gamma^{-1/2}\langle : \hat{\Pi} \partial_k \hat{\phi}: \rangle   \Big] \Big[\gamma^{-1} \tilde{\epsilon}^{lmn} \langle : \partial_i \hat{\phi} \partial_l \hat{\phi}: \rangle \langle : \partial_j \hat{\phi} \partial_m \hat{\phi}: \rangle \langle : \partial_k \hat{\phi} \partial_n \hat{\phi}: \rangle  \\  + \frac{1}{2}  \gamma^{-1/2}\tilde{\epsilon}_{lmn} \gamma^{la}\gamma^{mb} \frac{N^n}{N}\langle : \hat{\Pi} \partial_i \hat{\phi}: \rangle  \langle : \partial_a \hat{\phi} \partial_j \hat{\phi}: \rangle \langle : \partial_b \hat{\phi} \partial_k \hat{\phi}: \rangle    \Big]\,  ,
 \end{multline}
 where $\tilde{\epsilon}$ denotes is the totally anti-symmetric symbol.
\subsection{Geometry of tangent bundles: definitions and identities \label{defAndId}}
The expressions involved in \eqref{F00BeforeInt} to \eqref{F11BeforeInt}, which appear before integrating over the spatial tangent space (associated to the hypersurface $\Sigma_t$ at a common space-point $x^{\mu}$), can be expressed in terms of the geometry of tangent bundles which is covered for example in \cite{Sarbach:2013uba} and \cite{de2011methods}. 
We use the general notation
\be
X_R = X^{k}(x^{\mu}, r^k) \frac{\partial}{\partial r^k}\, , \quad  X_E = X^{k}(x^{\mu}, r^k) e_k = X^k(x^{\mu}, r^k) \Big[ \partial_k - r^m {^{(3)} \Gamma^l_{\; k m}}(x^{\mu})  \frac{\partial}{\partial r^l} \Big]\, ,
\ee
where the vector $X_R \in T T \Sigma_t^V$ lies in the vertical part of the tangent space  $ TT\Sigma_t$  of the tangent bundle $ T\Sigma_t$ and the vector $X_E$ in the remaining horizontal tangent space. The associated derivative operators will act on functions $f^{\pm} \in T \Sigma_t$ on the tangent bundle. These functions are obtained by translating the function $f \in \Sigma_t$ on the spatial hypersurface along a spatial geodesic with initial tangent vector $r^k$, 
\be
f^{\pm} (x^{\mu}, r^k) = \exp{ \Big[\pm \frac{1}{2} r^k {^{(3)}\nabla_k^H} \Big]} f(x^{\mu})= \exp{ \Big[\pm \frac{1}{2} \mathcal{L}_{r_E} \Big]} f(x^{\mu}) \, , f \in C^{\infty} (\Sigma_t) \, ,
\ee
where in local coordinates
\be
r_E = r^k e_k = r^k \Big( \partial_k - r^m {^{(3)} \Gamma^l_{\; k m}} \frac{\partial}{\partial r^l} \Big)\, .
\ee
We made use of the horizontal lift ${^{(3)}\nabla_k^H}$ of the covariant derivative ${^{(3)}\nabla}$ (induced on $\Sigma_t$ via the 3+1 decomposition) to the tangent bundle $\text{T}\Sigma_t$  that we  introduced in \eqref{horizontalLiftCovDer},
\be
{^{(3)}{\nabla}_k^{\text{H}}}  = {^{(3)}\nabla}_k - r^l {^{(3)}\Gamma^{n}_{\; k l }} \frac{\partial}{\partial r^n} \label{horLifCotangent} \, .
\ee
Wigner transforming dertivative operators will also give rise to the horizontal lift of the covariant derivative on the cotangent bundle of spatial hypersurfaces which we will denote as
\be
{D}_k := {^{(3)} \nabla}_{k} + p_l {^{(3)} \Gamma^l_{\; k j}} \frac{\partial}{ \partial p_j}\, , \quad  D_k p_j = 0\, .
\ee
If we want to calculate the differential operators appearing in \eqref{F00BeforeInt} to \eqref{F11BeforeInt}, we see that we need to rewrite differential operators acting on individual fields $\hat{u}^{\pm}$ that are translated geodesically in opposite directions in such a way that will act on the product, schematically
\be
 \langle (\mathcal{D} \hat{u}^+) \hat{u}^-  \rangle \rightarrow \mathcal{D}_*^+ ( \langle \hat{u}^+ \hat{u}^- \rangle).
\ee
In order to achieve this, we need to find annihilation operators $\mathcal{P}^{\pm}$  such that 
\be
\big[ X_{E,R} f^{\pm} \big](x^{\mu}, r^k) =  \big[ \mathcal{P}^{\pm} \big[ X_{E,R} \big] f^{\pm} \big](x^{\mu}, r^k)\,  \quad  \text{where} \; \mathcal{P}^{\pm}\big[ X_{E,R} \big] f^{\mp} = 0\,.
\ee
Since derivative with respect to the tangent space coordinate $\partial_r$ annihilate functions on the spatial hypersurface $f (x^{\mu})$, we can commute the exponential shift operator with any vertical derivative operator  $X_R$ to obtain the representation
\be
X_R  f^{\pm}  = - \sum_{n=1}^{\infty}\frac{1}{(\pm 2)^n} \frac{1}{n!} \underbrace{\Big[r_E, \Big[...,\Big[r_E ,  X_R \Big] ... \Big] \Big]}_\text{n times}f^{\pm}\, .
\ee
We find 
\bea
\Big[r_E ,  X_R \Big] &=& {^{(3)} \nabla_{r_E}^H} X_R   - X_E \, ,\\
\Big[r_E ,  X_E \Big] &=& {^{(3)} \nabla_{r_E}^H} X_E   +  R_R \big[X \big] \, ,
\eea
where the horizontal lift of the covariant derivative acting on $X_R$ and $X_E$ reads in components
\bea
{^{(3)} \nabla_{r_E}^H} X_R   &=& r^k  \Big[{^{(3)}\nabla_k} X^m-  r^n {^{(3)} \Gamma^l_{\; k n}} \frac{\partial}{\partial r^l} X^m \Big] \frac{\partial}{\partial r^m}\, , \\
{^{(3)} \nabla_{r_E}^H} X_E  &=& r^k  \Big[{^{(3)}\nabla_k} X^m -  r^n {^{(3)} \Gamma^l_{\; k n}} \frac{\partial}{\partial r^l} X^m \Big] e_m\, ,
\eea
and the vector field $R_R \big[X \big]$ is defined as
\be
R_R \big[X\big] = {^{(3)}R^l_{\; i k j }} r^i r^j X^k \frac{\partial}{\partial r^l} \, .
\ee
We then compute
\be
\Big[ r_E, \Big[r_E ,  X_R \Big]\Big] = \big( {^{(3)} \nabla_{r_E}^H}\big)^2 X_R  -  2{^{(3)} \nabla_{r_E}^H} X_E   - R_R \big[ X \big] \, .
\ee
\begin{multline}
\Big[ r_E,\Big[ r_E,  \Big[r_E ,  X_R \Big]\Big]\Big] = \big({^{(3)} \nabla_{r_E}^H} \big)^3 X_R  -3 \big({^{(3)} \nabla_{r_E}^H}\big)^2 X_E \\  - 2 R_R \big[ {^{(3)} \nabla_{r_E}^H} X \big]
-{^{(3)} \nabla_{r_E}^H} \big( R_R \big[X \big] \big) +  R_E \big[X \big]  \ \, .
\end{multline}
\begin{multline}
\Big[ r_E,\Big[ r_E,\Big[ r_E,  \Big[r_E ,  X_R \Big]\Big]\Big]\Big] = \big({^{(3)} \nabla_{r_E}^H} \big)^4 X_R
 -  4 \big({^{(3)} \nabla_{r_E}^H}\big)^3 X_E -  3  R_R \big[ \big({^{(3)} \nabla_{r_E}^H}\big)^2 X \big] \\
 -2 {^{(3)} \nabla_{r_E}^H} \Big( R_R \big[{^{(3)} \nabla_{r_E}^H} X\big] \Big) -2 R_E\big[{^{(3)} \nabla_{r_E}^H} X\big]  \\
-\big( {^{(3)} \nabla_{r_E}^H} \big)^2 \big( R_R \big[X \big] \big)+2{^{(3)} \nabla_{r_E}^H} \big( R_E \big[X \big] \big)  +  R_R \big[ R   \big[X \big] \big] \, .
\end{multline}
Further commutators go beyond the order of derivatives that we want to keep track of in the spatial gradient expansion.
We now set
\be
H_n^{\pm}  \big[X \big] := \frac{1}{(\pm 2)^n} \frac{1}{n!} \underbrace{\Big[r_E, \Big[...,\Big[r_E ,  X_R \Big] ... \Big] \Big]}_\text{n times}  \, ,
\ee
and find
\be
\Big[ X_R   \pm \frac{1}{2} \big( {^{(3)} \nabla_{r_E}^H} X_R   - X_E \big) +  \sum_{n=2}^{\infty}H_n^{\pm}\big[X\big] \Big]f^{\pm}  = 0 \, .
\ee
\be
\Big[\frac{1}{2} X_E \mp  X_R  - \frac{1}{2} {^{(3)} \nabla_{r_E}^H} X_R  \mp  \sum_{n=2}^{\infty} H_n^{\pm} \big[X \big]\Big]f^{\pm}  = 0 \, .
\ee
The last two identities can be used to define the annihilation operators we were looking for.
We define
\be
\mathcal{P}^{\pm} \big[X\big] := \Big[\frac{1}{2} X_E \pm  X_R  - \frac{1}{2} {^{(3)} \nabla_{r_E}^H} X_R  \pm  \sum_{n=2}^{\infty} H_n^{\mp}\big[X\big] \Big] \, , \; \mathcal{P}^{\pm} \big[X\big] f^{\mp} =0 \, .
\ee
\begin{multline}
\mathcal{P}^{\pm} \big[X\big] = \frac{1}{2} X_E \pm  X_R  - \frac{1}{2} {^{(3)} \nabla_{r_E}^H} X_R   \pm  \frac{1}{8} \Big[  \big( {^{(3)} \nabla_{r_E}^H}\big)^2 X_R  -  2{^{(3)} \nabla_{r_E}^H} X_E   - R_R \big[ X \big] \Big] \\ - \frac{1}{48} \Big[\big({^{(3)} \nabla_{r_E}^H} \big)^3 X_R  -3 \big({^{(3)} \nabla_{r_E}^H}\big)^2 X_E   - 3 R_R \big[ {^{(3)} \nabla_{r_E}^H} X \big]
-\big({^{(3)} \nabla R} \big)_R \big[ X \big] +  R_E \big[X \big]  \Big]  \\ \pm  \frac{1}{192}  \Big[
\big({^{(3)} \nabla R} \big)_E \big[ X \big] -  2 \big({^{(3)} \nabla_{r_E}^H}\big)^3 X_E    \Big] + \mathcal{O}\big( {^{(3)}R^2} \big)\,.
\end{multline}
We then have
\bea
X_E &=& \mathcal{P}^+ \big[X\big] + \mathcal{P}^- \big[X\big]+{^{(3)} \nabla_{r_E}^H} X_R  +2 \sum_{n=1}^{\infty} H_{2n+1}^{+}  \big[X\big]\, ,\\ 
X_R &=& \frac{1}{2} \Big[ \mathcal{P}^+ \big[X\big] - \mathcal{P}^- \big[X\big]- 2\sum_{n=1}^{\infty}  H_{2n}^{+} \big[X\big]   \Big] \, .
\eea
Since we are only interested in expansions that include terms of order $({^{(3)}\nabla})^3$ for the gradient expansion, we can re-express the higher order derivative coefficients by iteration
\begin{multline}
X_E = \mathcal{P}^+ \big[X\big] + \mathcal{P}^- \big[X\big] + \frac{1}{2} \Big[ \mathcal{P}^+ \big[{^{(3)} \nabla_{r_E}^H} X\big] - \mathcal{P}^- \big[{^{(3)} \nabla_{r_E}^H} X\big] \Big] \\
-\frac{1}{8} \Big[\big( {^{(3)} \nabla_{r_E}^H}\big)^3 X_R  -  2\big( {^{(3)} \nabla_{r_E}^H}\big)^2 X_E   - R_R \big[{^{(3)} \nabla_{r_E}^H} X \big]   \Big]  \\  + \frac{1}{24} \Big[ \big({^{(3)} \nabla_{r_E}^H} \big)^3 B_R  -3 \big({^{(3)} \nabla_{r_E}^H}\big)^2 B_E \\  - 2 R_R \big[ {^{(3)} \nabla_{r_E}^H} B \big]
-{^{(3)} \nabla_{r_E}^H} \big( R_R \big[B \big] \big) +  R_E \big[B \big]  \Big]
+ \mathcal{O}\big( {^{(3)}R^2} \big) \, .
\end{multline}
\begin{multline}
X_E = \mathcal{P}^+ \big[X\big] + \mathcal{P}^- \big[X\big] + \frac{1}{2} \Big[ \mathcal{P}^+ \big[{^{(3)} \nabla_{r_E}^H} X\big] - \mathcal{P}^- \big[{^{(3)} \nabla_{r_E}^H} X\big] \Big] \\
-\frac{1}{12}\big( {^{(3)} \nabla_{r_E}^H}\big)^3 X_R  + \frac{1}{8} \big( {^{(3)} \nabla_{r_E}^H}\big)^2 X_E   - \frac{1}{24} {^{(3)} \nabla_{r_E}^H} \big( R_R \big[X \big] \big)   \\  + \frac{1}{24} R_R \big[ {^{(3)} \nabla_{r_E}^H} X \big]
+  \frac{1}{24} R_E \big[X \big]  
+ \mathcal{O}\big( {^{(3)}R^2} \big) \, .
\end{multline}
We define
\be
\big({^{(3)} \nabla R} \big)_R \big[ X \big] := r^s \big( {^{(3)} \nabla_s} {^{(3)} R^l_{\; ikj} } \big) X^k r^i r^j \frac{\partial}{\partial r^l}
\ee
and get
\begin{multline}
X_E = \mathcal{P}^+ \big[X\big] + \mathcal{P}^- \big[X\big] + \frac{1}{2} \Big[ \mathcal{P}^+ \big[{^{(3)} \nabla_{r_E}^H} X\big] - \mathcal{P}^- \big[{^{(3)} \nabla_{r_E}^H} X\big] \Big] \\
+ \frac{1}{8}\Big[ \mathcal{P}^+ \big[ \big( {^{(3)} \nabla_{r_E}^H}\big)^2 X\big] + \mathcal{P}^- \big[ \big( {^{(3)} \nabla_{r_E}^H}\big)^2 X \big] \Big] 
 + \frac{1}{48} \Big[ \mathcal{P}^+ \big[\big( {^{(3)} \nabla_{r_E}^H}\big)^3 X \big] - \mathcal{P}^- \big[\big( {^{(3)} \nabla_{r_E}^H}\big)^3 X \big] \Big]  \\ 
 +  \frac{1}{24} \Big[ \mathcal{P}^+ \big[R \big[X \big] \big] + \mathcal{P}^- \big[R \big[X \big] \big] \Big]   +  \frac{1}{48} \Big[ \mathcal{P}^+ \big[   
  R \big[{^{(3)} \nabla^H_{r_E}} X \big]\big] - \mathcal{P}^- \big[
  R \big[{^{(3)} \nabla^H_{r_E}}  X  \big] \big] \Big] 
+ \mathcal{O}\big( {^{(3)}R^2} \big) \, .
\end{multline}
Similarly, we get
\begin{multline}
X_R = \frac{1}{2} \Big[ \mathcal{P}^+ \big[X\big] - \mathcal{P}^- \big[X\big] \Big]
+\frac{1}{4} \Big[ \mathcal{P}^+ \big[{^{(3)} \nabla_{r_E}^H} X\big] + \mathcal{P}^- \big[{^{(3)} \nabla_{r_E}^H} X\big] \Big]  
\\  +\frac{1}{16} \Big[ \mathcal{P}^+ \big[\big( {^{(3)} \nabla_{r_E}^H}\big)^2 X \big] - \mathcal{P}^- \big[\big( {^{(3)} \nabla_{r_E}^H}\big)^2 X\big] \Big] 
  + \frac{1}{16} \Big[ \mathcal{P}^+ \big[R \big[ X \big]\big] - \mathcal{P}^- \big[R \big[ B \big]\big] \Big] \\ +  \frac{1}{32}\Big[ \mathcal{P}^+ \big[ \big({^{(3)} \nabla R} \big) \big[ X \big]\big] + \mathcal{P}^- \big[\big({^{(3)} \nabla R} \big) \big[ X \big]\big]\Big]   + \mathcal{O}\big( {^{(3)}R^2} \big) \, .
\end{multline}
Since we will eventually act on functions shifted in the $\pm r$ direction, we introduce the notation
\begin{multline}
X_E^{\pm} = \mathcal{P}^{\pm} \big[X\big]  \pm \frac{1}{2}  \mathcal{P}^{\pm} \big[{^{(3)} \nabla_{r_E}^H} X\big]  
+ \frac{1}{8} \mathcal{P}^{\pm} \big[ \big( {^{(3)} \nabla_{r_E}^H}\big)^2 X\big]
 \pm \frac{1}{48}  \mathcal{P}^{\pm} \big[\big( {^{(3)} \nabla_{r_E}^H}\big)^3 X \big]   \\ 
 +  \frac{1}{24}  \mathcal{P}^{\pm} \big[R \big[X \big] \big]   \pm  \frac{1}{48}  \mathcal{P}^{\pm} \big[
  R \big[{^{(3)} \nabla^H_{r_E}}   X \big]\big] + \mathcal{O}\big( {^{(3)}R^2} \big) \, . \label{XE}
\end{multline}
\begin{multline}
X_R^{\pm} =  \pm \frac{1}{2} \mathcal{P}^{\pm} \big[X\big] 
+\frac{1}{4} \mathcal{P}^{\pm} \big[{^{(3)} \nabla_{r_E}^H} X\big] 
 \pm \frac{1}{16}  \mathcal{P}^{\pm} \big[\big( {^{(3)} \nabla_{r_E}^H}\big)^2 X \big] 
  \pm \frac{1}{16}  \mathcal{P}^{\pm} \big[R \big[ X \big]\big]  \\ +  \frac{1}{32}  \mathcal{P}^{\pm} \big[  \big({^{(3)} \nabla R} \big) \big[ X \big]\big]   + \mathcal{O}\big( {^{(3)}R^2} \big) \, . \label{XR}
\end{multline}
The operators \eqref{XE} and \eqref{XR} can now be pulled out from individual factors to act on the whole product as in the scheme
\be
 \langle (\mathcal{D}^{+}_{E,R} \hat{u}^+) \hat{u}^-  \rangle \rightarrow \mathcal{D}_{E,R}^+ ( \langle \hat{u}^+ \hat{u}^- \rangle).
\ee
After the differential operators act on the product, we want replace the projection operators again in terms of concrete expressions and we find,
\begin{multline}
X_E^{\pm} = \frac{1}{2} X_E \pm  X_R   
\pm \frac{1}{32} \big({^{(3)} \nabla_{r_E}^H} \big)^3 X_E  
 \mp  \frac{1}{12}  R_R \big[ X \big] 
\mp \frac{1}{96} R_E \big[ {^{(3)} \nabla_{r_E}^H} X \big] 
 \mp \frac{1}{192} \big({^{(3)} \nabla R} \big)_E \big[ X \big] + \mathcal{O}\big( {^{(3)}R^2} \big) \, ,
\end{multline}
\begin{multline}
X_R^{\pm} = \pm \frac{1}{4} X_E + \frac{1}{2} X_R \mp \frac{1}{96} \big({^{(3)} \nabla_{r_E}^H} \big)^3 X_R  \pm \frac{1}{48}  R_E \big[ X \big] +\frac{1}{384} \big({^{(3)} \nabla R} \big)_E \big[ X \big]  \pm \frac{1}{96} \big({^{(3)} \nabla R} \big)_R \big[ X \big]\\ -\frac{1}{48} R_E \big[ {^{(3)} \nabla_{r_E}^H} X \big]
\mp \frac{1}{32} R_R \big[ {^{(3)} \nabla_{r_E}^H} X \big] + \mathcal{O}\big( {^{(3)}R^2} \big) \, .
\end{multline}
We will also have to commute the spatial curved Laplace operator with the spatially covariant shift operators and rewrite it in terms of the annihilation operators. 
In order to commute the spatial Laplace operator, we apply a trick.
We realize that once more that all operations we perform in the integrals that define the Wigner transformed operators take place in the spatial tangent bundle $T\Sigma_t$ specified by the time-slicing. This bundle can be equipped with some natural metric $\widetilde{\gamma}$ that satisfies
\bea
\widetilde{\gamma} \big( e_i, e_j \big) &=& \gamma_{ij}\, ,\\
\widetilde{\gamma} \big(e_i , \frac{\partial}{\partial r^j} \big) &=& 0\, .
\eea
The important point is that this metric will satisfy
\be
\big[{^{(3)}\nabla^H_{e_k}} \widetilde{\gamma}\big] \big( e_i , e_j \big) = 0\,, 
\ee
which follows from ${^{(3)}\nabla^H_{e_k}} e_i = {^{(3)}\Gamma^j_{\;k i}} e_j$ and so we can use the usual techniques for computations with a metric compatible covariant derivative.
{The spatial Laplace operator acting on fields that depend only on the submanifold  $\Sigma_t$ can be rewritten in terms of the horizontal lift of the covariant derivative to the tangent bundle $T\Sigma_t$
\be
\gamma^{ij} {^{(3)}\nabla_i }{^{(3)}\nabla_j } f(x^{\mu}) = 
\widetilde{\gamma}^{ij}{^{(3)}\nabla^H_{e_i}}{^{(3)}\nabla^H_{e_j}} f(x^{\mu}) \, .
\ee
On the other hand, we can rewrite 
\be
\mathcal{L}_{r_E}^n = \Big[ r^i {^{(3)}\nabla_{e_i}^H } \Big]^n = r^{i_1} ... r^{i_n}{^{(3)}\nabla_{e_{i_1}}^H }{^{(3)}\nabla_{e_{i_n}}^H }\, ,  
\ee
since ${^{(3)}\nabla_{r_E} } r_E = 0$.
It is now a matter of computing commutation relations for the horizontal lift of the covariant derivative to the tangent bundle to find the first correction terms that result from commuting the spatial Laplacian acting on $\hat{\phi} (x^{\mu})$ or $\hat{\Pi}(x^{\mu})$ with the shift exponentials and we give the concrete expressions in the following appendix. 
Note that since the basis $e_k$ is not a coordinate basis ($\big[e_i , e_j \big] \neq 0$), the commutator of horizontal covariant derivatives acting on functions is also non-zero
\be
\Big[ {^{(3)}\nabla_{e_i}^H }\, , {^{(3)}\nabla_{e_j}^H } \Big] f(x^{\mu}, r^k) = - 
r^n {^{(3)} R^l_{\; n ij }} \frac{\partial}{\partial r^l}f(x^{\mu}, r^k)\, .
\ee
We find for the adapted basis vectors
\bea
\Big[ {^{(3)}\nabla_{e_i}^H }\, , {^{(3)}\nabla_{e_j}^H } \Big] \big( e_k  \big) &=&  
 {^{(3)} R^l_{\;k  ij }}  e_l\, , \\
 \Big[ {^{(3)}\nabla_{e_i}^H }\, , {^{(3)}\nabla_{e_j}^H } \Big] \Big( \frac{\partial}{\partial r^k}  \Big) &=&  
 {^{(3)} R^l_{\;k  ij }} \frac{\partial}{\partial r^l} \, ,
\eea
and opposite signs for the dual adapted basis. Note that due to the propoperty
\be
{^{(3)} \nabla^H_{\frac{\partial}{\partial r^k}}} \frac{\partial}{\partial r^j} = {^{(3)} \nabla^H_{\frac{\partial}{\partial r^k}}} e_j = 0\, ,
\ee
the partial derivative $\frac{\partial}{\partial r^k}$ acts itself covariantly and its commutator with the basis vectors $e_k$ can computed to be zero also when acting on functions due to non-vanishing commutator of the partial derivatives $\big[e_i, \frac{\partial}{\partial r^k} \big] \neq 0$,
\be
\Big[ {^{(3)}\nabla_{e_i}^H }\, , {^{(3)}\nabla_{\frac{\partial}{\partial r^j}}^H } \Big] f(x^{\mu}, r^k) = 0\, .
\ee
\subsection{$\mathcal{T}^{\pm}_*$, $\mathcal{M}^{\pm}_*$, $\big({^{(3)} \square}\big)^{\pm}_*$ and $({^{(3)} \nabla} N)^{\pm}_*$ \label{commis}}
The general expressions from the previous appendix will now be used to compute the concrete expressions that appear in \eqref{F00BeforeInt} to \eqref{F11BeforeInt}. We will need the commutator
\be
\mathcal{T}^{\pm} := \Big[ \partial_t , \exp \Big[ \pm \frac{1}{2} r^k {^{(3) \nabla_k^H}} \Big] \Big], ,
\ee 
which can be computed by using the three tensor  ${^{(3)} \dot{\Gamma}^l_{\; i k }}$ on the spatial hypersurface,
\begin{multline}
 {^{(3)} \dot{\Gamma}^l_{\; i k }} = {^{(3)} \nabla^l} \big[N K_{ik} \big] -2 {^{(3)} \nabla_{(i}} \big[N K_{k)}^{\; l} \big]\\ +\Big[ {^{(3)} \nabla_{(i}} {^{(3)} \nabla_{k)}}N^l  +{^{(3)} \nabla_{(i}} {^{(3)} \nabla^{l}}N_{k)} -{^{(3)} \nabla^{l}} {^{(3)} \nabla_{(i}}N_{k)}     \Big]\, ,
\end{multline}
 with
\be
T_{(ik)} = \frac{T_{ik} + T_{ki}}{2} \, .
\ee
We can define a vertical vector field
\be
\dot{\Gamma}_R := r^i r^k  {^{(3)} \dot{\Gamma}^l_{\; i k }} \frac{\partial}{\partial r^l} =\Big[ r_E , \partial_t \Big] \,,
\ee
in terms of which we have
\begin{multline}
\mathcal{T}^{\pm} = \mp \frac{1}{2} \dot{\Gamma}_R - \frac{1}{8} \Big[ r_E ,  \dot{\Gamma}_R \Big]  \mp \frac{1}{48} \Big[r_E, \Big[ r_E ,  \dot{\Gamma}_R \Big] \Big]+  \frac{1}{384}  \Big[r_E , \Big[r_E ,\Big[ r_E ,  \dot{\Gamma}_R \Big] \Big] \Big]   + \mathcal{O}\big( {^{(3)}R^2} \big)\\
= \mp \frac{1}{2} \dot{\Gamma}_R + \frac{1}{8} \Big[ \dot{\Gamma}_E  -  {^{(3)} \nabla^H_{r_E}}   \dot{\Gamma}_R  \Big]  \mp \frac{1}{48} \Bigg[ \big( {^{(3)} \nabla_{r_E}^H}\big)^2 \dot{\Gamma}_R  -  2{^{(3)} \nabla_{r_E}^H} \dot{\Gamma}_E   - R_R \big[ \dot{\Gamma} \big] \Bigg] \\ +  \frac{1}{384} \Bigg[  -3 \big({^{(3)} \nabla_{r_E}^H}\big)^2 \dot{\Gamma}_E   +  R_E \big[\dot{\Gamma} \big]  \Bigg]  + \mathcal{O}\big( {^{(3)}R^2} \big)\,.
\end{multline}
Acting on functions translated in the $\pm r$ direction leads us to define the operator
\begin{multline}
\mathcal{T}^{\pm}_*    
:=  \mp \frac{1}{2} \dot{\Gamma}_R^{\pm} + \frac{1}{8} \Big[ \dot{\Gamma}_E^{\pm}  -  {^{(3)} \nabla^H_{r_E}}   \dot{\Gamma}_R^{\pm}  \Big]  \mp \frac{1}{48} \Bigg[ \big( {^{(3)} \nabla_{r_E}^H}\big)^2 \dot{\Gamma}_R^{\pm}  -  2{^{(3)} \nabla_{r_E}^H} \dot{\Gamma}_E^{\pm}   - R_R^{\pm} \big[ \dot{\Gamma} \big] \Bigg] \\ +  \frac{1}{384} \Bigg[  -3 \big({^{(3)} \nabla_{r_E}^H}\big)^2 \dot{\Gamma}_E^{\pm}   +  R_E^{\pm} \big[\dot{\Gamma} \big]  \Bigg]  + \mathcal{O}\big( {^{(3)}R^2} \big) \,.
\end{multline}
Since we will be dealing only with the sum of this operator with opposite signs, we have
\be
\mathcal{T}^{+}_*  +\mathcal{T}^{-}_*  = - \frac{1}{8} \dot{\Gamma}_E - \frac{1}{24}  {^{(3)} \nabla_{r_E}^H} \dot{\Gamma}_R  + \mathcal{O}\big( {^{(3)}R^2} \big) 
\ee
We will also need the following commutator
\be
\widetilde{\mathcal{M}}^{\pm} := \Big[ M_E , \exp \Big[ \pm \frac{1}{2} r^k {^{(3)}\nabla^H_k} \Big] \Big]\, ,
\ee
with
\be
M_E := N^k e_k \, .
\ee
We find
\begin{multline}
\widetilde{\mathcal{M}}^{\pm}
 =
 \mp \frac{1}{2} \Big[ {^{(3)} \nabla^H_{r_E}}M_E + R_R [M]   \Big] 
 - \frac{1}{8}  \big({^{(3)} \nabla^H_{r_E}} \big)^2 M_E
 - \frac{1}{8}  R_R \big[ {^{(3)} \nabla^H_{r_E}} M \big] \\
  - \frac{1}{8} \Big[ {^{(3)} \nabla^H_{r_E}} \big(R_R [M]  \big) 
  - R_E [M]   \Big] \\
 \mp \frac{1}{48} \Big[  \big({^{(3)} \nabla^H_{r_E}} \big)^3 M_E -R_E \big[ {^{(3)} \nabla^H_{r_E}} M \big]-2 {^{(3)} \nabla^H_{r_E}} \big(R_E [M]  \big) \Big] + \mathcal{O}\big( {^{(3)}R^2} \big) 
  \, ,
\end{multline}
and define
\begin{multline}
\widetilde{\mathcal{M}}^{\pm}_*  :=
 \mp \frac{1}{2} \Big[ {^{(3)} \nabla^H_{r_E}}M_E^{\pm} + R_R^{\pm} [M]   \Big] 
 - \frac{1}{8}  \big({^{(3)} \nabla^H_{r_E}} \big)^2 M_E^{\pm}
 - \frac{1}{8}  R_R^{\pm} \big[ {^{(3)} \nabla^H_{r_E}} M \big] \\
  - \frac{1}{8} \Big[ {^{(3)} \nabla^H_{r_E}} \big(R_R [M]  \big)^{\pm} 
  - R_E^{\pm} [M]   \Big] \\
 \mp \frac{1}{48} \Big[  \big({^{(3)} \nabla^H_{r_E}} \big)^3 M_E^{\pm} -R_E^{\pm} \big[ {^{(3)} \nabla^H_{r_E}} M \big]-2 {^{(3)} \nabla^H_{r_E}} \big(R_E [M]  \big)^{\pm} \Big] + \mathcal{O}\big( {^{(3)}R^2} \big) 
  \, .
\end{multline}
It is convenient to define to following operator
\be
\mathcal{M}_*^{\pm} :=  M_E^{\pm}  - \widetilde{\mathcal{M}}^{\pm}_*\, ,
\ee
and we have
\begin{multline}
\mathcal{M}_*^{+} +\mathcal{M}_*^{-} = M_E   + {^{(3)} \nabla^H_{r_E}}M_R   + \frac{1}{8}  \big({^{(3)} \nabla^H_{r_E}} \big)^2 M_E   + \frac{1}{24}  \big({^{(3)} \nabla^H_{r_E}} \big)^3 M_R \\ + \frac{1}{8}R_E [M]   + \frac{1}{24}{^{(3)} \nabla^H_{r_E}} \big(R_R [M]  \big)   + \mathcal{O}\big( {^{(3)}R^2} \big) \, .
\end{multline}
It is now the sum $\mathcal{T}^{+}_* + \mathcal{T}^{+}_* + \mathcal{M}^{+}_* + \mathcal{M}^{-}_*$ that will enter the dynamical equations \eqref{F00BeforeInt} to \eqref{F11BeforeInt}, and this sums happens to reduce to a rather short expressions to second order in the spatial gradient after performing the Wigner transformation (and dropping boundary terms). Making us of the horizontal lift of the spatial covariant derivative to the cotangent bundle \eqref{horLifCotangent}, we find for example in \eqref{F00BeforeInt} up to boundary terms, 
\begin{multline}
\gamma^{1/2} \int_{\Sigma_t } d^3r  e^{- \frac{i}{\hbar} r^k p_k } \big[ \mathcal{M}_*^{+} +\mathcal{M}_*^{-}+ \mathcal{T}_*^{+} +\mathcal{T}_*^{-} \big]\big[ {:\hat{u}^+ \hat{u}^-:} \big]  
  \\  =
  \Bigg[N^k  D_k   - N^k_{\; ; k}  -
  p_k N^k_{\; ; m} \frac{\partial}{\partial p_m}     
  - \hbar^2 \frac{1}{12} N^k_{\; ; k q m}  \frac{\partial^2}{\partial p_q \partial p_m}  
    \\
  +    \frac{\hbar^2}{4}\Bigg( \frac{1}{2} {^{(3)} \nabla^l} \big[N K_{ik} \big] - {^{(3)} \nabla_{i}} \big[N K_{k}^{\; l} \big]  \Bigg) \frac{\partial^2}{\partial p_k \partial p_i} D_l  
\\ - \frac{\hbar^2}{12} \Bigg( \frac{1}{2} {^{(3)} \nabla_l}{^{(3)} \nabla^l} \big[N K_{ik} \big] - {^{(3)} \nabla_l}{^{(3)} \nabla_{i}} \big[N K_{k}^{\; l} \big]  -{^{(3)} \nabla_i}{^{(3)} \nabla_{k}} \big[N K \big]  \Bigg) \frac{\partial^2}{\partial p_k \partial p_i}
\\ - \frac{\hbar^2}{12} p_l \Bigg( \frac{1}{2}{ ^{(3)} \nabla_s}{^{(3)} \nabla^l} \big[N K_{ik} \big] - {^{(3)} \nabla_s}{^{(3)} \nabla_{i}} \big[N K_{k}^{\; l} \big]\Bigg)\frac{\partial^3}{\partial p_k \partial p_i \partial p_s} + \mathcal{O} \big( \hbar^4 \big) \Bigg] \hat{F}_{\phi \phi} \, .
\end{multline}
Similarly, we will have to commute the operator $\gamma^{ij} \partial_i N \partial_j$ and so we define
\be
\big({\nabla^{(3)}}{N}\big)_*^{\pm} :=  \big({\nabla^{(3)}}{N}\big)_E^{\pm}  - \widetilde{\big({\nabla^{(3)}}{N}\big)^{\pm}}_*\, ,
\ee
again with
\be
\widetilde{\big({\nabla^{(3)}}{N}\big)^{\pm}} := \Big[ \big({\nabla^{(3)}}{N}\big)_E , \exp \Big[ \pm \frac{1}{2} r^k {^{(3)}\nabla^H_k} \Big] \Big]\, ,
\ee
where
\be
\big({\nabla^{(3)}}{N}\big)_E := \gamma^{ij} \partial_i N e_j \, .
\ee
We find
\begin{multline}
\big({\nabla^{(3)}}{N}\big)_*^{\pm} = \big({\nabla^{(3)}}{N}\big)_E^{\pm} -\Bigg[ \mp \frac{1}{2} \Big[ {^{(3)} \nabla^H_{r_E}}\big({\nabla^{(3)}}{N}\big)_E^{\pm} + R_R^{\pm} [\big({\nabla^{(3)}}{N}\big)]   \Big] \\
 - \frac{1}{8}  \big({^{(3)} \nabla^H_{r_E}} \big)^2 \big({\nabla^{(3)}}{N}\big)_E^{\pm}
 - \frac{1}{8}  R_R^{\pm} \big[ {^{(3)} \nabla^H_{r_E}} \big({\nabla^{(3)}}{N}\big) \big] \\
  - \frac{1}{8} \Big[ {^{(3)} \nabla^H_{r_E}} \big(R_R [\big({\nabla^{(3)}}{N}\big)]  \big)^{\pm} 
  - R_E^{\pm} [\big({\nabla^{(3)}}{N}\big)]   \Big] \\
 \mp \frac{1}{48} \Big[  \big({^{(3)} \nabla^H_{r_E}} \big)^3 \big({\nabla^{(3)}}{N}\big)_E^{\pm} -R_E^{\pm} \big[ {^{(3)} \nabla^H_{r_E}} \big({\nabla^{(3)}}{N}\big) \big]-2 {^{(3)} \nabla^H_{r_E}} \big(R_E [\big({\nabla^{(3)}}{N}\big)]  \big)^{\pm} \Big] \Bigg] + \mathcal{O}\big( {^{(3)}R^2} \big) 
   \, . 
\end{multline}

We will be dealing with the sum and differences of those operators. Since they already come with a spatial derivative themselves we have
\be
\big({\nabla^{(3)}}{N}\big)_*^{+} +\big({\nabla^{(3)}}{N}\big)_*^{-} = \big({\nabla^{(3)}}{N}\big)_E   + {^{(3)} \nabla^H_{r_E}}\big({\nabla^{(3)}}{N}\big)_R     + \mathcal{O}\big( {^{(3)}R^2} \big) \, .
\ee
\begin{multline}
\big({\nabla^{(3)}}{N}\big)_*^{+} -\big({\nabla^{(3)}}{N}\big)_*^{-} = 2 \big({\nabla^{(3)}}{N}\big)_R  + \frac{1}{2}{^{(3)} \nabla^H_{r_E}}\big({\nabla^{(3)}}{N}\big)_E + \frac{1}{4} \big({^{(3)} \nabla^H_{r_E} }\big)^2 \big({\nabla^{(3)}}{N}\big)_R \\
+ \frac{1}{12} R_R\big[\big({\nabla^{(3)}}{N}\big) \big]     + \mathcal{O}\big( {^{(3)}R^2} \big) \, .
\end{multline}
We find for the expressions appearing in \eqref{F+BeforeInt} and \eqref{F-BeforeInt} to second order in the spatial gradient expansion and up to boundary terms,
\begin{multline}
\gamma^{1/2} \int_{\Sigma_t} d^3r  e^{- \frac{i}{\hbar} r^k p_k } \big[\big({\nabla^{(3)}}{N}\big)_*^{+} +\big({\nabla^{(3)}}{N}\big)_*^{-} \big]\big[{: \hat{u}^+ \hat{u}^- :}\big]  
  \\  =\gamma^{1/2} \int_{\Sigma_t} d^3r  e^{- \frac{i}{\hbar} r^k p_k } 
  \Bigg[N_{;}^{\; k} {^{(3)} \nabla_{e_k}^H}   + r^l N_{; \; \; l}^{\; k } \frac{\partial}{\partial r^k} + \mathcal{O}\big( {^{(3)}R^2} \big) \Bigg] \Big[{: \hat{u}^+ \hat{u}^- :}\Big]
  \\  =
  \Bigg[N_{; k} {^{(3)} D^k}  - N_{; k}^{\; \; k} - p_k N_{; \; \; l}^{\; k } \frac{\partial}{\partial p_l}  + \mathcal{O} \big( \hbar^4 \big)\Bigg] \hat{F}_{\phi \phi} \, ,
\end{multline}
\begin{multline}
\gamma^{1/2} \int_{\Sigma_t} d^3r  e^{- \frac{i}{\hbar} r^k p_k } \big[\big({\nabla^{(3)}}{N}\big)_*^{+} -\big({\nabla^{(3)}}{N}\big)_*^{-} \big]\big[{ : \hat{u}^+ \hat{u}^- :} \big]  
  \\  =\gamma^{1/2} \int_{\Sigma_t} d^3r  e^{- \frac{i}{\hbar} r^k p_k } 
  \Bigg[ 2 N_{;}^{\; k} \frac{\partial}{\partial r^k} + \frac{1}{2}r^l N_{; \; \; l}^{\; k }{^{(3)} \nabla_{e_k}^H} + \frac{1}{4} r^l r^m N_{; \; \; l m}^{\; k } \frac{\partial}{\partial r^k} \\
- \frac{1}{12} r^m r^k {^{(3)} R^l_{\; k m s }} N_{;}^{\; s} \frac{\partial}{\partial r^l} \big] + \mathcal{O}\big( {^{(3)}R^2} \big)  \Bigg] \Big[ {: \hat{u}^+ \hat{u}^- :}\Big]
  \\  =\frac{i}{\hbar}  
  \Bigg[2  \gamma^{lk} p_l N_{; k}  + \frac{1}{2} N_{;k l} \frac{\partial}{\partial p_l} D^k  - \frac{\hbar^2}{4} p_k  \gamma^{ks} N_{; s l m} \frac{\partial^2}{\partial p_l \partial p_m} - \frac{\hbar^2}{2} N_{; \; \; k m}^{\; k}  \frac{\partial}{\partial p_m} \\- \frac{\hbar^2}{6} {^{(3)}R^{m}_{\; l}} N_{; m}  \frac{\partial}{\partial p_l}
  +\frac{\hbar^2}{12} p_l{^{(3)}R^{l}_{\; k m s}}N_{;}^{\; s}  \frac{\partial^2}{\partial p_k \partial p_m}+ \mathcal{O} \big( \hbar^4 \big) \Bigg] \hat{F}_{\phi \phi} \, .
\end{multline}
Finally, we commute the spatial Laplace operator appearing in \eqref{F+BeforeInt} to \eqref{F11BeforeInt} and let it act on the product of the geodesically shifted canonical operators 
\be
\Big( N^+ \big({^{(3)} \square}\big)^+_* \pm N^- \big({^{(3)} \square}\big)^-_*   \Big)\big[ f^+ g^- \big] := \Big[  N {^{(3)} \square} f \Big]^+ g^- \pm f^+ \Big[  N {^{(3)} \square} g \Big]^- \,, \label{lapExp2}
\ee
where $f^+$ and $g^-$ are placeholders for the operators $\hat{u}^{\pm}$ and $\hat{v}^{\pm}$.
The first term that appears in the series when commuting the spatial Laplacian with the shift exponentials is the horizontally lifted operator itself,
\begin{multline}
N^+ \gamma^{ij} \Big[{^{(3)} \nabla_{e_i }^H}{^{(3)} \nabla_{e_j }^H} f^+ \Big] \hat{u}^-  \pm N^- \gamma^{ij}  f^+ \Big[{^{(3)} \nabla_{e_i }^H}{^{(3)} \nabla_{e_j }^H} g^- \Big]  
\\ = 
\Big[ N^+ \gamma^{ij} {^{(3)} \nabla_{e_i }^H} e_j^+ \pm  N^- \gamma^{ij} {^{(3)} \nabla_{e_i }^H} e_j^- \Big]\Big[ f^+ g^- \Big]- \gamma^{ij}\Big[ N^+ e_i^+ e_j^- \pm N^- e_i^+ e_j^- \Big] \Big[ f^+ g^- \Big] \\
+\gamma^{ij} \Big( N^+ \pm N^- \Big) f^+ \Big[ \big[ e_i^+ \, , e_j^- \big] g^-  \Big]\, . \label{laplaceExp}
\end{multline}
The second part is not manifestly spatially covariant, but our calculation should now show that it actually is.
We will need the following identities
\be
\gamma^{ij} \big[ e_i^+, e_j^- \big]^- = - \gamma^{ij} {^{(3)} \Gamma^l_{\; ij}} \Big[- \frac{1}{4} e_l + \frac{1}{2} \frac{\partial}{\partial r^l} \Big] - \gamma^{ij}\frac{1}{12}\Big[e_i \, ,r^m  r^k  {^{(3)}R^l_{\; k m j}} \frac{\partial}{\partial r^l} \Big]^- \, ,
\ee
\begin{multline}
- \gamma^{ij}\Big[ N^+  \Bigg[\frac{1}{2} e_i + \frac{\partial}{\partial r^i}    
 +  \frac{1}{12}  r^k r^m {^{(3)}R^l_{\; k m i}} \frac{\partial}{\partial r^l}
 \Bigg]  \Bigg[\frac{1}{2} e_j - \frac{\partial}{\partial r^j}    
 -  \frac{1}{12}  r^k r^m {^{(3)}R^l_{\; k m j}} \frac{\partial}{\partial r^l}
 \Bigg]\\  \pm N^-\Bigg[\frac{1}{2} e_i + \frac{\partial}{\partial r^i}    
 +  \frac{1}{12}  r^k r^m {^{(3)}R^l_{\; k m i}} \frac{\partial}{\partial r^l}
 \Bigg]   \Bigg[\frac{1}{2} e_j - \frac{\partial}{\partial r^j}    
 -  \frac{1}{12}  r^k r^m {^{(3)}R^l_{\; k m j}} \frac{\partial}{\partial r^l}
 \Bigg] \Big] \Big[ f^+ g^- \Big] \\
+\gamma^{ij} \Big( N^+ \pm N^- \Big) f^+ \Big[ \big[ e_i^+ \, , e_j^- \big] g^-  \Big] 
\\= 
- \gamma^{ij}\Big[ N^+  \Bigg[\frac{1}{2} {^{(3)} \nabla^H_{e_i}}  + \frac{\partial}{\partial r^i}    
 +  \frac{1}{12}  r^k r^m {^{(3)}R^l_{\; k m i}} \frac{\partial}{\partial r^l}
 \Bigg]  \Bigg[\frac{1}{2} {^{(3)} \nabla^H_{e_j}}  - \frac{\partial}{\partial r^j}    
 -  \frac{1}{12}  r^k r^m {^{(3)}R^l_{\; k m j}} \frac{\partial}{\partial r^l}
 \Bigg]\\  \pm N^-\Bigg[\frac{1}{2} {^{(3)} \nabla^H_{e_i}}  + \frac{\partial}{\partial r^i}    
 +  \frac{1}{12}  r^k r^m {^{(3)}R^l_{\; k m i}} \frac{\partial}{\partial r^l}
 \Bigg]   \Bigg[\frac{1}{2} {^{(3)} \nabla^H_{e_j}}  - \frac{\partial}{\partial r^j}    
 -  \frac{1}{12}  r^k r^m {^{(3)}R^l_{\; k m j}} \frac{\partial}{\partial r^l}
 \Bigg] \Big] \Big[ f^+ g^- \Big] \\
-\gamma^{ij} \Big( N^+ \pm N^- \Big) \Big[ \frac{1}{24}r^k r^m {^{(3)}R^l_{\; k m j;i}} \frac{\partial}{\partial r^l}    \Big] \Big[ f^+  g^-  \Big] \, .
\end{multline}
Thus the term \eqref{laplaceExp} takes the form 
\begin{multline}
N^+ \gamma^{ij} \Big[{^{(3)} \nabla_{e_i }^H}{^{(3)} \nabla_{e_j }^H} f^+ \Big] g^-  \pm N^- \gamma^{ij}  f^+ \Big[{^{(3)} \nabla_{e_i }^H}{^{(3)} \nabla_{e_j }^H} g^- \Big]  
\\ = 
\Big[ N^+ \gamma^{ij} {^{(3)} \nabla_{e_i }^H} \Bigg[\frac{1}{2} {^{(3)} \nabla^H_{e_j}}  + \frac{\partial}{\partial r^j}    
 +  \frac{1}{12}  r^k r^m {^{(3)}R^l_{\; k m j}} \frac{\partial}{\partial r^l}
 \Bigg]    \\ \pm  N^- \gamma^{ij} {^{(3)} \nabla_{e_i }^H} \Bigg[\frac{1}{2} {^{(3)} \nabla^H_{e_j}}  - \frac{\partial}{\partial r^j}    
 -  \frac{1}{12}  r^k r^m {^{(3)}R^l_{\; k m j}} \frac{\partial}{\partial r^l}
 \Bigg]   \Big]\Big[f^+ g^- \Big] \\
- \gamma^{ij}\Big[ N^+  \Bigg[\frac{1}{2} {^{(3)} \nabla^H_{e_i}}  + \frac{\partial}{\partial r^i}    
 +  \frac{1}{12}  r^k r^m {^{(3)}R^l_{\; k m i}} \frac{\partial}{\partial r^l}
 \Bigg]  \Bigg[\frac{1}{2} {^{(3)} \nabla^H_{e_j}}  - \frac{\partial}{\partial r^j}    
 -  \frac{1}{12}  r^k r^m {^{(3)}R^l_{\; k m j}} \frac{\partial}{\partial r^l}
 \Bigg]\\  \pm N^-\Bigg[\frac{1}{2} {^{(3)} \nabla^H_{e_i}}  + \frac{\partial}{\partial r^i}    
 +  \frac{1}{12}  r^k r^m {^{(3)}R^l_{\; k m i}} \frac{\partial}{\partial r^l}
 \Bigg]   \Bigg[\frac{1}{2} {^{(3)} \nabla^H_{e_j}}  - \frac{\partial}{\partial r^j}    
 -  \frac{1}{12}  r^k r^m {^{(3)}R^l_{\; k m j}} \frac{\partial}{\partial r^l}
 \Bigg] \Big] \Big[ f^+ g^- \Big] \\
-\gamma^{ij} \Big( N^+ \pm N^- \Big) \Big[ \frac{1}{24}r^k r^m {^{(3)}R^l_{\; k m j;i}} \frac{\partial}{\partial r^l}    \Big] \Big[ f^+  g^-  \Big]\\
= \gamma^{ij}\Bigg[   \big( N^+ \mp N^- \big) \frac{\partial}{\partial r^i} {^{(3)} \nabla^H_{e_j}}  + \big( N^+ \pm N^- \big)\Big[ \frac{1}{4} {^{(3)} \nabla^H_{e_i}} {^{(3)} \nabla^H_{e_j}} + \frac{\partial}{\partial r^i}\frac{\partial}{\partial r^j}  \Big] \\
+\frac{1}{12} \big( N^+ \mp N^- \big) r^k r^m {^{(3)}R^l_{\; k m j; i}} \frac{\partial}{\partial r^l}
+\frac{1}{12} \big( N^+ \mp N^- \big) r^k r^m {^{(3)}R^l_{\; k m j}} \frac{\partial}{\partial r^l} {^{(3)} \nabla^H_{e_i}} \\
+\frac{1}{6} \big( N^+ \pm N^- \big) r^k r^m {^{(3)}R^l_{\; k m j}} \frac{\partial}{\partial r^l}\frac{\partial}{\partial r^i} \Bigg]\Big[ f^+  g^-  \Big]\\
+ \Big( N^+ \pm N^- \Big) \Big[ \frac{1}{12} r^m {^{(3)}R^l_{\;  m }} \frac{\partial}{\partial r^l}  \Big] \Big[ f^+  g^-  \Big]\, .
\end{multline}
Let us now include the commutator of the spatial Laplacian with the shift exponentials.  We have 
\begin{multline}
\Big[ {^{(3)}{\square}} \, , \exp \big[ \pm \frac{1}{2}  r^i {^{(3)}\nabla_{e_i}^H }  \big] \Big] f(x^{\mu}) \\ =
 - \sum_{n=1}^{\infty} \frac{1}{(\pm 2)^n} \Bigg[  r^{i_1} {^{(3)}\nabla_{e_{i_1}}^H } \, , ... \, , 
 \Big[ r^{i_n} {^{(3)}\nabla_{e_{i_n}}^H } \, ,\widetilde{\gamma}^{ij}{^{(3)}\nabla^H_{e_i}}{^{(3)}\nabla^H_{e_j}} \Big] \Bigg]\exp \big[ \pm \frac{1}{2} r^i {^{(3)}\nabla_{e_i}^H }  \big]f(x^{\mu}) 
 \\ =
 - {\gamma}^{ij}\sum_{n=1}^{\infty} \frac{r^{i_1}...r^{i_n}}{(\pm 2)^n n!} \Bigg[  {^{(3)}\nabla_{e_{i_1}}^H } \, , ... \, , 
 \Big[  {^{(3)}\nabla_{e_{i_n}}^H } \, ,{^{(3)}\nabla^H_{e_i}}{^{(3)}\nabla^H_{e_j}} \Big] \Bigg]\exp \big[ \pm \frac{1}{2} r^i {^{(3)}\nabla_{e_i}^H }  \big]f(x^{\mu}) 
 \\ =\pm \frac{1}{2}r^k \gamma^{ij} \Bigg[    r^m  {^{(3)}R^{l}_{\; mk j ; i}} \frac{\partial}{\partial r^l} +  {^{(3)}R^{l}_{\; i k j}}  {^{(3)}\nabla_{e_l}^H} +  2r^m  {^{(3)}R^{l}_{\; mk j }} \frac{\partial}{\partial r^l} {^{(3)}\nabla_{e_i}^H}  + \mathcal{O} \big( {^{(3)}R^2} \big)  \Bigg] \exp \big[ \pm \frac{1}{2} r^i {^{(3)}\nabla_{e_i}^H } \big] f(x^{\mu}) \\
  +  \frac{1}{8}r^k r^s \gamma^{ij} \Bigg[    {^{(3)}R^{l}_{\; i k j; s}}  {^{(3)}\nabla_{e_l}^H}+  2r^m  {^{(3)}R^{l}_{\; mk j;s }} \frac{\partial}{\partial r^l} {^{(3)}\nabla_{e_i}^H}   + \mathcal{O} \big( {^{(3)}R^2} \big) \Bigg] \exp \big[ \pm \frac{1}{2} r^i {^{(3)}\nabla_{e_i}^H } \big] f(x^{\mu})\, ,
\end{multline}
which will result in the following expression that we have to include in \eqref{lapExp2},
\begin{multline}
N^+\Bigg[  \frac{1}{2}r^k \gamma^{ij} \Bigg[    r^m  {^{(3)}R^{l}_{\; mk j ; i}} \frac{\partial}{\partial r^l} +  {^{(3)}R^{l}_{\; i k j}}  {^{(3)}\nabla_{e_l}^H} +  2r^m  {^{(3)}R^{l}_{\; mk j }} \frac{\partial}{\partial r^l} {^{(3)}\nabla_{e_i}^H} \Bigg]f^+ \\
  +  \frac{1}{8}r^k r^s \gamma^{ij} \Bigg[    {^{(3)}R^{l}_{\; i k j; s}}  {^{(3)}\nabla_{e_l}^H}  +  2r^m  {^{(3)}R^{l}_{\; mk j;s }} \frac{\partial}{\partial r^l} {^{(3)}\nabla_{e_i}^H} \Bigg] f^+ \Bigg] g^-  \\\mp N^-
  f^+ \Bigg[  \frac{1}{2}r^k \gamma^{ij} \Bigg[    r^m  {^{(3)}R^{l}_{\; mk j ; i}} \frac{\partial}{\partial r^l} +  {^{(3)}R^{l}_{\; i k j}}  {^{(3)}\nabla_{e_l}^H} +  2r^m  {^{(3)}R^{l}_{\; mk j }} \frac{\partial}{\partial r^l} {^{(3)}\nabla_{e_i}^H} \Bigg] g^-  \\
  -  \frac{1}{8}r^k r^s \gamma^{ij} \Bigg[    {^{(3)}R^{l}_{\; i k j; s}}  {^{(3)}\nabla_{e_l}^H}  +  2r^m  {^{(3)}R^{l}_{\; mk j;s }} \frac{\partial}{\partial r^l} {^{(3)}\nabla_{e_i}^H} \Bigg]  g^- \Bigg] 
  \\ =  \Bigg[ \frac{3}{8}\big( N^+ \mp N^- \big)r^k r^m \gamma^{ij}      {^{(3)}R^{l}_{\;  m k j;i}} \frac{\partial}{\partial r^l}+\frac{1}{8}\big( N^+ \mp N^- \big)r^s r^k r^m \gamma^{ij}      {^{(3)}R^{l}_{\;  m k j;s}} \frac{\partial}{\partial r^l}\frac{\partial}{\partial r^i}\\
+  \frac{1}{4}\big( N^+ \mp N^- \big)r^k \gamma^{ij}      {^{(3)}R^{l}_{\; i k j}}  {^{(3)}\nabla_{e_l}^H}+  \frac{1}{2}\big( N^+ \pm N^- \big)r^k \gamma^{ij}      {^{(3)}R^{l}_{\; i k j}}  \frac{\partial}{\partial r^l} \\
+  \frac{1}{4}\big( N^+ \mp N^- \big)r^k \gamma^{ij}   r^m   {^{(3)}R^{l}_{\; m k j}}   \Big[ \frac{\partial}{\partial r^l}{^{(3)}\nabla_{e_i}^H}+{^{(3)}\nabla_{e_l}^H} \frac{\partial}{\partial r^i} \Big]  \\
+\frac{1}{2}\big( N^+ \pm  N^- \big) r^k r^m \gamma^{ij} {^{(3)}R^{l}_{\; m k j}}  \frac{\partial}{\partial r^l} \frac{\partial}{\partial r^i} \Bigg] \big[f^+ g^- \big]\, .
\end{multline}
Combining all contributions and throwing away those terms that vanishing when integrating over the tangent space variable $r^k$, we end up with the following expression to second order in the spatial gradient expansion,
\begin{multline}
\Big( N^+ \big({^{(3)} \square}\big)^+_* \pm N^- \big({^{(3)} \square}\big)^-_*   \Big)\big[ f^+ g^- \big] \\=
\gamma^{ij}\Bigg[   \big( N^+ \mp N^- \big) \frac{\partial}{\partial r^i} {^{(3)} \nabla^H_{e_j}}  + \big( N^+ \pm N^- \big)\Big[ \frac{1}{4} {^{(3)} \nabla^H_{e_i}} {^{(3)} \nabla^H_{e_j}} + \frac{\partial}{\partial r^i}\frac{\partial}{\partial r^j}  \Big] \\
-\frac{7}{24} \big( N^+ \mp N^- \big) r^k r^m {^{(3)}R^l_{\; k m j; i}} \frac{\partial}{\partial r^l}
-\frac{1}{6} \big( N^+ \mp N^- \big) r^k r^m {^{(3)}R^l_{\; k m j}} \frac{\partial}{\partial r^l} {^{(3)} \nabla^H_{e_i}} \\
-\frac{1}{3} \big( N^+ \pm N^- \big) r^k r^m {^{(3)}R^l_{\; k m j}} \frac{\partial}{\partial r^l}\frac{\partial}{\partial r^i}
-\frac{5}{12} \big( N^+ \pm N^- \big) r^m {^{(3)}R^l_{\; i m j}} \frac{\partial}{\partial r^l} \\
-\frac{1}{8}\big( N^+ \mp N^- \big)r^s r^k r^m      {^{(3)}R^{l}_{\;  m k j;s}} \frac{\partial}{\partial r^l}\frac{\partial}{\partial r^i}  + \mathcal{O} \big( {^{(3)}R^2} \big) \Bigg]\Big[ f^+  g^-  \Big]\\
-\Bigg[ 
  \frac{1}{4}\big( N^+ \mp N^- \big)r^k \gamma^{ij}      {^{(3)}R^{l}_{\; i k j}}  {^{(3)}\nabla_{e_l}^H}\\
+  \frac{1}{4}\big( N^+ \mp N^- \big)r^k \gamma^{ij}   r^m   {^{(3)}R^{l}_{\; m k j}}   {^{(3)}\nabla_{e_l}^H} \frac{\partial}{\partial r^i}   + \mathcal{O} \big( {^{(3)}R^2} \big)  \Bigg] \big[f^+ g^- \big]\, .
\end{multline}
Performing the $r$-integral of the covariant Wigner transform yields the following expressions appearing in \eqref{F+BeforeInt} and \eqref{F-BeforeInt} to second order in the spatial gradient expansion and up to boundary terms,
\begin{multline}
\gamma^{1/2} \int_{\Sigma_t} d^3r \,  e^{- \frac{i}{\hbar} r^k p_k}\Big( N^+ \big({^{(3)} \square}\big)^+_* + N^- \big({^{(3)} \square}\big)^-_*   \Big)\big[ {: \hat{u}^+ \hat{u}^- :} \big] \\=\gamma^{1/2} \int_{\Sigma_t} d^3r \,  e^{- \frac{i}{\hbar} r^k p_k}
\gamma^{ij}\Bigg[   \big( r^k {^{(3)} \nabla_k N} \big) \frac{\partial}{\partial r^i} {^{(3)} \nabla^H_{e_j}}  + 2 N \Big[ \frac{1}{4} {^{(3)} \nabla^H_{e_i}} {^{(3)} \nabla^H_{e_j}} + \frac{\partial}{\partial r^i}\frac{\partial}{\partial r^j}  \Big] \\
+ \Big(\frac{r^k r^l}{4} {^{(3)} \nabla_k }{^{(3)} \nabla_l } N  \Big)  \frac{\partial}{\partial r^i}\frac{\partial}{\partial r^j}  
-\frac{2}{3} N r^k r^m {^{(3)}R^l_{\; k m j}} \frac{\partial}{\partial r^l}\frac{\partial}{\partial r^i}
- \frac{5}{6} N r^m {^{(3)}R^l_{\; i m j}} \frac{\partial}{\partial r^l} + \mathcal{O} \big( {^{(3)}R^2} \big)  \Bigg]\big[ {: \hat{u}^+ \hat{u}^- :} \big]\\
=
 \Bigg[ - p_i \big(  {^{(3)} \nabla_k N} \big)   \frac{\partial}{\partial p_k }   D^i  - \big( {^{(3)} \nabla_i N} \big)  D^i + 2 N \Big[ \frac{1}{4} D_i D^i - \gamma^{ij} \frac{p_i p_j}{\hbar^2}  \Big] \\
+ \gamma^{ij}  \frac{p_j p_j}{4 }\Big( {^{(3)} \nabla_k }{^{(3)} \nabla_l } N  \Big)  \frac{\partial}{\partial p_k}\frac{\partial}{\partial p_l} 
+  \frac{ p_i}{2 }\Big( {^{(3)} \nabla^i }{^{(3)} \nabla_k } N  \Big)  \frac{\partial}{\partial p_k}
+ \frac{1}{2 } \Big({^{(3)} \nabla^i }{^{(3)} \nabla_i } N  \Big) \\
-\frac{2}{3} N p_l p_i \gamma^{ij}   {^{(3)}R^l_{\; k m j}} \frac{\partial}{\partial p_k}\frac{\partial}{\partial p_m}  -\frac{1}{6} p_i N {^{(3)}R^i_{\; k}} \frac{\partial}{\partial p_k} + \frac{1}{6} N {^{(3)}R} + \mathcal{O} \big( \hbar^4 \big)  \Bigg]\hat{F}_{\phi \phi}  \, ,
\end{multline}
\begin{multline}
\gamma^{1/2} \int_{\Sigma_t} d^3r \,  e^{- \frac{i}{\hbar} r^k p_k}\Big( N^+ \big({^{(3)} \square}\big)^+_* - N^- \big({^{(3)} \square}\big)^-_*   \Big)\big[ {: \hat{u}^+ \hat{u}^- :} \big] \\=\gamma^{1/2} \int_{\Sigma_t} d^3r \,  e^{- \frac{i}{\hbar} r^k p_k}
\gamma^{ij}\Bigg[   \big( 2N + \frac{r^k r^l}{4} {^{(3)} \nabla_k }{^{(3)} \nabla_l  } N   \big) \frac{\partial}{\partial r^i} {^{(3)} \nabla^H_{e_j}}  + \big( r^k{^{(3)} \nabla_k } N  \big)\Big[ \frac{1}{4} {^{(3)} \nabla^H_{e_i}} {^{(3)} \nabla^H_{e_j}} + \frac{\partial}{\partial r^i}\frac{\partial}{\partial r^j}  \Big] \\
 + \frac{1}{24}\big( r^k r^l r^m {^{(3)} \nabla_k }{^{(3)} \nabla_l } {^{(3)} \nabla_m } N \big)\frac{\partial}{\partial r^i}\frac{\partial}{\partial r^j}  
-\frac{7}{12} N r^k r^m {^{(3)}R^l_{\; k m j; i}} \frac{\partial}{\partial r^l}
-\frac{1}{3}N r^k r^m {^{(3)}R^l_{\; k m j}} \frac{\partial}{\partial r^l} {^{(3)} \nabla^H_{e_i}} \\
-\frac{1}{3} \big( r^s {^{(3)} \nabla_s }N  \big) r^k r^m {^{(3)}R^l_{\; k m j}} \frac{\partial}{\partial r^l}\frac{\partial}{\partial r^i}
-\frac{5}{12}\big( r^s {^{(3)} \nabla_s }N  \big)  r^m {^{(3)}R^l_{\; i m j}} \frac{\partial}{\partial r^l} \\
-\frac{1}{4} N r^s r^k r^m      {^{(3)}R^{l}_{\;  m k j;s}} \frac{\partial}{\partial r^l}\frac{\partial}{\partial r^i} + \mathcal{O} \big( {^{(3)}R^2} \big)\Bigg]\big[ {: \hat{u}^+ \hat{u}^- :} \big]\\
-\Bigg[ 
+  \frac{1}{2} N r^k \gamma^{ij}      {^{(3)}R^{l}_{\; i k j}}  {^{(3)}\nabla_{e_l}^H}
+  \frac{1}{2} N r^k \gamma^{ij}   r^m   {^{(3)}R^{l}_{\; m k j}}   {^{(3)}\nabla_{e_l}^H} \frac{\partial}{\partial r^i}  + \mathcal{O} \big( {^{(3)}R^2} \big) \Bigg] \big[ {: \hat{u}^+ \hat{u}^- :} \big]
\\ =  \Bigg[
 \frac{i}{\hbar}\Big( 2N  p_j - \frac{\hbar^2}{2} \big( {^{(3)} \nabla_k }{^{(3)} \nabla^j  } N   \big) \frac{\partial}{\partial p_k}     -  \frac{\hbar^2}{4} p_j  \big( {^{(3)} \nabla_k }{^{(3)} \nabla_l  } N   \big) \frac{\partial^2}{\partial p_k \partial p_l} \Big)  D^j   \\
+ \frac{i}{\hbar} \big({^{(3)} \nabla_k }N  \big)  \Big( \frac{\hbar^2}{4} D^j D_j - \gamma^{jl} p_l p_j \Big)\frac{\partial }{\partial p_k} - 2
\frac{i}{\hbar} p_k \big({^{(3)} \nabla^k }N  \big)  \\
+ \frac{i \hbar}{24} p_r p_s \gamma^{rs}  \big( {^{(3)} \nabla_k }{^{(3)} \nabla_l } {^{(3)} \nabla_m } N \big)
 \frac{\partial^3}{\partial p_k \partial p_l \partial p_m }  
 + \frac{i \hbar}{8}  p_j \big( {^{(3)} \nabla_k }{^{(3)} \nabla_l } {^{(3)} \nabla^j} N \big)
 \frac{\partial^2}{\partial p_k \partial p_l }  \\
  + \frac{i \hbar}{4} \big( {^{(3)} \nabla_k }{^{(3)} \nabla_j } {^{(3)} \nabla^j} N \big)
 \frac{\partial}{\partial p_k  } 
  +  \frac{i \hbar}{12} {^{(3)} R^l_{ \;  k }} \big( {^{(3)} \nabla_l} N \big) \frac{\partial}{\partial p_k  }  \\
 + i\hbar N \Big(\frac{1}{12} p_j \gamma^{jl}  {^{(3)} R^s_{\; k m l; s}} \frac{\partial^2}{\partial p_k \partial p_m} - \frac{1}{3}  {^{(3)} R^j_{\; k ; j }} \frac{\partial}{\partial p_k} - \frac{1}{4} p_i p_l \gamma^{ij} {^{(3)} R^l_{\; m k  j; s }} \frac{\partial^3}{\partial p_s \partial p_k \partial p_m}  + \frac{1}{4} p_j  {^{(3)} R^j_{\; m ; s }}  \frac{\partial^2}{\partial p_s \partial p_m} \Big) \\
 + i \hbar N \Big( \frac{5}{6} p_l  {^{(3)} R^l_{\;  k m j }} \frac{\partial^2}{\partial p_k \partial p_m} D^j - \frac{1}{3}  {^{(3)} R^j_{\;  k  }} \frac{\partial}{\partial p_k } D_j  \Big) \\
 + i \hbar  \Big( - \frac{1}{3} \gamma^{ij} p_i p_l {^{(3)} \nabla^s} N{^{(3)} R^l_{\;  k m j  }} \frac{\partial^3}{\partial p_s \partial p_k \partial p_m} +
 \frac{5}{12}{^{(3)} \nabla_j} N {^{(3)} R^j_{\;  k  }} \frac{\partial}{\partial p_k} 
 + \frac{1}{12}{^{(3)} \nabla^j} N {^{(3)} R} \frac{\partial}{\partial p_j}  \Big)\\
 + i \hbar  \Big( - \frac{5}{8} \gamma^{ij}  p_l {^{(3)} \nabla_i} N  {^{(3)} R^l_{\;  k m j }}  \frac{\partial^2}{\partial p_k \partial p_m} + \frac{5}{12} p_j {^{(3)} \nabla_s} N{^{(3)} R^j_{ \; k }}  \frac{\partial^2}{\partial p_k \partial p_s}  \Big) + \mathcal{O} \big( \hbar^4 \big) 
 \Bigg] \hat{F}_{\phi \phi} \, .
\end{multline}
\subsection{Dynamics of Wigner transformed two-point functions $F_{\phi \phi}$, $F_{\phi \Pi}$, $F_{\Pi \phi}$ and $F_{\Pi \Pi}$ \label{dynWig}}
Performing the $r$-integral in \eqref{F00BeforeInt} to \eqref{F11BeforeInt} leads to the following equations in phase-space to next-to-leading order in the spatial gradient expansion where we also dropped boundary terms, anomalous contributions and simplified to self-interaction terms by assuming a Gaussian state truncation with vanishing one-point functions, 
\begin{multline}
 \partial_t {F}_{\phi \phi} 
=   \Big[ N - \frac{\hbar^2}{8}  N_{; k l} \frac{\partial^2}{\partial p_k \partial p_l}  \Big] \Big[ {F}_{\Pi \phi} + {F}_{\phi \Pi } \Big] 
- \frac{ \hbar}{2i} \Big[  N_{;k} \frac{\partial}{\partial p_k}  - \frac{\hbar^2}{24}N_{; klm} \frac{\partial^3}{\partial p_k \partial p_l \partial p_m}  \Big] \Big[ {F}_{\Pi \phi} - {F}_{\phi \Pi } \Big]\\
+ \Bigg[N^k  D_k    -
  p_k N^k_{\; ; m} \frac{\partial}{\partial p_m}     - NK 
  - \frac{\hbar^2 }{12} N^k_{\; ; k q m}  \frac{\partial^2}{\partial p_q \partial p_m}  
  +    \frac{\hbar^2}{4}\Bigg( \frac{1}{2} \big[N K_{ik} \big]_{;l} - \big[N K_{k l} \big]_{;i}  \Bigg) \frac{\partial^2}{\partial p_k \partial p_i} D^l   
\\ - \frac{\hbar^2}{12} \Bigg( \frac{1}{2}  \big[N K_{ik} \big]_{; \; \; l}^{ \; l} - \big[N K_{k}^{\; l} \big]_{; i l}  -  \big[N K \big]_{; ki}  \Bigg) \frac{\partial^2}{\partial p_k \partial p_i}
- \frac{\hbar^2}{12} p_l \Bigg( \frac{1}{2} \big[N K_{ik} \big]_{; \; \; s}^{\; l} -  \big[N K_{k}^{\; l} \big]_{; i s}\Bigg)\frac{\partial^3}{\partial p_k \partial p_i \partial p_s}\Bigg]  {F}_{\phi \phi}  \, .
\end{multline}
\begin{multline}
\frac{1}{2}\partial_t \big( {F}_{\Pi \phi} +{F}_{\phi \Pi} \big)
=   \Big[ N - \frac{\hbar^2}{8}  N_{; k l} \frac{\partial^2}{\partial p_k \partial p_l}  \Big] {F}_{\Pi \Pi } \\
+\frac{1}{2} \Bigg[N^k  D_k    -
  p_k N^k_{\; ; m} \frac{\partial}{\partial p_m}     
  - \frac{\hbar^2 }{12} N^k_{\; ; k q m}  \frac{\partial^2}{\partial p_q \partial p_m}  
  +    \frac{\hbar^2}{4}\Bigg( \frac{1}{2} \big[N K_{ik} \big]_{;l} - \big[N K_{k l} \big]_{;i}  \Bigg) \frac{\partial^2}{\partial p_k \partial p_i} D^l  
\\ - \frac{\hbar^2}{12} \Bigg( \frac{1}{2}  \big[N K_{ik} \big]_{; \; \; l}^{ \; l} - \big[N K_{k}^{\; l} \big]_{; i l}  + \frac{1}{2} \big[N K \big]_{; ki}  \Bigg) \frac{\partial^2}{\partial p_k \partial p_i} \\
- \frac{\hbar^2}{12} p_l \Bigg( \frac{1}{2} \big[N K_{ik} \big]_{; \; \; s}^{\; l} -  \big[N K_{k}^{\; l} \big]_{; i s}\Bigg)\frac{\partial^3}{\partial p_k \partial p_i \partial p_s}\Bigg] \big( {F}_{\Pi \phi} +{F}_{\phi \Pi} \big)
 \\
-\frac{\hbar}{4i}\Big[  \big(NK\big)_{;j} \frac{\partial}{\partial p_j}  - \frac{\hbar^2}{24}\big(NK\big)_{; jlm} \frac{\partial^3}{\partial p_j \partial p_l \partial p_m}  \Big]\big( {F}_{\Pi \phi}-{F}_{\phi \Pi} \big)
\\
-\Bigg[  \Big[\frac{m^2}{\hbar^2}+\gamma^{kj} \frac{p_k p_j}{\hbar^2}\Big] \Big[ N - \frac{\hbar^2}{8}  N_{; k l} \frac{\partial^2}{\partial p_k \partial p_l}  \Big]+\xi  NR 
\\+\frac{1}{2} \frac{{\lambda}}{\hbar} \Big[ N \int \frac{d^3 q }{\gamma^{1/2}} {F}_{\phi \phi} (q) - \frac{\hbar^2}{8} \big[ N \int \frac{d^3 q }{\gamma^{1/2}} {F}_{\phi \phi} (q)\big]_{; k l} \frac{\partial^2}{\partial p_k \partial p_l}  \Big] - \frac{1}{2} p_j  N_{;k}  \frac{\partial}{\partial p_k }   D^j   -  \frac{1}{4}N D_j D^j    \\
+\frac{1}{3} N p_l p_i \gamma^{ij}   {^{(3)}R^l_{\; k m j}} \frac{\partial}{\partial p_k}\frac{\partial}{\partial p_m}  +\frac{1}{12}N p_i {^{(3)}R^i_{\; k}} \frac{\partial}{\partial p_k} - \frac{1}{6} N {^{(3)}R} \Bigg]{F}_{\phi \phi}  \, .
\end{multline}
\begin{multline}
\frac{i}{2} \partial_t \big( {F}_{\Pi \phi} -{F}_{\phi \Pi} \big) 
= 
\frac{ \hbar}{2} \Big[  N_{;k} \frac{\partial}{\partial p_k}  - \frac{\hbar^2}{24}N_{; klm} \frac{\partial^3}{\partial p_k \partial p_l \partial p_m}  \Big] {F}_{\Pi \Pi } \\
+\frac{i}{2} \Bigg[N^k  D_k    -
  p_k N^k_{\; ; m} \frac{\partial}{\partial p_m}     
  - \frac{\hbar^2 }{12} N^k_{\; ; k q m}  \frac{\partial^2}{\partial p_q \partial p_m}  
  +    \frac{\hbar^2}{4}\Bigg( \frac{1}{2} \big[N K_{ik} \big]_{;l} - \big[N K_{k l} \big]_{;i}  \Bigg) \frac{\partial^2}{\partial p_k \partial p_i} D^l 
\\ - \frac{\hbar^2}{12} \Bigg( \frac{1}{2}  \big[N K_{ik} \big]_{; \; \; l}^{ \; l} - \big[N K_{k}^{\; l} \big]_{; i l}  + \frac{1}{2}  \big[N K \big]_{; ki}  \Bigg) \frac{\partial^2}{\partial p_k \partial p_i} \\
- \frac{\hbar^2}{12} p_l \Bigg( \frac{1}{2} \big[N K_{ik} \big]_{; \; \; s}^{\; l} -  \big[N K_{k}^{\; l} \big]_{; i s}\Bigg)\frac{\partial^3}{\partial p_k \partial p_i \partial p_s}\Bigg] \big( {F}_{\Pi \phi} -{F}_{\phi \Pi} \big)
 \\
-\frac{\hbar}{4}\Big[  \big(NK\big)_{;j} \frac{\partial}{\partial p_j}  - \frac{\hbar^2}{24}\big(NK\big)_{; jlm} \frac{\partial^3}{\partial p_j \partial p_l \partial p_m}  \Big]\big( {F}_{\Pi \phi}+{F}_{\phi \Pi} \big)\\
-\frac{1}{2}\Bigg[
 \frac{1}{\hbar}\Big( 2N  p_j    -  \frac{\hbar^2}{4} p_j  N_{; l k} \frac{\partial^2}{\partial p_k \partial p_l} \Big)  D^j  + \frac{1}{\hbar} \big({^{(3)} \nabla_k }N  \big)  \Big( \frac{\hbar^2}{4} D^j D_j \Big)\frac{\partial }{\partial p_k} \\
- \frac{1}{ \hbar} \Big[ m^2+  p_r p_s \gamma^{rs}\Big]\Big[N_{;k} \frac{\partial}{\partial p_k} - \frac{\hbar^2}{24}  N_{;mlk} \frac{\partial^3}{\partial p_k \partial p_l \partial p_m }  \Big]- \hbar \xi (NR)_{;k} \frac{\partial}{\partial p_k}  \\
-\frac{1}{2} \frac{{\lambda}}{\hbar} \Big[ \hbar \big[N \int \frac{d^3 q }{\gamma^{1/2}} {F}_{\phi \phi} (q)\big]_{;k} \frac{\partial}{\partial p_k}  - \frac{\hbar^3}{24} \big[ N \int \frac{d^3 q }{\gamma^{1/2}} {F}_{\phi \phi} (q)\big]_{;mlk} \frac{\partial^3}{\partial p_k \partial p_l \partial p_m } \Big]\\
 + \hbar N \Big(\frac{1}{12} p_j \gamma^{jl}  {^{(3)} R^s_{\; k m l; s}} \frac{\partial^2}{\partial p_k \partial p_m} - \frac{1}{3}  {^{(3)} R^j_{\; k ; j }} \frac{\partial}{\partial p_k} - \frac{1}{4} p_i p_l \gamma^{ij} {^{(3)} R^l_{\; m k  j; s }} \frac{\partial^3}{\partial p_s \partial p_k \partial p_m}  + \frac{1}{4} p_j  {^{(3)} R^j_{\; m ; s }}  \frac{\partial^2}{\partial p_s \partial p_m} \Big) \\
 +  \hbar N \Big( \frac{5}{6} p_l  {^{(3)} R^l_{\;  k m j }} \frac{\partial^2}{\partial p_k \partial p_m} D^j - \frac{1}{3}  {^{(3)} R^j_{\;  k  }} \frac{\partial}{\partial p_k } D_j  \Big) \\
 +  \hbar  \Big( - \frac{1}{3} \gamma^{ij} p_i p_l N_{;}^{\; s}{^{(3)} R^l_{\;  k m j  }} \frac{\partial^3}{\partial p_s \partial p_k \partial p_m}  
 + \frac{1}{12} N_{;}^{\; j} {^{(3)} R} \frac{\partial}{\partial p_j}  \Big) +  \frac{ \hbar}{12} {^{(3)} R^l_{ \;  k }} N_{;l} \frac{\partial}{\partial p_k  }  \\
 +  \hbar  \Big(  + \frac{5}{12} p_j N_{; s}{^{(3)} R^j_{ \; k }}  \frac{\partial^2}{\partial p_k \partial p_s}  \Big) - \frac{\hbar}{8} p_k   N^{\; k}_{; \; \;  l m} \frac{\partial^2}{\partial p_l \partial p_m} - \frac{\hbar}{4} N_{; \; \; k m}^{\; k}  \frac{\partial}{\partial p_m} \\- \frac{\hbar}{4} {^{(3)}R^{m}_{\;\; l}} N_{; m}  \frac{\partial}{\partial p_l}
  + \hbar \frac{13}{48} p_l{^{(3)}R^{l}_{\; k m s}}N_{;}^{\; s}  \frac{\partial^2}{\partial p_k \partial p_m}
 \Bigg] {F}_{\phi \phi}  \, .
\end{multline}
\begin{multline}
 \partial_t  {F}_{\Pi \Pi}
=\Bigg[N^k  D_k +NK   -
  p_k N^k_{\; ; m} \frac{\partial}{\partial p_m}     
  - \frac{\hbar^2 }{12} N^k_{\; ; k q m}  \frac{\partial^2}{\partial p_q \partial p_m}  
 \\ +    \frac{\hbar^2}{4}\Bigg( \frac{1}{2} \big[N K_{ik} \big]_{;l} - \big[N K_{k l} \big]_{;i}  \Bigg) \frac{\partial^2}{\partial p_k \partial p_i} D^l   - \frac{\hbar^2}{12} \Bigg( \frac{1}{2}  \big[N K_{ik} \big]_{; \; \; l}^{ \; l} - \big[N K_{k}^{\; l} \big]_{; i l}  +2  \big[N K \big]_{; ki}  \Bigg) \frac{\partial^2}{\partial p_k \partial p_i} \\
- \frac{\hbar^2}{12} p_l \Bigg( \frac{1}{2} \big[N K_{ik} \big]_{; \; \; s}^{\; l} -  \big[N K_{k}^{\; l} \big]_{; i s}\Bigg)\frac{\partial^3}{\partial p_k \partial p_i \partial p_s}\Bigg] {F}_{\Pi \Pi}\\
-\frac{i}{2}\Bigg[
 \frac{1}{\hbar}\Big( 2N  p_j    -  \frac{\hbar^2}{4} p_j  N_{; l k} \frac{\partial^2}{\partial p_k \partial p_l} \Big)  D^j  + \frac{1}{\hbar} \big({^{(3)} \nabla_k }N  \big)  \Big( \frac{\hbar^2}{4} D^j D_j \Big)\frac{\partial }{\partial p_k}  \\
- \frac{1}{ \hbar} \Big[ m^2+  p_r p_s \gamma^{rs}\Big]\Big[N_{;k} \frac{\partial}{\partial p_k} - \frac{\hbar^2}{24}  N_{;mlk} \frac{\partial^3}{\partial p_k \partial p_l \partial p_m }  \Big] - { \hbar} \xi  (NR)_{;k} \frac{\partial}{\partial p_k}  \\
-\frac{1}{2} \frac{{\lambda}}{\hbar} \Big[ \hbar \big[N \int \frac{d^3 q }{\gamma^{1/2}} {F}_{\phi \phi} (q)\big]_{;k} \frac{\partial}{\partial p_k}  - \frac{\hbar^3}{24} \big[ N \int \frac{d^3 q }{\gamma^{1/2}} {F}_{\phi \phi} (q)\big]_{;mlk} \frac{\partial^3}{\partial p_k \partial p_l \partial p_m } \Big]\\
 + \hbar N \Big(\frac{1}{12} p_j \gamma^{jl}  {^{(3)} R^s_{\; k m l; s}} \frac{\partial^2}{\partial p_k \partial p_m} - \frac{1}{3}  {^{(3)} R^j_{\; k ; j }} \frac{\partial}{\partial p_k} \\ - \frac{1}{4} p_i p_l \gamma^{ij} {^{(3)} R^l_{\; m k  j; s }} \frac{\partial^3}{\partial p_s \partial p_k \partial p_m}  + \frac{1}{4} p_j  {^{(3)} R^j_{\; m ; s }}  \frac{\partial^2}{\partial p_s \partial p_m} \Big) \\
 +  \hbar N \Big( \frac{5}{6} p_l  {^{(3)} R^l_{\;  k m j }} \frac{\partial^2}{\partial p_k \partial p_m} D^j - \frac{1}{3}  {^{(3)} R^j_{\;  k  }} \frac{\partial}{\partial p_k } D_j  \Big) \\
 +  \hbar  \Big( - \frac{1}{3} \gamma^{ij} p_i p_l N_{;}^{\; s}{^{(3)} R^l_{\;  k m j  }} \frac{\partial^3}{\partial p_s \partial p_k \partial p_m}  
 + \frac{1}{12} N_{;}^{\; j} {^{(3)} R} \frac{\partial}{\partial p_j}  \Big) +  \frac{ \hbar}{12} {^{(3)} R^l_{ \;  k }} N_{;l} \frac{\partial}{\partial p_k  }  \\
 +  \hbar  \Big(  + \frac{5}{12} p_j N_{; s}{^{(3)} R^j_{ \; k }}  \frac{\partial^2}{\partial p_k \partial p_s}  \Big) - \frac{\hbar}{8} p_k   N^{\; k}_{; \; \;  l m} \frac{\partial^2}{\partial p_l \partial p_m} - \frac{\hbar}{4} N_{; \; \; k m}^{\; k}  \frac{\partial}{\partial p_m} \\- \frac{\hbar}{4} {^{(3)}R^{m}_{\;\; l}} N_{; m}  \frac{\partial}{\partial p_l}
  + \hbar \frac{13}{48} p_l{^{(3)}R^{l}_{\; k m s}}N_{;}^{\; s}  \frac{\partial^2}{\partial p_k \partial p_m}
 \Bigg]\Big[ \hat{F}_{\Pi \phi} -  \hat{F}_{\phi \Pi}\Big]  \\
-\Bigg[  \Big[\frac{m^2}{\hbar^2}+\gamma^{kj} \frac{p_k p_j}{\hbar^2}  \Big] \Big[ N - \frac{\hbar^2}{8}  N_{; k l} \frac{\partial^2}{\partial p_k \partial p_l}  \Big]+ \xi NR \\
+\frac{1}{2} \frac{{\lambda}}{\hbar} \Big[  N \int \frac{d^3 q }{\gamma^{1/2}} {F}_{\phi \phi} (q) - \frac{\hbar^2}{8} \big[ N \int \frac{d^3 q }{\gamma^{1/2}} {F}_{\phi \phi} (q)\big]_{; k l} \frac{\partial^2}{\partial p_k \partial p_l}  \Big]\\ - \frac{1}{2} p_j  N_{;k}  \frac{\partial}{\partial p_k }   D^j   -  \frac{1}{4}N D_j D^j    \\
+\frac{1}{3} N p_l p_i \gamma^{ij}   {^{(3)}R^l_{\; k m j}} \frac{\partial}{\partial p_k}\frac{\partial}{\partial p_m}  +\frac{1}{12}N p_i {^{(3)}R^i_{\; k}} \frac{\partial}{\partial p_k} - \frac{1}{6} N {^{(3)}R} \Bigg] \Big[ {F}_{\Pi \phi} +  {F}_{\phi \Pi}\Big]  \, .
\end{multline}
\bibliographystyle{JHEP}
\bibliography{Biblio}
\nocite{*}
\end{document}